\documentclass{article}

\usepackage{microtype}
\usepackage{graphicx}
\usepackage{booktabs} 
\usepackage{hyperref}

\usepackage[accepted]{icml2024}
\usepackage{amsmath}
\usepackage{amssymb}
\usepackage{multirow}
\usepackage{multicol}
\usepackage{mathtools}
\usepackage{amsthm}
\usepackage[capitalize,noabbrev]{cleveref}
\usepackage{bbm}
\newcommand{\X}{{\mathcal X}}
\usepackage{amsmath}
\usepackage[hang,flushmargin]{footmisc}
\usepackage{enumitem,bbm}
\usepackage{subfig}
\def\balign#1\ealign{\begin{align}#1\end{align}}
\def\baligns#1\ealigns{\begin{align*}#1\end{align*}}
\def\balignat#1\ealign{\begin{alignat}#1\end{alignat}}
\def\balignats#1\ealigns{\begin{alignat*}#1\end{alignat*}}
\newenvironment{talign*}
 {\csname align*\endcsname}
 {\endalign}
\newenvironment{talign}
 {\csname align\endcsname}
 {\endalign}

\def\balignst#1\ealignst{\begin{talign*}#1\end{talign*}}
\def\balignt#1\ealignt{\begin{talign}#1\end{talign}}
\newcommand\setItemnumber[1]{\setcounter{enumi}{\numexpr#1-1\relax}}

\theoremstyle{plain}
\newtheorem{theorem}{Theorem}[section]

\theoremstyle{definition}
\newtheorem{definition}[theorem]{Definition}

\newtheorem{example}[theorem]{Example}
\theoremstyle{remark}

\AtBeginDocument{}
\newcommand{\BF}{{BF}}

\usepackage{dsfont}
\usepackage{float}

\icmltitlerunning{Scarce Resource Allocations That Rely On Machine Learning Should Be Randomized}
\begin{document}

\twocolumn[
\icmltitle{Scarce Resource Allocations That Rely On Machine Learning\\Should Be Randomized}
\begin{icmlauthorlist}
\icmlauthor{Shomik Jain}{mit_idss}
\icmlauthor{Kathleen Creel}{northeastern}
\icmlauthor{Ashia Wilson}{mit_idss,mit_eecs}
\end{icmlauthorlist}

\icmlaffiliation{mit_idss}{Institute for Data, Systems, and Society, MIT}
\icmlaffiliation{northeastern}{Department of Philosophy \& Religion and Khoury College of Computer Sciences, Northeastern University}
\icmlaffiliation{mit_eecs}{Department of Electrical Engineering and Computer Science, MIT}

\icmlcorrespondingauthor{}{shomikj@mit.edu}

\icmlkeywords{fairness, randomness, systemic exclusion}

\vskip 0.3in
]

\printAffiliationsAndNotice{}

\begin{abstract}
Contrary to traditional deterministic notions of algorithmic fairness, this paper argues that fairly allocating scarce resources using machine learning often requires randomness. We address \textit{why}, \textit{when}, and {\em how} to randomize by proposing stochastic procedures that more adequately account for all of the claims that individuals have to allocations of social goods or opportunities.
\end{abstract}


\section{Introduction}\label{sec:intro}
Sometimes resources or opportunities are scarce: jobs, welfare benefits, or life-saving medicines cannot be divided among all those who deserve them.  Worse yet, it is often unclear which individuals are most deserving. Perhaps they all are. Decision-makers hope to use algorithmic systems to allocate scarce resources and goods fairly. But without careful attention, it is easy for algorithms to replicate or amplify the biases and inequalities in their training data.  

The fair machine learning community has developed sophisticated theoretical and formal tools to reduce algorithmic bias, increase fairness, and promote justice. However, these tools are almost exclusively deterministic. For example, employers with more qualified applicants than job openings often rely on hiring algorithms to screen applicants for interviews~\citep{raghavan2020hiring}. These algorithms assign a score or ranking to candidates. Employers then threshold these scores or rankings to deterministically pick candidates to interview. Similarly, healthcare providers often have a limited supply of life-saving medical resources such as ventilators, therapeutics, or organs. Patients are often triaged 
based on algorithms that predict their survival rate or life expectancy post-treatment~\citep{chin2023medicine}. Most existing work on algorithmic fairness relies on deterministic algorithms to incorporate fairness. Once algorithmic bias has been reduced to the extent possible, the algorithm allocates resources to the top candidate(s). If Alice is the top-ranked candidate for every job or has the most expected quality-adjusted life-years, she should deterministically receive the job offer or organ every time.

Recent works on arbitrariness and fairness suggest that even \textit{counterfactual} non-determinism can be unfair. If there exist many possible models with similar predictive performance but slightly different decisions on individuals, a state of affairs called ``predictive multiplicity''~\citep{marx2020predictive} or ``model multiplicity''~\citep{black2022model}, it is unfair to naively pick one of the models for our decision-making algorithm~\citep{hsu2022rashomon}. Instead, we should reduce multiplicity by altering the training process to reduce the variance that leads to diverging predictions~\citep{cooper2023variance}, especially on under-represented individuals~\citep{ganesh2023impact}, iterate until predictions agree about individuals~\citep{roth2023reconciling} or even abstain from making predictions on some people altogether~\cite{cooper2023variance}.

While sharing the goal of reducing bias and increasing fairness, this work argues that the fair machine learning community has underutilized non-determinism and randomization as tools to achieve fairness. In some settings that involve algorithmic decision-making, we contend that non-determinism is \textit{required} for fair outcomes. In what follows, we first motivate \textit{why} and \textit{when} fairness requires randomization. We adopt philosopher John Broome's concept of the value of lotteries in fairness to argue that randomization is needed in scarce resource settings to respect the \textit{claims} that individuals have to resources, even if they do not receive them, by giving each person with a claim a chance.  

Second, we argue that because algorithmic predictions involve uncertainty, it is unfair to those on whom we make mistakes to deterministically commit to those mistakes. This is especially true in multi-shot contexts in which each individual is affected by multiple decision-makers or a series of decisions over time. 
Across an ecosystem of multiple decision-makers, consistently allocating scarce resources and goods to the \textit{same} candidate(s) can be sub-optimal, as it prevents the decision-makers from learning~\citep{kleinberg2021monoculture, peng2023monoculture}, or unfair, as it means that many decision-makers reject or make mistakes on the same individuals~\citep{ajunwa2021, creel_hellman_2022, bommasani2022homogenization, toups2023ecosystem, jain2023pluralism}. And when individuals receive a series of decisions over time, biased early allocations often affect the data available for the next decision-maker.  An initial negative judgement leads to a series of further negative judgements, forming a ``patterned inequality'' or compounding injustice~\citep{eidelson2021patterned, hellman2018compounding}. For these reasons, deterministic judgements under uncertainty can reinforce structural injustices.  Therefore, we take the position that \textbf{many scarce resource allocations that rely on machine learning\footnote{We consider any data-driven algorithmic decision-making process to be under the umbrella of the term ``machine learning''.} should be randomized}.

Having motivated randomizing allocations~(\autoref{sec:why_randomize}), we then formalize \textit{how} to randomize to bring about more fair outcomes.  We consider two settings: when claims are known  (\autoref{sec:claims_known}), and when claims are uncertain (\autoref{sec:claims_unknown}). Finally, we discuss why existing deterministic methods may not be enough to achieve both fair and efficient allocations (\autoref{sec:discussion}).

\section{Related Work}

The idea that randomizing decisions  might promote fairness when resources are scarce is not new. Lotteries have been used to admit students to some public schools~\citep{ hastings2006} and medical schools~\citep{CohenSchotanus2006}. In healthcare, lotteries were also used to allocate COVID-19 treatments~\citep{McCreary2023}.


We build on work that extends this concept to machine learning and advocates for using randomness to increase fairness in algorithmic decision-making. For example, concerned with decision quality and the loss of diversity in the decision-making process that comes from relying on machine learning, \citet{grgic2017ensemble} propose randomizing among models in classifier ensembles obtained by retraining multiple times. We second their concern, connecting it to the growing literature on the homogenization of outcomes that results from automated decision-making~\cite{kleinberg2021monoculture, ajunwa2021, creel_hellman_2022, bommasani2022homogenization, toups2023ecosystem, jain2023pluralism} and from the use of foundation models~\citep{bommasani2021foundation}. We extend the idea of randomizing among models with similar performance to the setting of fair allocations based on claims. 

\citet{agarwal2022power} are concerned with the loss of accuracy that results when we impose fairness constraints and introduce a randomized framework for classification to address this concern. We follow them in demonstrating that many randomization procedures have minimal impact on accuracy while improving fairness. Furthermore, \citet{singh2021fairness} argue that evaluating candidate merit without accounting for uncertainty is unfair in rankings. To address this unfairness, they introduce the notion of a ``posterior merit distribution,'' suggesting that goods should be allocated based on the probability that an individual is among the top $k$ candidates. We agree that fairness sometimes requires quantifying uncertainty and extend the argument to specify \textit{when} and \textit{how} to incorporate randomness based on uncertainty.   

The reinforcement learning literature also considers randomization under uncertainty in multi-shot contexts. However, these works focus on stochastic methods that can help a single decision-maker learn and improve their own utility over time~\citep{agrawal2012analysis, joseph2016fairness, li2020hiring}. We consider the broader multi-shot context that may involve multiple decision-makers making any number of allocations, and center fairness and individual claims as our motivation for randomness. Past work also centers individual fairness under uncertainty \citep{Roth2016} and the distribution of errors across individuals \citep{Roth2019}.



\section{Why and When To Randomize Allocations}\label{sec:why_randomize}
In this section, we motivate randomization by arguing that (weighted) lotteries better achieve fairness than deterministic allocation algorithms in certain settings. We give two reasons.  First, relying on John Broome's characterization of fairness \citep{broome1990fairness}, we show that when more individuals deserve a good than can receive it, the best way to respect each person's \textit{claim} to that good is to give them a chance to receive it by holding a lottery.

Second, deterministic decision-making over-represents the certainty of our predictions. 
Many ML-based social allocation problems have uncertain parameters. We cannot be sure that we have formulated the problem-to-be-solved well, that we have chosen appropriate variables or parameters, or that our data is accurate~\citep{passi2019problem}.  Indeed, in many cases we suspect that our problem formulation, variable choice, and data gathering all may have been systematically skewed by social conditions.
Our uncertainty leads us to underestimate the claims of some to the good and overestimate the claims of others, which has moral implications in situations when not all claims can be satisfied. 
\subsection{Fairness and Individual Claims}

To motivate \textit{why} lotteries are sometimes needed to ensure fairness, we turn to John Broome's influential theory of fairness~\citep{broome1990fairness}. Broome's theory is based on the moral concept of a ``claim.'' An individual has a claim to a good, resource, or opportunity when she is owed it for reasons of fairness~\citep[p.96]{broome1990fairness}.  For example, if a kidney is being allocated and there are two individuals who have been on the waiting list for an equal amount of time and are in all other respects equivalent, both individuals have a \textit{claim} on the kidney.

Some claims stem from \textit{desert}: the person deserves the good~\citep[p.93]{broome1990fairness}. All claimants on kidneys deserve the chance at life, so all have a claim, even if there is disagreement about what factors attenuate the strength of their claims.  While claims based on a right to life always exist, other claims only arise when a good is being allocated~\citep[p.97]{broome1990fairness}. The best candidate for a job does not have a claim to work at a company with no openings, but once there is an opening she may have stronger claims than others based on her \textit{merit}. Desert, need, and merit can all ground claims.   

Claims are different from both the more familiar utility calculations and side-constraints such as rights. Utilities can be weighed against each other: a kidney is allocated according to utilitarian principles such that its allocation produces the greatest overall benefit. Once the allocation is calculated, nothing remains to be said about the individual who did not receive a kidney. Unlike utilities, however, claims linger. The otherwise-identical individual who did not receive a kidney still had a claim to that kidney, and she was not fairly treated if her claims was simply ``overridden'' by a deterministic allocation~\citep[p.98]{broome1990fairness}. Claims are also unlike side-constraints such as rights. If someone has a right to a good, that right cannot be discharged or outweighed: it ``directly ... determines what ought to be done''~\citep[p.91]{broome1990fairness}. Rights are not comparative between individuals: they simply mandate what must be done to respect the right. 

Claims, by contrast, are essentially comparative, as they are a matter of fairness. If both people in need of a kidney are equal in all morally relevant senses, then they have equally strong claims on the kidney. It would be unfair to deterministically allocate the kidney to one person because doing so would \textit{override} or ignore the equivalent claim that the other patient has on the same kidney~\citep[p.95]{broome1990fairness}. But what if Person $a$ has a slightly stronger claim than Person $b$? Broome argues that if ``fairness requires everyone to have an equal chance when their claims are exactly equal, then it is implausible it should require some people to have no chance at all when their claims fall only a little below equality''~\citep[p.99]{broome1990fairness}.  In other words, $b$'s claim does not go away just because $a$'s claim is marginally stronger.
This motivates Broome's proposal to allocate scarce and indivisible goods using a lottery weighted by the strength of claims. A weighted lottery allows stronger claims to have a proportionately stronger chance while not overriding weak claims. By giving everyone with a claim a chance, lotteries and other randomization techniques give a ``surrogate satisfaction'' to the claimants -- the next best thing to actually receiving the good~\citep[p.99]{broome1990fairness}. In summary, according to Broome, the following two conditions should be met in order for an allocation to be considered \textit{fair}:
\begin{enumerate}[label={BF.\arabic*},topsep=-0.75ex, itemsep=-0.75ex]
\item The chance of a positive outcome should be greater for those with stronger claims. \label{BF1}
\item Stronger claims should not completely override weaker claims. \label{BF2}
\end{enumerate}



\subsection{Multi-Shot Contexts: Systemic Denial of Claims}

The perspective of individual fairness concerns whether each allocation respects individual claims, but we should also consider whether the structure of allocations as a whole is fair. Concerns of structural injustice arise when certain individuals find their claims repeatedly denied, whether by multiple decision-makers at the same time (systemic exclusion) or by multiple decision-makers across time (patterned inequality). In conditions of systemic exclusion, decision-makers across an ecosystem are correlated in their decisions such that they make mistakes on the same people~\citep{creel_hellman_2022, bommasani2022homogenization, toups2023ecosystem}. For example, different companies attempting to hire candidates in the same sector often rely on the same third-party vendors for their automated hiring tools. In fact, over 60\% of Fortune 100 companies use the same vendor (HireVue)~\citep{hirevueArticle}. Relying on the same vendor can lead to identical outcomes if different companies use the same underlying model to rank candidates, or correlated outcomes if each company personalizes their model. In either case, correlation between decision-makers can lead to the same individuals being ``algorithmically blackballed'' and excluded from opportunities~\citep[681]{ajunwa2021}. 

Patterned inequality~\citep{eidelson2021patterned} is another form of systemic injustice. It occurs when receiving one allocation increases your likelihood of receiving future allocations (and likely also your claims to those allocations) such that clear patterns in social inequality emerge as a result of initial conditions. This situation is also referred to as the ``Matthew effect,'' in which the rich (or otherwise advantaged) get richer over time due to their starting condition of advantage. Patterned inequality is visible in domains such as healthcare, where allocations of life-saving medical services are often conditioned on projections of life expectancy or evaluations of current health. Both of these, however, are influenced by past access to treatment and health insurance, for which there are well-documented inequalities 
between socioeconomic groups~\citep{schmidt2021rationing}. Algorithmic decision-making can exacerbate these inequalities by recognizing the current gap in health without recognizing the unequal starting conditions of health, wealth, and stability that gave rise to them~\citep{eidelson2021patterned, hellman2018compounding, jain2023pluralism}. Though both these forms of structural injustice involve a myriad of dynamics, we argue that randomization can help to address both concerns.

\subsection{Inherent Uncertainties in Predicting Claims}

While Broome argues that a lottery weighted on the strength of claims is fair, he acknowledges that it is not always clear which particular reasons are claims and which are not~\citep[p.93]{broome1990fairness}. This makes it difficult to compare the strength of claims, leading some to reject the claims framework altogether~\cite{kirkpatrick2015broome}. However, this so-called ``calculation objection'' applies to any problem formulation for allocating resources, including utilitarian and rights-based frameworks~\cite{passi2019problem, mitchell2021algorithmic}. 
Given the inherent uncertainty in any problem formulation, deterministic allocations on any basis will be unfair to some individuals, especially if we acknowledge that claims exist and some will not have been fairly calculated. 

Let us assume there exists some problem formulation for the strength of claims (e.g. worker productivity in hiring, life-expectancy in healthcare). Many formulations require certainty about what an individual \textit{will} do (e.g. individual risk). However, individual future outcomes are fundamentally unknowable, especially since the events in question are typically realized only once~\citep{dawid2017individual,dwork2021outcome, roth2023reconciling}. Instead, decision-makers often must estimate what an individual is \textit{likely} to do based on their features and data about what other people with similar features have done in the past. For instance, in the kidney allocation, we might determine that a patient's claim should be based on how much longer they are expected to live. If we have data on prior patients, we could develop an algorithm to gauge the strength of a patient's claim based on features such as their age, medical history, and lifestyle. 

When posed as a supervised learning problem, the choice of features, training data, and model class each contribute additional uncertainty to our estimates of the strength of claims. First, a person's features may or may not be predictive of their claim or even measurable. In many social settings, the vast majority of people remain inseparable on the basis of the features that can be measured (also referred to as there being no ``margin''). For example, in the canonical New Adult Census dataset, 95\% of individuals have feature representations for which there exist examples in both prediction classes of high and low income. 
An individual could have a strong claim that is predicted to be weak because there were only a few examples of similar individuals in the dataset and they all happened to have weak claims. Deterministically picking the strongest predicted claims may constitute a kind of stereotyping in this situation. Likewise, people may ``look risky'' because the features measured in the data are not adequate to evaluate their claims. In both cases, people are systemically denied opportunities they deserve. We argue that in these situations randomizing can increase the fairness of allocations. 


Moreover, the uncertainty in predictions of claims may be higher for some individuals than for others. 
Consider the phenomenon of \textit{predictive multiplicity}, wherein there exist multiple models with similar accuracy that yield different predictions for certain individuals~\citep{marx2020predictive, black2022model}. For individuals with high variance in their predictions, it seems unfair to weight their chances based on the prediction of a naively chosen model. The related concept of \textit{leave-one-out unfairness} highlights how some individuals can receive radically different predictions due to the inclusion or removal of a single other person in the training data~\cite{black2021leave,broderick2020automatic}. This may be due to the fact that some individuals are outliers based on the selected features or under-represented in the training data. Uncertainty quantification methods such as conformal prediction can help to identify these individuals~\citep{angelopoulos2021conformal}. As we describe further in~\autoref{sec:claims_unknown}, these methods offer a way to account for varying levels of uncertainty in a weighted lottery using predicted claims.



\section{How To Randomize Allocations}\label{sec:how_to_randomize}
We now formalize \textit{how} randomization can help to address the ethical demand of satisfying individual claims in algorithmic decision-making. Specifically, we propose different methods for randomization when claims are known or uncertain and also show how these methods can help alleviate the structural concerns of systemic exclusion and patterned inequality. We also discuss the potential tradeoff between randomization and accuracy and how to interpolate between them when the tradeoff exists. 

\subsection{When Claims Are Known}\label{sec:claims_known}
Consider a setting in which there are $n$ individuals and each individual $i$ is assigned a score $c_i \in [0,1]$ in perfect accordance with their claim. We say that individual $i$ has a stronger claim than individual $j$ if $c_i > c_j$. A decision-maker allocates outcomes $o_i \in \{0, 1\}$ to each individual $i$. Importantly, there is scarcity in that not all claims can be satisfied: only $k$ out of $n$ individuals can receive positive outcomes with $k\ll n$, for a selection rate of $k/n$.

\begin{definition}
\label{def:lottery}
An \textbf{iterative weighted selection} chooses one individual in each round $t$ \textit{without} replacement until $k$ individuals are selected. Specifically, an individual $i$ in round $t$ has probability $w_{i,t}$ of being selected. For all rounds $t \in \{1,\ldots,k\}$, we require $\sum_{j=1}^{n - t + 1} w_{j,t} = 1$ so that exactly one individual is selected per round.
\end{definition} 

Note that the formulation above encapsulates many kinds of selections. For example, deterministically selecting the top $k$ claims\footnote{This is equivalent to selecting all claims greater than or equal to the threshold $T = c_{(k)}$ where $c_{(k)}$ is the k-th largest claim.} would take $w_{i,t} = \mathds{1}[c_i = \max_{j \in \{1...n-t+1)\}} c_j ]$. Recall that Broome's notion of fairness calls for weights that are chosen in proportion to a claim's strength. 

\begin{definition}
\label{def:fairness}
An \textbf{allocation} $A$ involves the assignment of outcomes $o_i$ through an iterative weighted selection based on weights $w_{i,t}$ for each claim $c_i$ and each round $t$. It satisfies \textbf{Broome-Fairness} (\BF) if for all rounds $t$ and all individuals $i,j$ not yet selected:
\begin{enumerate}[topsep=-1ex,itemsep=-1ex]
    \item $c_i > c_j \implies w_{i,t} > w_{j,t}\;$ (c.f.~\ref{BF1})
    \item $c_i > 0 \implies w_{i,t} > 0\;$ (c.f.~\ref{BF2})
\end{enumerate}
\end{definition}

\begin{example}[BF Lottery]
\label{ex:bf_lottery}
An allocation with $w_{i,t} = \frac{c_i}{C_t}$ satisfies BF, where $w_{i,t}$ is calculated among the remaining individuals $i$ in round $t$ and $C_t = \sum_{j=1}^{n-t+1} c_j$ represents the sum over claims not selected in previous rounds.
\end{example}

Importantly, deterministic allocations do not satisfy \BF{} because some weights are zero (violating~\ref{BF2}) and no distinction is made among rejected claims of varying strengths (violating~\ref{BF1}). An unweighted lottery also violates~\ref{BF1} by assigning the same weight to all individuals: $w_{i,t}=\frac{1}{n-t+1}$. 

\subsubsection{Systemic Harms}\label{sec:systemic_harms}
A lottery weighted by the strength of claims can help to alleviate the structural concerns of systemic exclusion and patterned inequality. Suppose there are $m>1$ decision-makers conducting allocations either concurrently or across time. Let $o_i^{(j)}$ denote the outcome for individual $i$ by decision-maker $j$. Our concern is with the proportion of individuals (or groups) who exclusively receive negative outcomes from all $m$ decision-makers.

\begin{definition}
\label{def:ser}
The \textbf{systemic exclusion rate} (SER)~\citep{bommasani2022homogenization} across $m>1$ decision-makers is:\begin{talign*}\mathbb{E}_i\left[\prod_{j=1}^m \mathds{1}[o_i^{(j)} = 0]\right] 
\end{talign*}
\end{definition}

To illustrate why randomization can help reduce SER, we begin by considering two stylized models of allocations. As our stylized models will suggest, if the existing SER is sufficiently high across the $m$ decision-makers or $m$ is sufficiently large, then in expectation, randomization will help both systemic exclusion and patterned inequality. 
 
\begin{example}[Systemic Exclusion]
\label{prop:1}
Suppose there are $m$ allocations at the same time and that there are many more individuals with similarly strong claims than available positive outcomes. If the SER is sufficiently high for the existing set of allocations, than any allocation satisfying \BF{} will decrease the SER in expectation. 
\end{example}

\begin{figure}[t]
\centering
\subfloat[\centering Many Possible Distributions of Claims]{\label{fig:claims_dist}
\includegraphics[width=0.77\columnwidth]{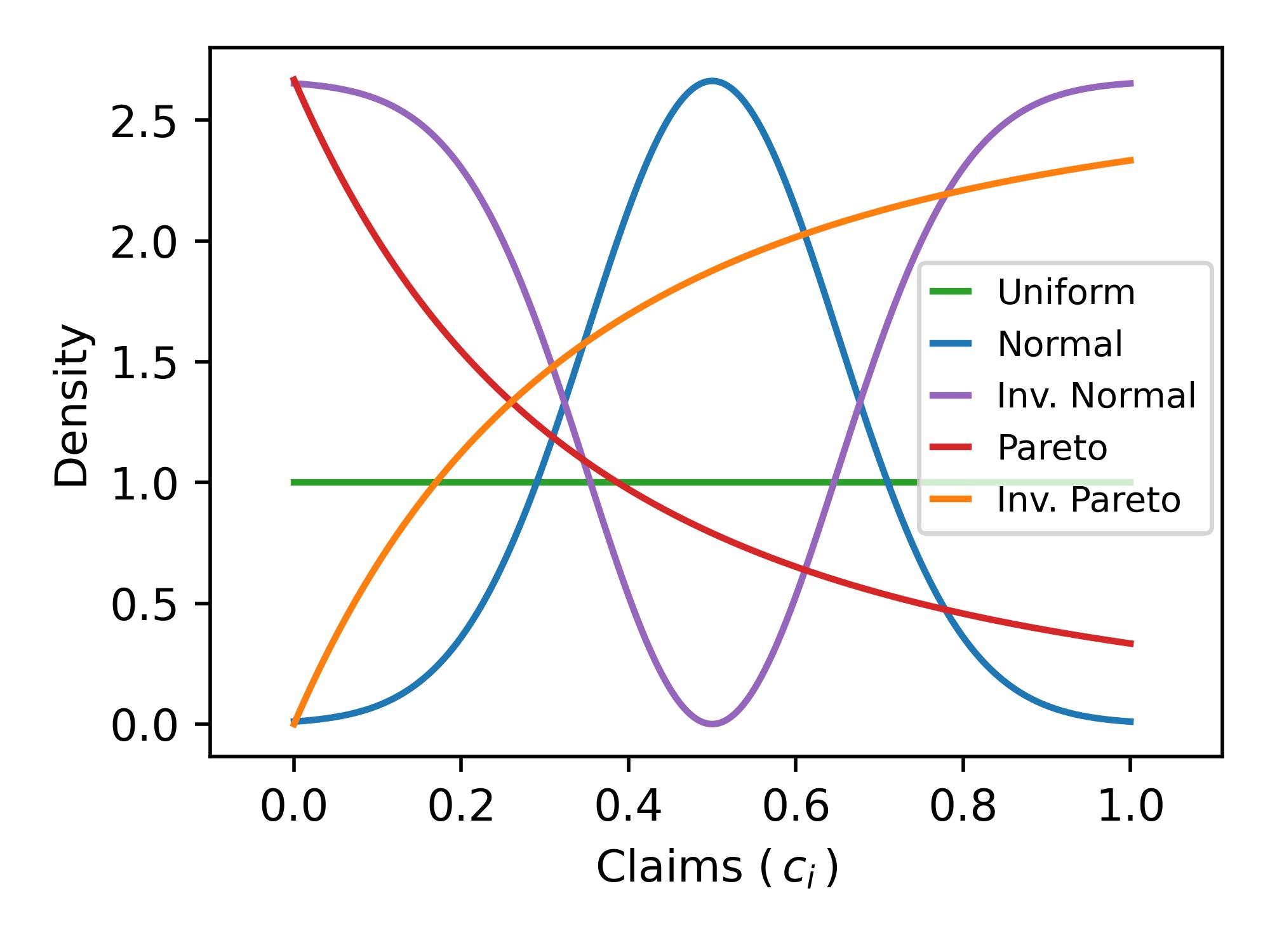}}
\end{figure}

\begin{figure}[t]
\setcounter{subfigure}{1}
\centering
\subfloat[\centering Reduction in SER using BF Lottery ($k/n$ = 0.25)]{\label{fig:claims_ser}
\includegraphics[width=0.77\columnwidth]{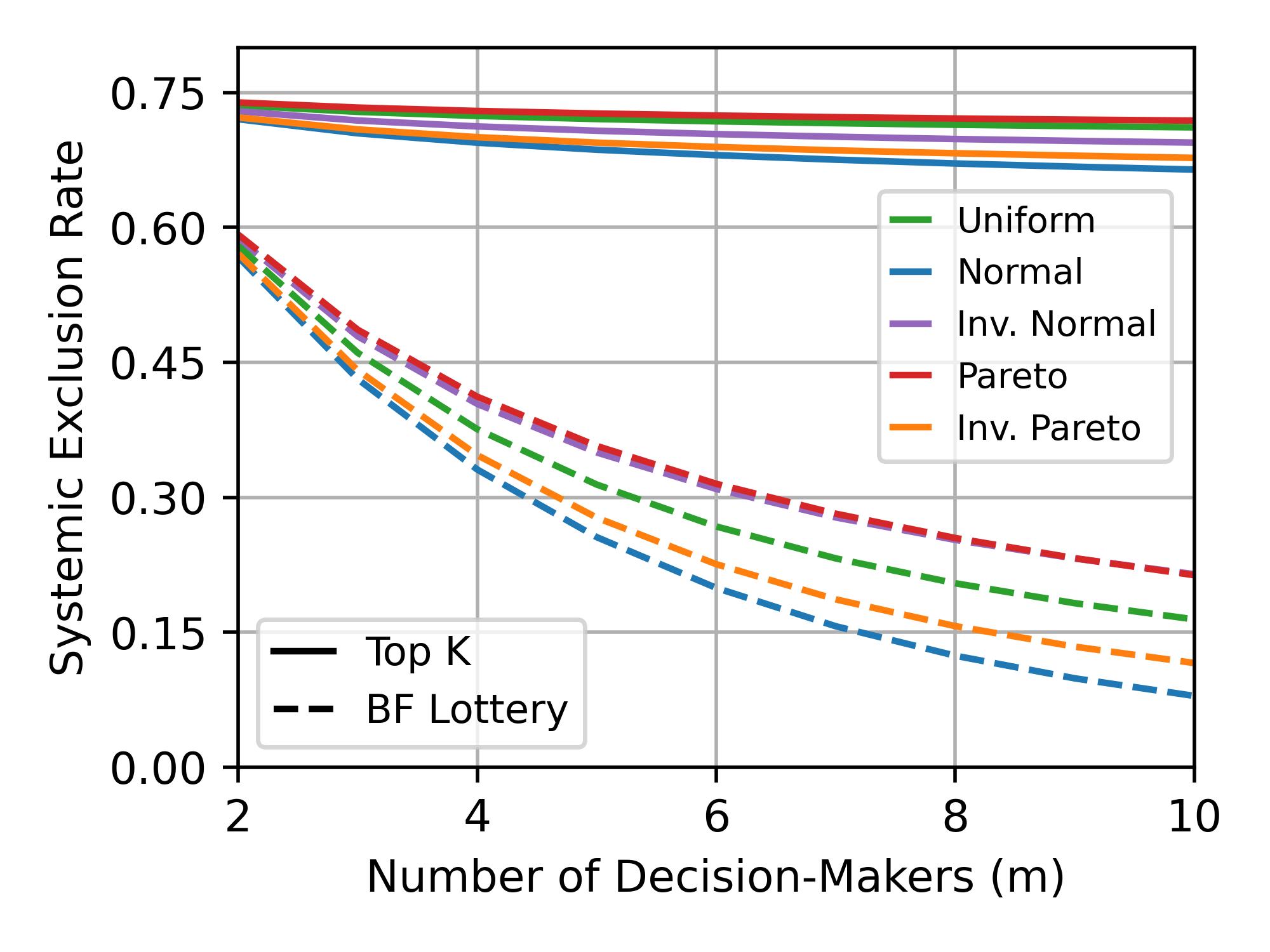}
}
\end{figure}

\begin{example}[Patterned Inequality]
\label{prop:2}
Suppose there are $m$ sequential allocations across time, and that receiving a positive outcome increases an individual's claim in the next allocation. Also assume that in the first allocation, there are many more individuals with similarly strong claims than available positive outcomes. Then any set of sequential allocations satisfying \BF{} will decrease the SER in expectation when compared to a set of deterministic allocations if the benefit from a positive outcome is sufficiently high. 
\end{example}

In general, the ability of allocations satisfying BF to reduce the SER will depend on the distribution of claims, correlation between allocations across decision-makers, and selection rate ($k/n$). We simulate how much randomization can reduce SER for various distributions of claims, when each decision-maker has a noisy estimation of these claims ($\pm\,N(0, \sigma^2)$). As Figure~\subref{fig:claims_dist} illustrates, we consider the following distributions:
\begin{itemize}[topsep=-1ex,itemsep=-1ex]
    \item Uniform: all claims equally likely  
    \item Normal: more average claims
    \item Inverted Normal: more strong and weak claims
    \item Pareto: more weak claims 
    \item Inverted Pareto: more strong claims
\end{itemize}

For all these distributions and many different selection rates and noise amounts, we observe a substantial reduction in SER if each decision-maker uses the weighted lottery satisfying BF in Example~\ref{ex:bf_lottery} rather than deterministically selecting their top $k$ claims. Figure~\subref{fig:claims_ser} provides a snapshot of our results in the setting where $k/n = 0.25$ and $\sigma=0.025$ (Appendix Figure~\ref{fig:ser_full_bf} shows other cases are similar).

\subsubsection{Utility and Randomization}\label{sec:utility}
Why might one want to allocate deterministically? From the viewpoint of risk-averse decision-makers, the objective is to maximize their own utility. In hiring, for instance, a company might allocate job interview slots based on a candidate's likelihood of being hired. We simplify to the case where each individual has some utility $o^*_i \in \{0,1\}$ (e.g. whether or not the candidate would be hired). 

\begin{definition}
\label{def:utility}
The \textbf{utility} of an allocation is $\frac{1}{k} \sum_{i=1}^{n} o^*_i \cdot \mathds{1}\{o_i = 1\}$, which is simply the {\em precision}: i.e. the proportion of selected individuals $k$ that provide utility.
\end{definition}

Note that $o^*_i$ can only be observed if the individual receives the resource being allocated ($o_i = 1$), which motivates the idea of \textit{expected} utility. 

\begin{definition}
\label{def:expected_utility}
Suppose an individual's chance of providing utility is $p_i = \mathbb{P}(o^*_i = 1)$. Accordingly, the \textbf{expected utility} of an allocation is $\frac{1}{k} \sum_{i=1}^{n} p_i \cdot \mathds{1}\{o_i = 1\}$.
\end{definition}

Utility aligns with the strength of claims in some merit-based allocations. The individuals with the strongest claims are those with the most ``merit'' or closest fit between their skills and the needs of the role. These candidates therefore are also the most likely to be hired by the company. Individuals may have other claims besides merit, such as claims based in desert or entitlement. However, we adopt the decision-maker's perspective because it is \textit{the least favorable} viewpoint to motivate randomization and results in the worst-possible tradeoffs between the decision-maker's notion of utility and desirable properties of randomization.

\begin{figure}[ht!]
\setcounter{subfigure}{2}
\centering
\subfloat[\centering  Expected Utility for Top $k$ v. BF Lotteries\newline (Partial BF: $k'$ = 0.5$\cdot k$, $n'$ = $k$)]{\label{fig:claims_utility}
\includegraphics[width=0.77\columnwidth]{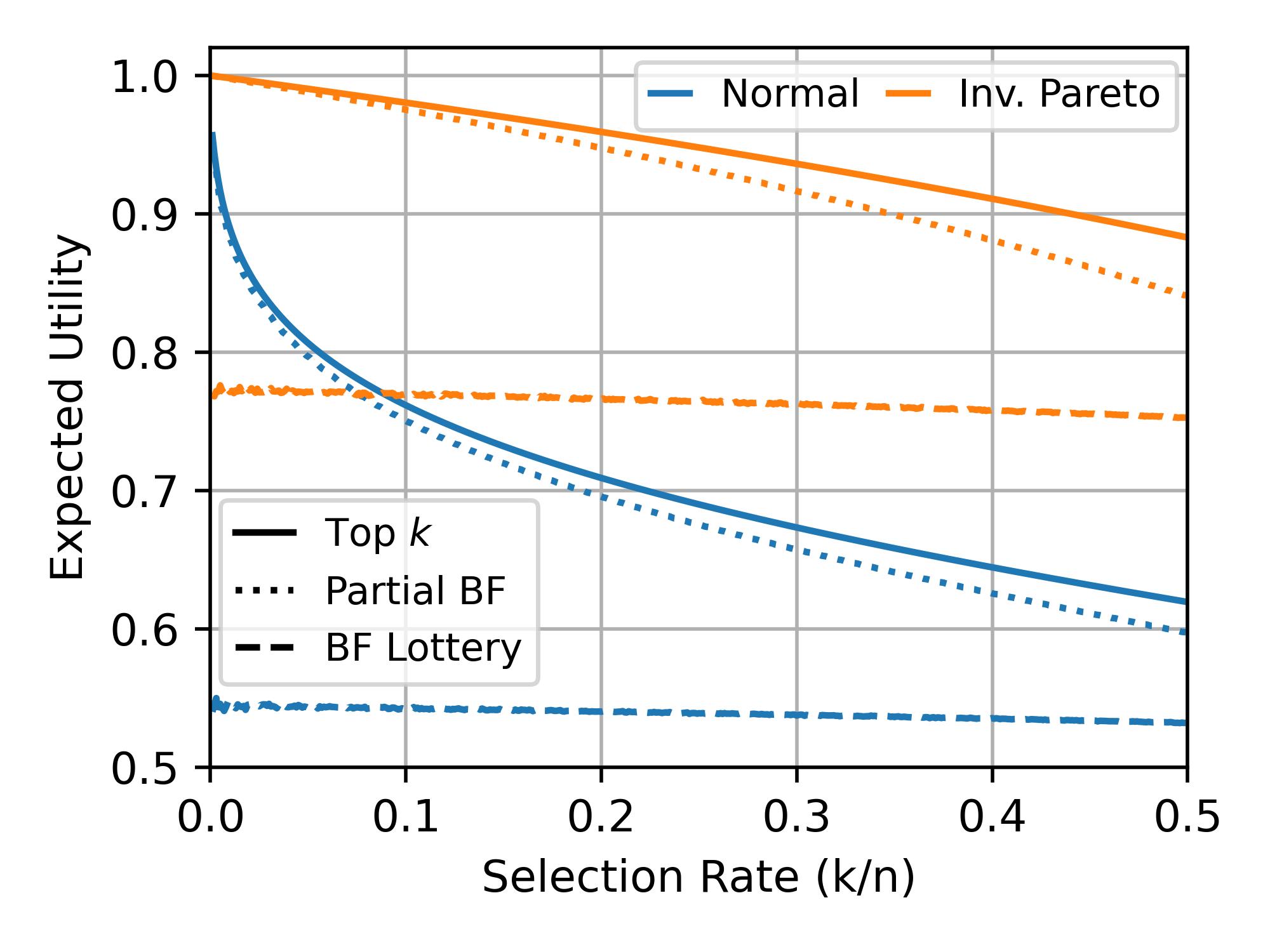}
} \\
\subfloat[\centering  Expected Utility for Varying Partial BF Randomization Rates (Normal Dist; $k/n$ = 0.25)]{\label{fig:claims_partial_bf}
\includegraphics[width=0.77\columnwidth]{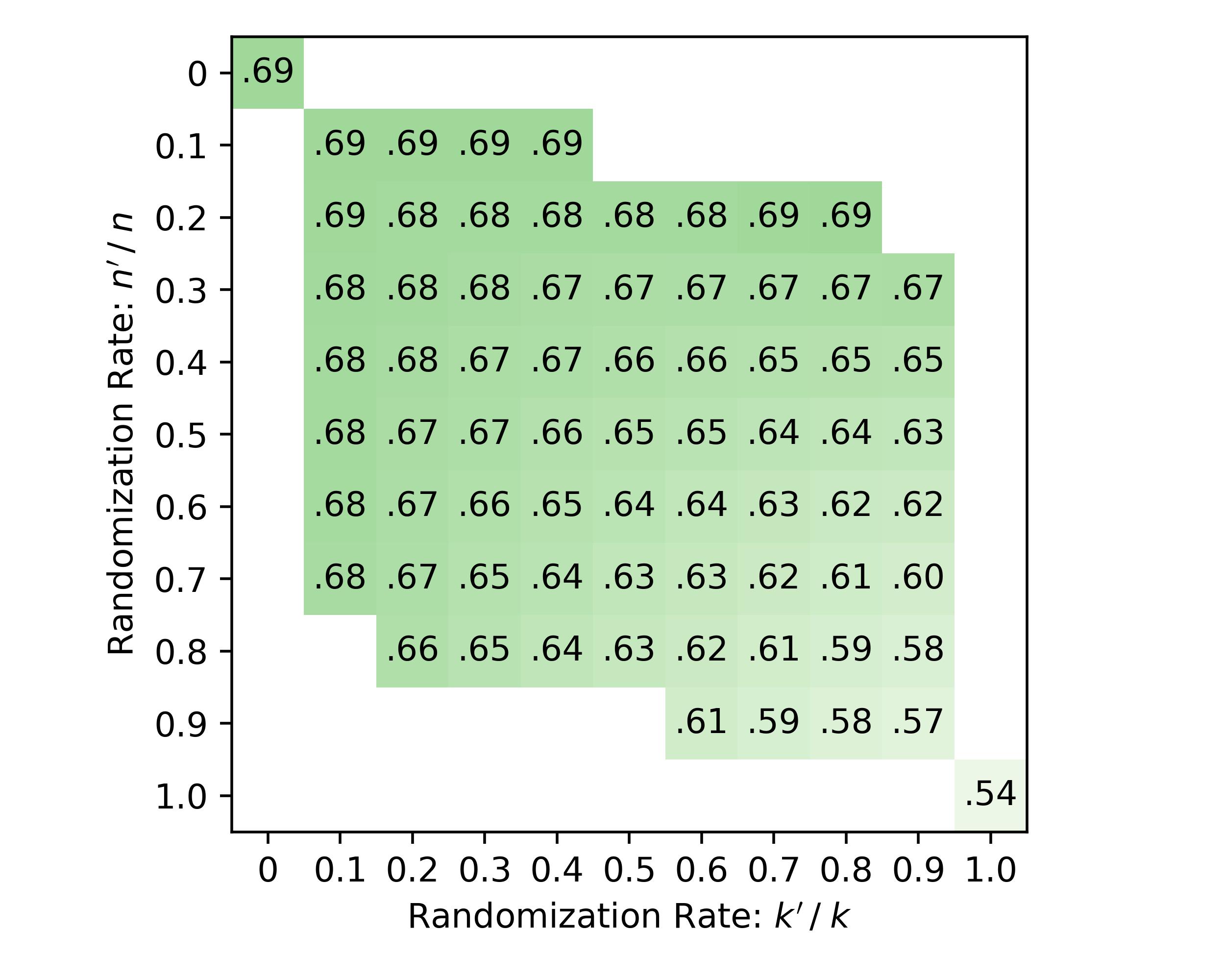}} \\
\subfloat[\centering SER v. Expected Utility Tradeoff for Varying\newline Partial BF Randomization Rates ($k/n$ = 0.25)]{\label{fig:claims_ser_partial}
\includegraphics[width=0.77\columnwidth]{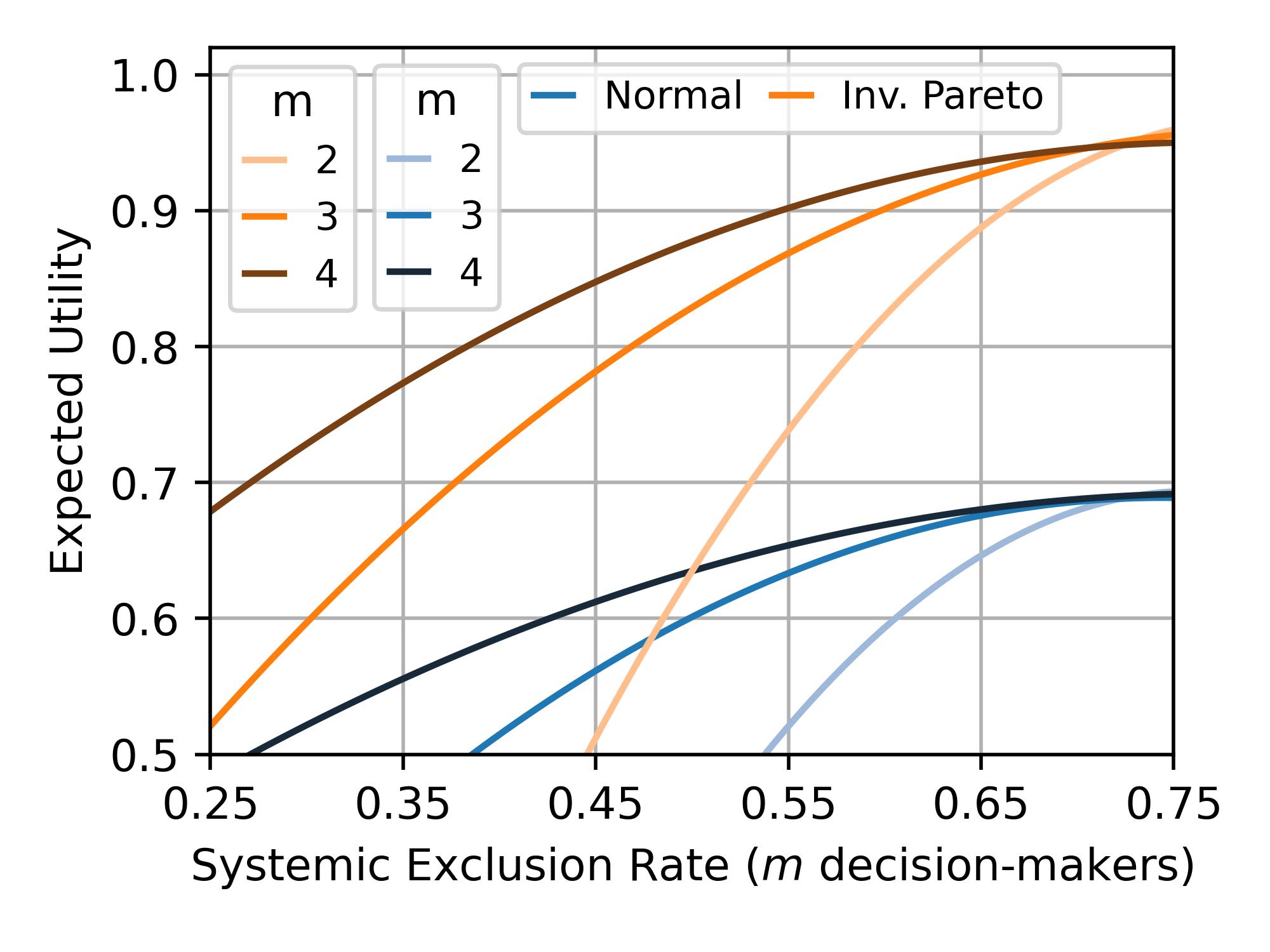}
}
\end{figure}

If claims are exactly the chance of providing utility, then deterministically selecting the $k$ strongest claims will maximize the  expected utility. But as we discussed above, this overrides the most claims and violates both \ref{BF1} and \ref{BF2}. On the other hand, any amount of randomization will lead to some probability of allocating resources to those who do not have the strongest claims, resulting in some sacrifice of expected utility in favor of respecting more claims. Figure~\subref{fig:claims_utility} illustrates the difference in expected utility between the top $k$ allocation and BF lottery in Example~\ref{ex:bf_lottery} for the normal and inverted Pareto distributions (other distributions are similar). Note that this tradeoff increases with scarcity (i.e. low selection rates).

Balancing this tradeoff requires a consideration of the fairness arguments on each side. Consider two patients who both have a claim to a single kidney. Patient $a$'s claim is based in their survival probability of 0.51, while patient $b$ has a survival probability of 0.49. If we deterministically chose to give the kidney to patient $a$, Broome would argue that this is unfair to patient $b$ since they have no chance at all to receive the kidney despite only having a slightly weaker claim~\citep[p.99]{broome1990fairness}. But what if we instead compare patient $c$, who has a survival rate of 0.99, to patient $d$, who only has a survival rate of 0.01? Brad Hooker points out in an objection to Broome that ``a great unfairness would occur'' if we held a weighted lottery and patient $d$ won given the comparative strength of patient $c$'s claim in this case~\citep{hooker2005fairness}.\footnote{Hooker's objection considers one patient (here, $c$) who would die without the medicine and another ($d$) who would only lose a finger~\citep{hooker2005fairness}.} This motivates the idea of not randomizing some very strong or weak claims while still conducting a weighted lottery for the remaining claims.

\begin{definition}
\label{def:partial_fairness}
An allocation satisfies \textbf{partial Broom-Fairness} when the criteria \ref{BF1} and \ref{BF2} are met for:
\begin{enumerate}[topsep=-1ex,itemsep=-1ex]
    \item a subset of resources: $k' \in (0, k]$
    \item a subset of claims: $n' \in (k', n - k + k']$
\end{enumerate}
\end{definition}

\begin{example}[Partial BF Lottery]
\label{ex:partial_bf_lottery}
The following allocation satisfies partial BF: Give $k-k'$ resources to the top claims. Then conduct the iterative weighted selection in Example~\ref{ex:bf_lottery} for the remaining $k'$ resources over the $n'$ claims closest to the $k$-th largest claim.
\end{example}

We discuss how a lottery satisfying partial BF can result in a much smaller tradeoff with expected utility, yet still substantially reduce SER. Figure~\subref{fig:claims_utility} shows that the expected utility difference is $<$ 0.05 across different selection rates for a partial BF lottery that uses $k'$ = 0.5$\cdot k$ and $n'$ = $k$. In other words, we first allocate half the available resources to the top $0.5k$ claims, then randomize over the next $k$ strongest claims for the other half of the available resources. Note that this has the effect of randomizing near the so-called ``decision-boundary,'' which represents the $k$-th largest claim in our framework. Figure~\subref{fig:claims_partial_bf} explores how varying partial BF randomization rates (i.e. different combinations of $k'$ and $n'$) change the difference in expected utility in the setting\footnote{Appendix~\ref{app:claims_known} replicates Figure~\subref{fig:claims_partial_bf} for different dist. and $k/n$.} where claims are normally distributed and $k/n$ = 0.25. We find that larger randomization rates have a disproportionately larger decrease in expected utility.

Figure~\subref{fig:claims_ser_partial} shows the lowest SER that we can achieve for a given tradeoff with expected utility. Specifically, we compare varying partial BF randomization rates for the same setting as Figure~\subref{fig:claims_ser} where $k/n = 0.25$ and the noise is $\sigma=0.025$. Consider, for instance, the 2\% difference in expected utility from the partial BF lottery in Figure~\subref{fig:claims_utility} when claims are normally distributed. This yields greater than a 20\% reduction in SER when there are $m>2$ decision-makers. See Appendix~\ref{app:claims_known} for many other examples across different distributions of claims, selection rates, and amounts of noise added for each decision-maker.

\subsection{When Claims Are Uncertain}\label{sec:claims_unknown}

A fundamental assumption in machine learning is that the targets of interest (in our case, claims) are predictable from a set of measurable features in some domain $\X$. While $p_i = \mathbb{P}(o^*_i = 1)$ might be unknowable, the conditional probability $p(x_i) = \mathbb{P}(o^*_i = 1\,|\,x_i)$ can be estimated from data. The validity of taking $p(x_i)$ to be an estimate of $p_i$ depends on the choice of features that are measured and predictability of the outcomes from those features. Putting these concerns aside, a machine learning model $\hat{p}: \X \rightarrow [0,1]$ maps an individual's features $x_i \in \X$ to a prediction $\hat{p}(x_i)$, which estimates the conditional probability $p(x_i) = \mathbb{P}(o^*_i = 1\,|\,x_i)$. In a healthcare allocation, $\hat{p}(x_i)$ might represent a model’s estimate based on prior patients in a hospital, whereas $p(x_i)$ represents the conditional probability if we could measure all possible patients represented in feature space $\X$.

Standard practice in machine learning would be to deterministically assign $o_i = 1$ to the individuals with the $k$ highest value of $\hat{p}(x_i)$. While this doesn't satisfy BF, it is unclear what implementable allocation does due to the distinction between $\hat{p}(x_i)$ and $p_i$. For example, we could use the weighted lotteries in Example~\ref{ex:bf_lottery} or~\ref{ex:partial_bf_lottery} by replacing $c_i$ with $\hat{p}(x_i)$. However, the estimation error in $\hat{p}(x_i)$ could be higher for certain individuals than others, potentially violating~\ref{BF1}. 

\begin{figure}[t!]
\setcounter{subfigure}{5}
\centering
\subfloat[\centering Distribution of Swiss Unemployment Predictions]{\label{fig:swiss_dist}
\includegraphics[width=0.77\columnwidth]{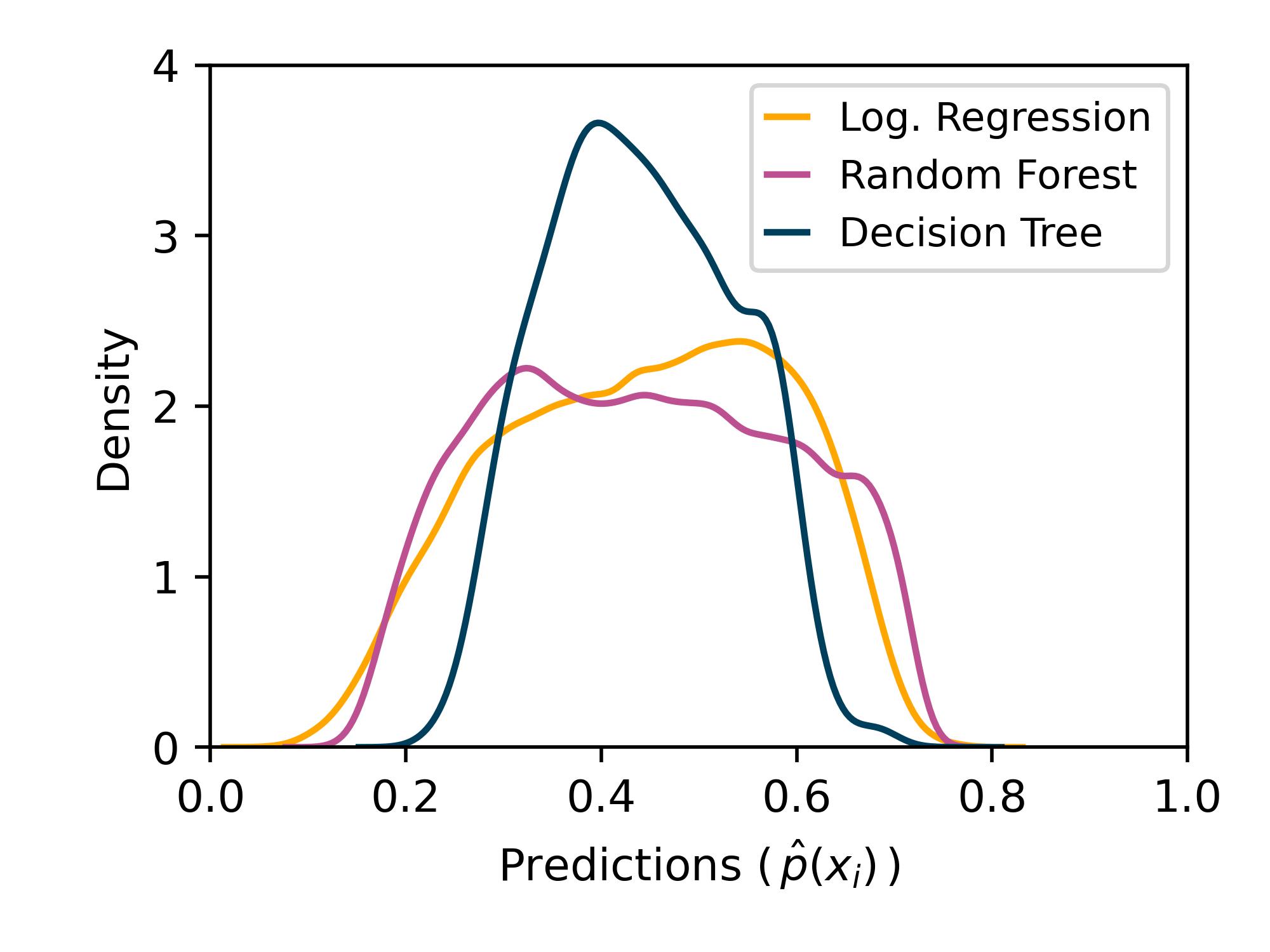}
}
\end{figure}


We take our working example\footnote{Appendix~\ref{app:claims_unknown} includes an additional example of income prediction using the \textit{New Adult Census} dataset~\citep{ding2021retiring}.} to be the 2003 \textit{Swiss Unemployment} dataset~\citep{swissData}. The goal is to allocate scarce unemployment assistance resources such as job search and training programs. Suppose an individual's true \textit{claim} to these benefits is how long they would remain unemployed without them, and that the programs want to target those who would have remained unemployed for at least 1 year. In this example, individual claims align with the decision-makers' notion of utility as long-term unemployment. We can only estimate an individual's probability of being long-term unemployed based on features such as their age, place of residence, education, previous job, prior income, etc. Figure~\subref{fig:swiss_dist} shows that predictions\footnote{We subset to individuals that did not receive an unemployment benefit ($n=78,294$) and use an 80-20 train-test split (with 5 repetitions). Randomization results avg. over 100 iterations.} appear to follow a normal distribution for 3 different model classes: logistic regression, random forests, and decision trees. 

For our main analysis, we use a selection rate of $k/n$ = 0.25 and explore other selection rates in the Appendix (which yield similar results). 22\% of individuals in the dataset received some form of unemployment assistance, although the most effective programs only had capacity for $<$5\% of individuals. Among those that did not receive assistance, 44\% of individuals remained long-term unemployed (at least 1 year). For our selection rate of $k/n$ = 0.25, standard practice would choose the top 25\% of predictions. This yields an (observed) utility of just 63.3\% on average across all 3 models. 

In what follows, we first explore using the partial weighted lottery in Example~\ref{ex:partial_bf_lottery} to randomize over predictions near the decision-boundary. We then propose two other randomization methods that quantify and incorporate the varying levels of uncertainty in predictions across individuals. We discuss how each method changes how many resources ($k'$) and what kinds of people ($n'$) are randomized, while maintaining a minimal loss in utility.

\begin{table}[t!]
\small
\caption{\textbf{Randomizing using variance} compared to randomizing near the decision-boundary \& the top $k$ allocation.}
\begin{tabular}{cccccc} 
\toprule
\multirow{3}{*}{Model} & \multicolumn{2}{c}{Random Rate} & \multicolumn{3}{c}{Utility} \\
\cmidrule(lr){2-3} \cmidrule(lr){4-6}
& \multirow{2}{*}{$k'/k$} & \multirow{2}{*}{$n'/n$} & \multirow{2}{*}{Variance} & Decision- & \multirow{2}{*}{Top $k$}\\
& & & & Boundary & \\
\toprule
LR & 14.0\% & 6.8\% & 62.9\% & 62.8\% & 63.1\% \\
RF & 32.2\% & 15.0\% & 64.1\% & 63.7\% & 64.3\% \\
DT & 73.7\% & 39.0\% & 61.5\% & 58.9\% & 62.9\% \\
\bottomrule
\end{tabular} 
\label{tab:results_variance}
\end{table}

\textbf{Randomizing Near Decision-Boundary.\hspace{3pt}} We first consider using the partial weighted lottery in Example~\ref{ex:partial_bf_lottery} by replacing $c_i$ with $\hat{p}(x_i)$. Recall that this has the effect of randomizing near the decision-boundary or $k$-th largest prediction. We find small tradeoffs with utility that are very similar to those for \textit{expected} utility that we saw for when claims are known and normally distributed (c.f. Figure~\ref{fig:claims_partial_bf}). For example, we observe just a 0.8\% drop in utility for partial randomization with $k'$ = 0.5$\cdot k$ and $n'$ = $k$, which randomizes half the available resources across the $k$ closest predictions to the decision-boundary on either side\footnote{In our working example with $k/n$ = 0.25, choosing $k'$ = 0.5$\cdot k$ and $n'$ = $k$ would first select the predictions above the 87.5-th percentile, and then randomize the remaining resources across people with predictions in the 62.5 to 87.5 percentile.}. Figure~\ref{fig:swiss} in the Appendix shows how utility is affected by different partial randomization rates (i.e. different $k'$ and $n'$).


\textbf{Randomizing Using Variance.\hspace{3pt}}  A variety of methods exist to estimate the variance of predictions~\citep{black2021leave, cooper2023variance, ganesh2023impact}. For example, \citet{cooper2023variance} propose re-training on bootstrapped sub-samples of the training data. Consider the set of predictions $(\hat{p}_{(1)},\ldots,\hat{p}_{(m)})$ across $m$ bootstrapped models\footnote{\citet{ganesh2023impact} show how to efficiently estimate the variance in predictions by changing the data order across epochs in a single training run.}. We contend that if \textit{any} of these models placed an individual among the top $k$ claims, then they should have a chance to receive $o_i=1$. Specifically, we propose directly assigning $o_i=1$ to individuals placed in the top $k$ by all models, and then conducting an iterative weighted selection among the remaining individuals, where the weights represent the proportion of models that placed them in the top $k$.

When compared to randomizing near the decision-boundary, we observe that randomizing using this estimation of variance results in a smaller utility loss for all model classes. Table~\ref{tab:results_variance} shows the randomization rates and utility that result from randomizing according to 11 bootstrapped models trained on 50\% of the available training data. For the same randomization rates, we compare the utility that results from randomizing near the decision-boundary, and also report the utility from no randomization (top $k$). Consider the random forest model as an example: randomizing using variance results in just a 0.2\% utility loss while randomizing 32\% of resources over 15\% of people. These randomization rates yield a 0.6\% utility loss for randomizing near the decision-boundary.


\begin{table}[t!]
\small
\centering
\caption{\textbf{Randomizing outliers} ($\alpha=0.2$) compared to randomizing near the decision-boundary \& the top $k$ allocation.}
\begin{tabular}{cccccc} 
\toprule
\multirow{3}{*}{Model} & \multicolumn{2}{c}{Random Rate} & \multicolumn{3}{c}{Utility} \\
\cmidrule(lr){2-3} \cmidrule(lr){4-6}
& \multirow{2}{*}{$k'/k$} & \multirow{2}{*}{$n'/n$} & \multirow{2}{*}{Outliers} & Decision- & \multirow{2}{*}{Top $k$}\\
& & & & Boundary & \\
\toprule
LR & 1.2\% & 20.1\% & 62.7\% & 63.0\% & 63.1\% \\ 
RF & 1.0\% & 20.1\% & 64.0\% & 64.3\% & 64.4\% \\
DT & 3.0\% & 20.1\% & 62.2\% & 62.8\% & 62.9\% \\
\bottomrule
\end{tabular} 
\label{tab:results_outliers}
\end{table}


\textbf{Randomizing Outliers.\hspace{3pt}}  Many out-of-the-box methods exist for outlier detection, which quantify the uncertainty in a prediction that stems from a lack of similar individuals in the training data~\citep{pimentel2014outlier}. For example, in the Swiss unemployment dataset there exists an individual $i$ (and $i'$) with very high (and very low) predicted value across all bootstrapped models, but with $o_i = 0$ (and $o_{i'} = 1$). Conformal prediction offers a way to assign a confidence measure to outlier detection methods, and produces low p-values for both individuals ($<$0.10). This motivates the use of conformal prediction to flag outliers~\citep{angelopoulos2021conformal} and then deploy a lottery for the resources that would have gone to ``outliers individuals'' based on a top $k$ allocation.


Specifically, consider the pool of individuals that we believe are outliers with high confidence ($\text{p-value}\leq \alpha$) for some small $\alpha$. If some of these individuals fall in the top $k$, then we propose to randomize those resources over the entire pool of ``outlier'' individuals using an unweighted lottery. Note that the pool of individuals that we believe are outliers is model-agnostic, since it is computed based on the features. How many of these ``outlier'' individuals would have ended up in the top $k$ depends on the model. 

Table~\ref{tab:results_outliers} shows the randomization rates and tradeoff with utility for $\alpha=0.2$, which is slightly more than the utility loss for randomizing near the decision-boundary. We end up randomizing just 2\% of the available resources over 20\% of the total people (note this directly corresponds to our choice of $\alpha=0.2$). This suggests that the individuals being randomized based on outlier detection are different than those near the decision-boundary or with high variance in predictions. Figure~\ref{fig:risk_scores_vs_uncertainty} in the Appendix visualizes how the predictions with high uncertainty are different for each method.

\begin{figure}[h]
\setcounter{subfigure}{10}
\centering
\subfloat[\centering SER v. Utility Tradeoff for each Randomization Method (Model Class: Decision Tree)]{\label{fig:swiss_homogenization}
\includegraphics[width=0.77\columnwidth]{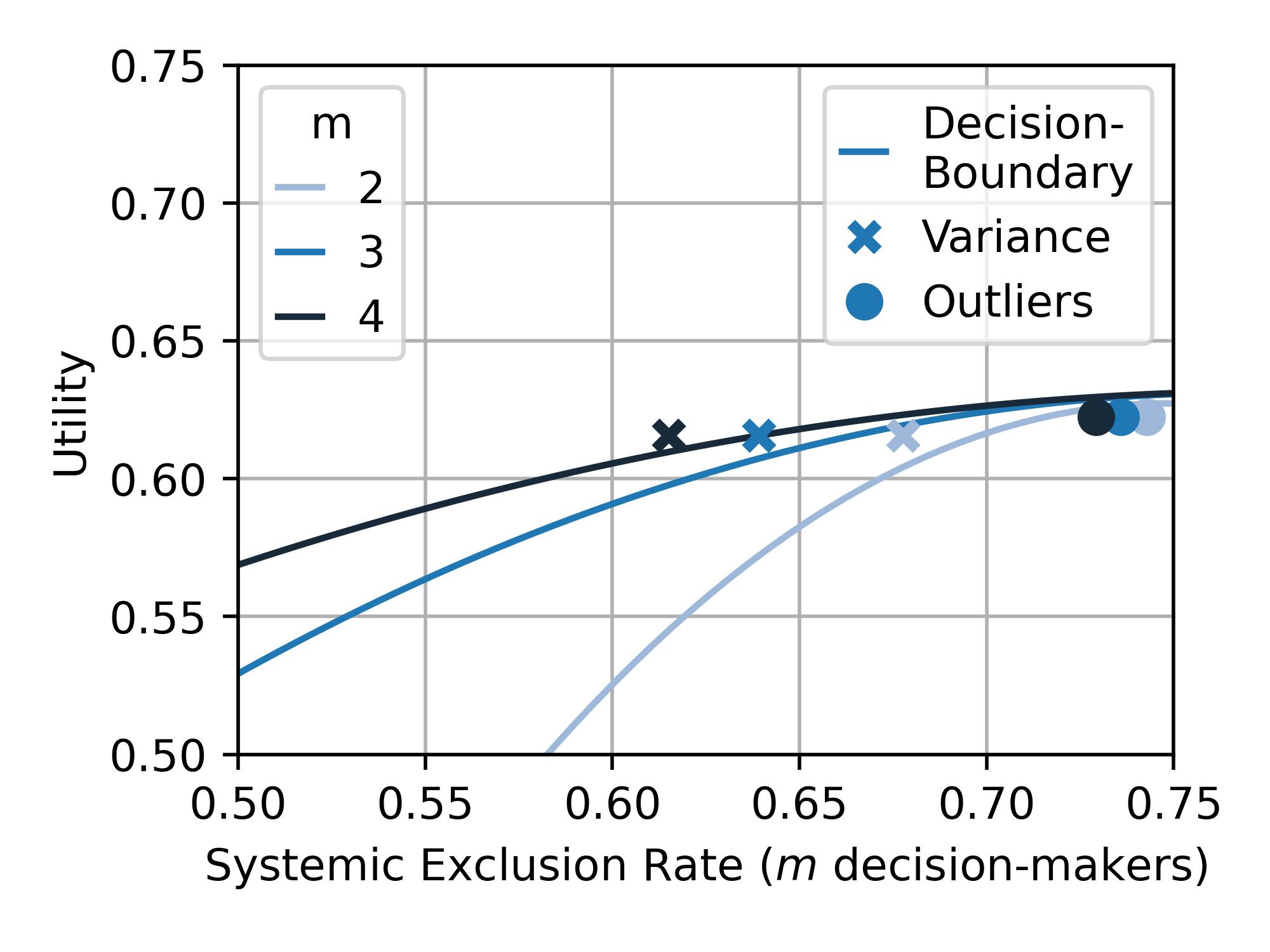}
}

\end{figure}

\textbf{Reduction in SER.\hspace{3pt}} Lastly, we turn to how much our randomization proposals could reduce the systemic exclusion rate (SER). Similar to the experiments when claims are known, we find that small tradeoffs with utility yield much larger reductions in SER. Figure~\subref{fig:swiss_homogenization} demonstrates our results for each randomization method using the decision tree model class. In this case, randomizing using variance has the best tradeoff, though results vary across model classes and selection rates (see Appendix~\ref{app:claims_unknown} for other cases). 


\section{Discussion}\label{sec:discussion}
We argued in \autoref{sec:why_randomize} that sometimes fairness requires randomizing allocations of scarce resources or opportunities, and in \autoref{sec:how_to_randomize} we provide randomization techniques that respect many claims while not losing significant predictive performance.  We now extend the argument and explore implications of these findings.

\textbf{Utility.\hspace{3pt}} When claims are known,  randomization sometimes trades off against expected utility or predictive success.  Although some may find this tradeoff hard to endorse, we suggest two things.  First, a claims-based moral framework holds that people's claims must be satisfied (or  acknowledged by the surrogate satisfaction of a lottery). Some reject claims and take utility to be the only currency of moral concern. However, anyone who agrees that it is more fair for a qualified candidate to have a chance than to never have had a chance can consider claims as an objective within a broader utility-maximization framework, such that overall utility can be improved by satisfying more claims. 
Second, our exploration of uncertainty suggests that what appears to be a tradeoff is, at times, a movement to a different point within the same bounds of uncertainty.  Over-optimizing for apparent utility ignores our true uncertainty about the facts and moral claims of the case.  Thus the utility we appear to give up in order to honor applicants' valid claims may be illusory: there may be no tradeoff at all. 

\textbf{Human Randomness.\hspace{3pt}} How does the intentional randomness we propose in this work compare to the natural variance of human decision-making? For example, despite being extensively trained in decision-making, judges who evaluate the same case~\cite{ludwig2021fragile} often disagree, and judges disagree with their past selves evaluating similar cases over time~\citep{Collins2008}.  This property should make human decision-making less homogeneous than algorithmic decision-making~\citep{creel_hellman_2022}.  
However, we do not find human randomness to be a satisfactory substitute for intentional randomization.  Although human decision-making is not consistent, its outcomes are not guaranteed to be distributed across people in accordance with their claims, as social biases concentrate bad outcomes on individuals from marginalized groups in many situations. Furthermore, we have showed above that fairness requires selecting the most appropriate form of randomness given the problem description and the underlying distribution of data.  Human inconsistency is not subject to these matching constraints.

\textbf{Scope.\hspace{3pt}}  We do not think that randomization is fair in all settings.  For example, criminal justice is served by respecting the procedural rights of defendants and attempting to determine whether the accusations they face are true.  Criminal justice is not a matter of \textit{comparative} claims: each defendant must be evaluated separately, not in comparison to others.  To randomize the outcomes would be unfair.  
However, we affirm the value of randomization in settings in which scarce resources must be fairly allocated on the basis of uncertain information. Since this encompasses many algorithmic decision-making contexts, we encourage the field of fair machine learning to consider randomization as an important element of fairness. 

\textbf{Code for Experiments.\hspace{3pt}} \url{https://github.com/shomikj/randomization_for_fairness}.

\section*{Impact Statement} 

This paper contributes to the literature on algorithmic fairness in two ways. (1) As a position paper, it encourages others to reconsider whether deterministic algorithms are always the right choice for fairness, arguing that randomization techniques are to be preferred in some settings. (2) In support of these arguments, the paper also presents concrete techniques that can be used to randomize and shows how they reduce systemic exclusion and patterned inequality. As such, we hope that it will have a positive impact in reducing \textit{bias and unfairness}.

However, it is also possible that a decision-maker might use the tools presented here to randomize outcomes in a domain that the authors warn would be unjust or inappropriate, such as the domain of criminal justice, and in doing so wrong decision subjects.  Since all of the randomization techniques that form the basis of the paper’s experiments are well established and easily implementable, the paper does not make improper use of these tools easier than it would have been before. But it is possible that its existence will suggest the idea to someone who might not have otherwise had it. 

The authors have attempted to prevent this outcome by making it clear which uses of randomization they believe are appropriate or inappropriate.

\bibliography{refs}

\begin{thebibliography}{46}
\providecommand{\natexlab}[1]{#1}
\providecommand{\url}[1]{\texttt{#1}}
\expandafter\ifx\csname urlstyle\endcsname\relax
  \providecommand{\doi}[1]{doi: #1}\else
  \providecommand{\doi}{doi: \begingroup \urlstyle{rm}\Url}\fi

\bibitem[Agarwal \& Deshpande(2022)Agarwal and Deshpande]{agarwal2022power}
Agarwal, S. and Deshpande, A.
\newblock On the power of randomization in fair classification and
  representation.
\newblock In \emph{Proceedings of the 2022 ACM Conference on Fairness,
  Accountability, and Transparency}, pp.\  1542--1551, 2022.

\bibitem[Agrawal \& Goyal(2012)Agrawal and Goyal]{agrawal2012analysis}
Agrawal, S. and Goyal, N.
\newblock Analysis of thompson sampling for the multi-armed bandit problem.
\newblock In \emph{Conference on learning theory}, pp.\  39--1. JMLR Workshop
  and Conference Proceedings, 2012.

\bibitem[Ajunwa(2021)]{ajunwa2021}
Ajunwa, I.
\newblock An auditing imperative for automated hiring systems.
\newblock \emph{Harvard Journal of Law \& Technology}, 34\penalty0 (2), 2021.

\bibitem[Angelopoulos \& Bates(2021)Angelopoulos and
  Bates]{angelopoulos2021conformal}
Angelopoulos, A.~N. and Bates, S.
\newblock A gentle introduction to conformal prediction and distribution-free
  uncertainty quantification.
\newblock \emph{arXiv preprint arXiv:2107.07511}, 2021.

\bibitem[Black \& Fredrikson(2021)Black and Fredrikson]{black2021leave}
Black, E. and Fredrikson, M.
\newblock Leave-one-out unfairness.
\newblock In \emph{Proceedings of the 2021 ACM Conference on Fairness,
  Accountability, and Transparency}, pp.\  285--295, 2021.

\bibitem[Black et~al.(2022)Black, Raghavan, and Barocas]{black2022model}
Black, E., Raghavan, M., and Barocas, S.
\newblock Model multiplicity: Opportunities, concerns, and solutions.
\newblock In \emph{Proceedings of the 2022 ACM Conference on Fairness,
  Accountability, and Transparency}, pp.\  850--863, 2022.

\bibitem[Bommasani et~al.(2021)Bommasani, Hudson, Adeli, Altman, Arora, von
  Arx, Bernstein, Bohg, Bosselut, Brunskill, et~al.]{bommasani2021foundation}
Bommasani, R., Hudson, D.~A., Adeli, E., Altman, R., Arora, S., von Arx, S.,
  Bernstein, M.~S., Bohg, J., Bosselut, A., Brunskill, E., et~al.
\newblock On the opportunities and risks of foundation models.
\newblock \emph{arXiv preprint arXiv:2108.07258}, 2021.

\bibitem[Bommasani et~al.(2022)Bommasani, Creel, Kumar, Jurafsky, and
  Liang]{bommasani2022homogenization}
Bommasani, R., Creel, K.~A., Kumar, A., Jurafsky, D., and Liang, P.~S.
\newblock Picking on the same person: Does algorithmic monoculture lead to
  outcome homogenization?
\newblock In \emph{Advances in Neural Information Processing Systems},
  volume~35, pp.\  3663--3678, 2022.

\bibitem[Broderick et~al.(2020)Broderick, Giordano, and
  Meager]{broderick2020automatic}
Broderick, T., Giordano, R., and Meager, R.
\newblock An automatic finite-sample robustness metric: when can dropping a
  little data make a big difference?
\newblock \emph{arXiv preprint arXiv:2011.14999}, 2020.

\bibitem[Broome(1990)]{broome1990fairness}
Broome, J.
\newblock Fairness.
\newblock In \emph{Proceedings of the Aristotelian Society}, volume~91, pp.\
  87--101, 1990.

\bibitem[Chin et~al.(2023)Chin, Afsar-Manesh, Bierman, Chang,
  Col{\'o}n-Rodr{\'\i}guez, Dullabh, Duran, Fair, Hernandez-Boussard,
  Hightower, et~al.]{chin2023medicine}
Chin, M.~H., Afsar-Manesh, N., Bierman, A.~S., Chang, C.,
  Col{\'o}n-Rodr{\'\i}guez, C.~J., Dullabh, P., Duran, D.~G., Fair, M.,
  Hernandez-Boussard, T., Hightower, M., et~al.
\newblock Guiding principles to address the impact of algorithm bias on racial
  and ethnic disparities in health and health care.
\newblock \emph{Journal of the American Medical Association}, 6\penalty0 (12),
  2023.

\bibitem[Cohen-Schotanus et~al.(2006)Cohen-Schotanus, Muijtjens, Reinders,
  Agsteribbe, van Rossum, and van~der Vleuten]{CohenSchotanus2006}
Cohen-Schotanus, J., Muijtjens, A. M.~M., Reinders, J.~J., Agsteribbe, J., van
  Rossum, H. J.~M., and van~der Vleuten, C. P.~M.
\newblock The predictive validity of grade point average scores in a partial
  lottery medical school admission system.
\newblock \emph{Medical Education}, 40\penalty0 (10):\penalty0 1012–1019,
  October 2006.
\newblock ISSN 1365-2923.

\bibitem[Collins(2008)]{Collins2008}
Collins, P.~M.
\newblock The consistency of judicial choice.
\newblock \emph{The Journal of Politics}, 70\penalty0 (3):\penalty0 861–873,
  July 2008.
\newblock ISSN 1468-2508.

\bibitem[Cooper et~al.(2023)Cooper, Barocas, De~Sa, and
  Sen]{cooper2023variance}
Cooper, A.~F., Barocas, S., De~Sa, C., and Sen, S.
\newblock Variance, self-consistency, and arbitrariness in fair classification.
\newblock \emph{arXiv preprint arXiv:2301.11562}, 2023.

\bibitem[Creel \& Hellman(2022)Creel and Hellman]{creel_hellman_2022}
Creel, K. and Hellman, D.
\newblock The algorithmic leviathan: Arbitrariness, fairness, and opportunity
  in algorithmic decision-making systems.
\newblock \emph{Canadian Journal of Philosophy}, 52\penalty0 (1):\penalty0
  26–43, 2022.
\newblock \doi{10.1017/can.2022.3}.

\bibitem[Dawid(2017)]{dawid2017individual}
Dawid, P.
\newblock On individual risk.
\newblock \emph{Synthese}, 194\penalty0 (9):\penalty0 3445--3474, 2017.

\bibitem[Ding et~al.(2021)Ding, Hardt, Miller, and Schmidt]{ding2021retiring}
Ding, F., Hardt, M., Miller, J., and Schmidt, L.
\newblock Retiring adult: New datasets for fair machine learning.
\newblock \emph{Advances in neural information processing systems},
  34:\penalty0 6478--6490, 2021.

\bibitem[Dwork et~al.(2021)Dwork, Kim, Reingold, Rothblum, and
  Yona]{dwork2021outcome}
Dwork, C., Kim, M.~P., Reingold, O., Rothblum, G.~N., and Yona, G.
\newblock Outcome indistinguishability.
\newblock In \emph{Proceedings of the 53rd Annual ACM SIGACT Symposium on
  Theory of Computing}, pp.\  1095--1108, 2021.

\bibitem[Eidelson(2021)]{eidelson2021patterned}
Eidelson, B.
\newblock Patterned inequality, compounding injustice, and algorithmic
  prediction.
\newblock \emph{American Journal of Law and Equality}, 1:\penalty0 252--276,
  2021.

\bibitem[Ganesh et~al.(2023)Ganesh, Chang, Strobel, and
  Shokri]{ganesh2023impact}
Ganesh, P., Chang, H., Strobel, M., and Shokri, R.
\newblock On the impact of machine learning randomness on group fairness.
\newblock In \emph{Proceedings of the 2023 ACM Conference on Fairness,
  Accountability, and Transparency}, pp.\  1789--1800, 2023.

\bibitem[Grgi{\'c}-Hla{\v{c}}a et~al.(2017)Grgi{\'c}-Hla{\v{c}}a, Zafar,
  Gummadi, and Weller]{grgic2017ensemble}
Grgi{\'c}-Hla{\v{c}}a, N., Zafar, M.~B., Gummadi, K.~P., and Weller, A.
\newblock On fairness, diversity and randomness in algorithmic decision making.
\newblock \emph{arXiv preprint arXiv:1706.10208}, 2017.

\bibitem[Hastings et~al.(2006)Hastings, Kane, and Staiger]{hastings2006}
Hastings, J., Kane, T., and Staiger, D.
\newblock Preferences and heterogeneous treatment effects in a public school
  choice lottery, 2006.
\newblock URL \url{http://dx.doi.org/10.3386/w12145}.

\bibitem[Hellman(2018)]{hellman2018compounding}
Hellman, D.
\newblock Indirect discrimination and the duty to avoid compounding injustice.
\newblock \emph{Foundations of Indirect Discrimination Law, Hart Publishing
  Company}, pp.\  2017--53, 2018.

\bibitem[Hooker(2005)]{hooker2005fairness}
Hooker, B.
\newblock Fairness.
\newblock \emph{Ethical theory and moral practice}, 8:\penalty0 329--352, 2005.

\bibitem[Hsu \& Calmon(2022)Hsu and Calmon]{hsu2022rashomon}
Hsu, H. and Calmon, F.
\newblock Rashomon capacity: A metric for predictive multiplicity in
  classification.
\newblock In Koyejo, S., Mohamed, S., Agarwal, A., Belgrave, D., Cho, K., and
  Oh, A. (eds.), \emph{Advances in Neural Information Processing Systems},
  volume~35, pp.\  28988--29000. Curran Associates, Inc., 2022.

\bibitem[Jain et~al.(2024)Jain, Suriyakumar, Creel, and
  Wilson]{jain2023pluralism}
Jain, S., Suriyakumar, V., Creel, K., and Wilson, A.
\newblock Algorithmic pluralism: A structural approach to equal opportunity.
\newblock In \emph{Proceedings of the 2024 ACM Conference on Fairness,
  Accountability, and Transparency}, FAccT '24, pp.\  197–206, 2024.
\newblock URL \url{https://doi.org/10.1145/3630106.3658899}.

\bibitem[Joseph et~al.(2016{\natexlab{a}})Joseph, Kearns, Morgenstern, and
  Roth]{Roth2016}
Joseph, M., Kearns, M., Morgenstern, J.~H., and Roth, A.
\newblock Fairness in learning: Classic and contextual bandits.
\newblock In Lee, D., Sugiyama, M., Luxburg, U., Guyon, I., and Garnett, R.
  (eds.), \emph{Advances in Neural Information Processing Systems}, volume~29.
  Curran Associates, Inc., 2016{\natexlab{a}}.
\newblock URL
  \url{https://proceedings.neurips.cc/paper_files/paper/2016/file/eb163727917cbba1eea208541a643e74-Paper.pdf}.

\bibitem[Joseph et~al.(2016{\natexlab{b}})Joseph, Kearns, Morgenstern, and
  Roth]{joseph2016fairness}
Joseph, M., Kearns, M., Morgenstern, J.~H., and Roth, A.
\newblock Fairness in learning: Classic and contextual bandits.
\newblock \emph{Advances in neural information processing systems}, 29,
  2016{\natexlab{b}}.

\bibitem[Kirkpatrick \& Eastwood(2015)Kirkpatrick and
  Eastwood]{kirkpatrick2015broome}
Kirkpatrick, J.~R. and Eastwood, N.
\newblock Broome's theory of fairness and the problem of quantifying the
  strengths of claims.
\newblock \emph{Utilitas}, 27\penalty0 (1):\penalty0 82--91, 2015.

\bibitem[Kleinberg \& Raghavan(2021)Kleinberg and
  Raghavan]{kleinberg2021monoculture}
Kleinberg, J. and Raghavan, M.
\newblock Algorithmic monoculture and social welfare.
\newblock \emph{Proceedings of the National Academy of Sciences}, 118\penalty0
  (22):\penalty0 e2018340118, 2021.

\bibitem[Lechner et~al.(2020)Lechner, Knaus, Huber, Frölich, Behncke, Mellace,
  and Strittmatter]{swissData}
Lechner, M., Knaus, M., Huber, M., Frölich, M., Behncke, S., Mellace, G., and
  Strittmatter, A.
\newblock Swiss active labor market policy evaluation [dataset].
\newblock Distributed by FORS, Lausanne, 2020.

\bibitem[Li et~al.(2020)Li, Raymond, and Bergman]{li2020hiring}
Li, D., Raymond, L.~R., and Bergman, P.
\newblock Hiring as exploration.
\newblock Technical report, National Bureau of Economic Research, 2020.

\bibitem[Ludwig \& Mullainathan(2021)Ludwig and
  Mullainathan]{ludwig2021fragile}
Ludwig, J. and Mullainathan, S.
\newblock Fragile algorithms and fallible decision-makers: lessons from the
  justice system.
\newblock \emph{Journal of Economic Perspectives}, 35\penalty0 (4):\penalty0
  71--96, 2021.

\bibitem[Marx et~al.(2020)Marx, Calmon, and Ustun]{marx2020predictive}
Marx, C., Calmon, F., and Ustun, B.
\newblock Predictive multiplicity in classification.
\newblock In \emph{International Conference on Machine Learning}, pp.\
  6765--6774. PMLR, 2020.

\bibitem[McCreary et~al.(2023)McCreary, Essien, Chang, Butler, Pathak,
  S\"{o}nmez, \"{U}nver, Steiner, Chrisman, Angus, and White]{McCreary2023}
McCreary, E.~K., Essien, U.~R., Chang, C.-C.~H., Butler, R.~A., Pathak, P.,
  S\"{o}nmez, T., \"{U}nver, M.~U., Steiner, A., Chrisman, M., Angus, D.~C.,
  and White, D.~B.
\newblock Weighted lottery to equitably allocate scarce supply of covid-19
  monoclonal antibody.
\newblock \emph{JAMA Health Forum}, 4\penalty0 (9):\penalty0 e232774, September
  2023.
\newblock ISSN 2689-0186.
\newblock \doi{10.1001/jamahealthforum.2023.2774}.
\newblock URL \url{http://dx.doi.org/10.1001/jamahealthforum.2023.2774}.

\bibitem[Mitchell et~al.(2021)Mitchell, Potash, Barocas, D'Amour, and
  Lum]{mitchell2021algorithmic}
Mitchell, S., Potash, E., Barocas, S., D'Amour, A., and Lum, K.
\newblock Algorithmic fairness: Choices, assumptions, and definitions.
\newblock \emph{Annual Review of Statistics and Its Application}, 8:\penalty0
  141--163, 2021.

\bibitem[Nawrat(2023)]{hirevueArticle}
Nawrat, A.
\newblock Inside hirevue's acquisition of modern hire.
\newblock
  \url{https://www.unleash.ai/hr-technology/inside-hirevues-acquisition-of-modern-hire/},
  2023.

\bibitem[Passi \& Barocas(2019)Passi and Barocas]{passi2019problem}
Passi, S. and Barocas, S.
\newblock Problem formulation and fairness.
\newblock In \emph{Proceedings of the conference on fairness, accountability,
  and transparency}, pp.\  39--48, 2019.

\bibitem[Peng \& Garg(2023)Peng and Garg]{peng2023monoculture}
Peng, K. and Garg, N.
\newblock Monoculture in matching markets.
\newblock \emph{arXiv preprint arXiv:2312.09841}, 2023.

\bibitem[Pimentel et~al.(2014)Pimentel, Clifton, Clifton, and
  Tarassenko]{pimentel2014outlier}
Pimentel, M.~A., Clifton, D.~A., Clifton, L., and Tarassenko, L.
\newblock A review of novelty detection.
\newblock \emph{Signal processing}, 99:\penalty0 215--249, 2014.

\bibitem[Raghavan et~al.(2020)Raghavan, Barocas, Kleinberg, and
  Levy]{raghavan2020hiring}
Raghavan, M., Barocas, S., Kleinberg, J., and Levy, K.
\newblock Mitigating bias in algorithmic hiring: Evaluating claims and
  practices.
\newblock In \emph{Proceedings of the 2020 Conference on Fairness,
  Accountability, and Transparency}, pp.\  469--481, 2020.

\bibitem[Roth et~al.(2023)Roth, Tolbert, and Weinstein]{roth2023reconciling}
Roth, A., Tolbert, A., and Weinstein, S.
\newblock Reconciling individual probability forecasts.
\newblock In \emph{Proceedings of the 2023 ACM Conference on Fairness,
  Accountability, and Transparency}, pp.\  101--110, 2023.

\bibitem[Schmidt et~al.(2021)Schmidt, Roberts, and
  Eneanya]{schmidt2021rationing}
Schmidt, H., Roberts, D.~E., and Eneanya, N.~D.
\newblock Rationing, racism and justice: advancing the debate around
  ``colourblind'' {COVID-19} ventilator allocation.
\newblock \emph{Journal of Medical Ethics}, 2021.

\bibitem[Sharifi-Malvajerdi et~al.(2019)Sharifi-Malvajerdi, Kearns, and
  Roth]{Roth2019}
Sharifi-Malvajerdi, S., Kearns, M., and Roth, A.
\newblock Average individual fairness: Algorithms, generalization and
  experiments.
\newblock In Wallach, H., Larochelle, H., Beygelzimer, A., d\textquotesingle
  Alch\'{e}-Buc, F., Fox, E., and Garnett, R. (eds.), \emph{Advances in Neural
  Information Processing Systems}, volume~32. Curran Associates, Inc., 2019.
\newblock URL
  \url{https://proceedings.neurips.cc/paper_files/paper/2019/file/0e1feae55e360ff05fef58199b3fa521-Paper.pdf}.

\bibitem[Singh et~al.(2021)Singh, Kempe, and Joachims]{singh2021fairness}
Singh, A., Kempe, D., and Joachims, T.
\newblock Fairness in ranking under uncertainty.
\newblock In \emph{Advances in Neural Information Processing Systems},
  volume~34, pp.\  11896--11908, 2021.

\bibitem[Toups et~al.(2023)Toups, Bommasani, Creel, Bana, Jurafsky, and
  Liang]{toups2023ecosystem}
Toups, C., Bommasani, R., Creel, K., Bana, S., Jurafsky, D., and Liang, P.
\newblock Ecosystem-level analysis of deployed machine learning reveals
  homogeneous outcomes.
\newblock In \emph{Advances in Neural Information Processing Systems}, 2023.

\end{thebibliography}
\bibliographystyle{icml2024}

\newpage
\appendix
\onecolumn

\section{Appendix}

\subsection{When Claims Are Known Experiments}
\label{app:claims_known}

\begin{figure*}[t!]
\centering
\caption{Normal Distribution of Claims}
\subfloat[\centering Distribution of Claims]{
\includegraphics[width=0.4\columnwidth]{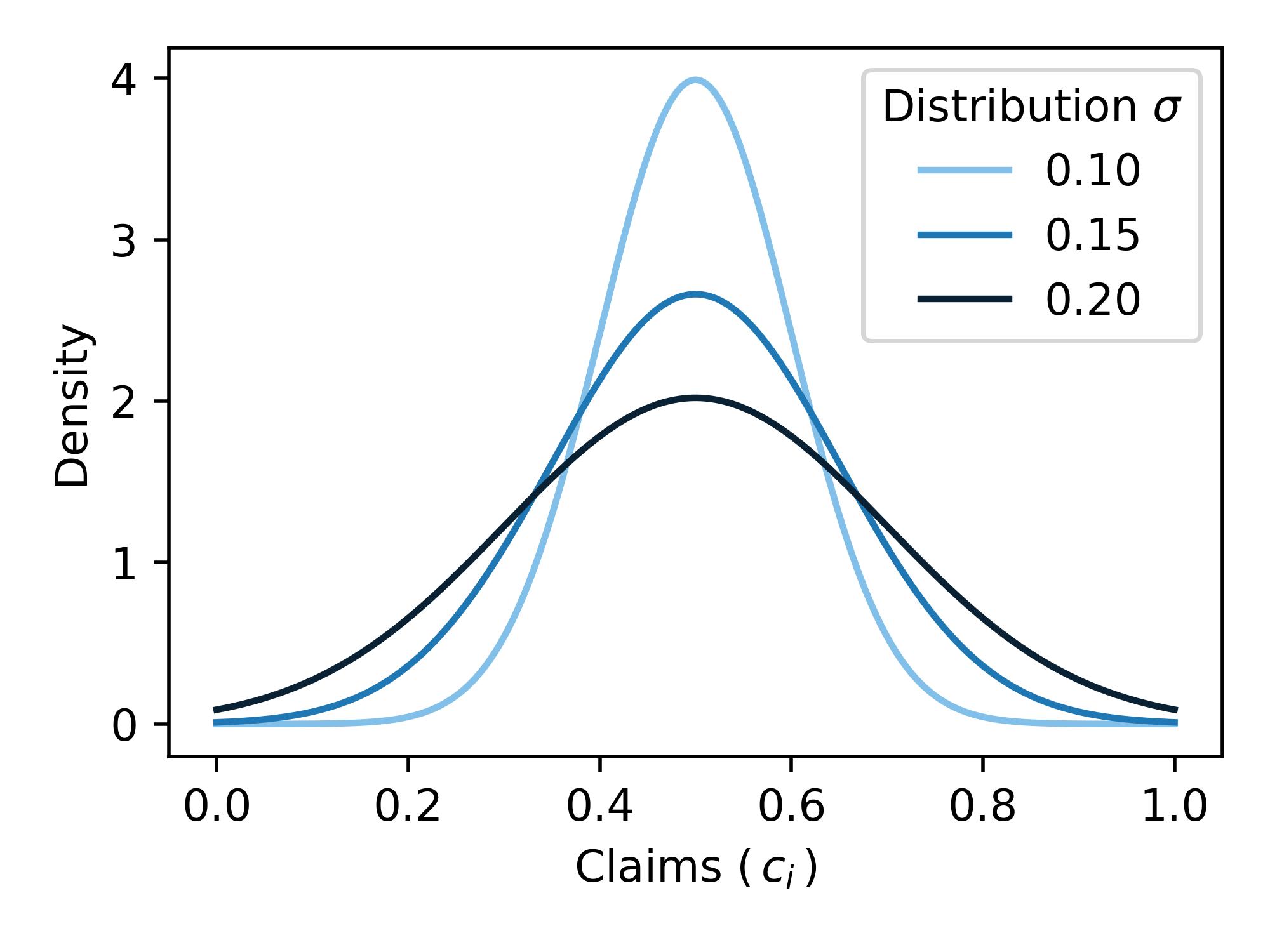}
} 
\subfloat[\centering Expected Utility for Top $k$ v. BF Lottery in Ex~\ref{ex:bf_lottery}]{
\includegraphics[width=0.4\columnwidth]{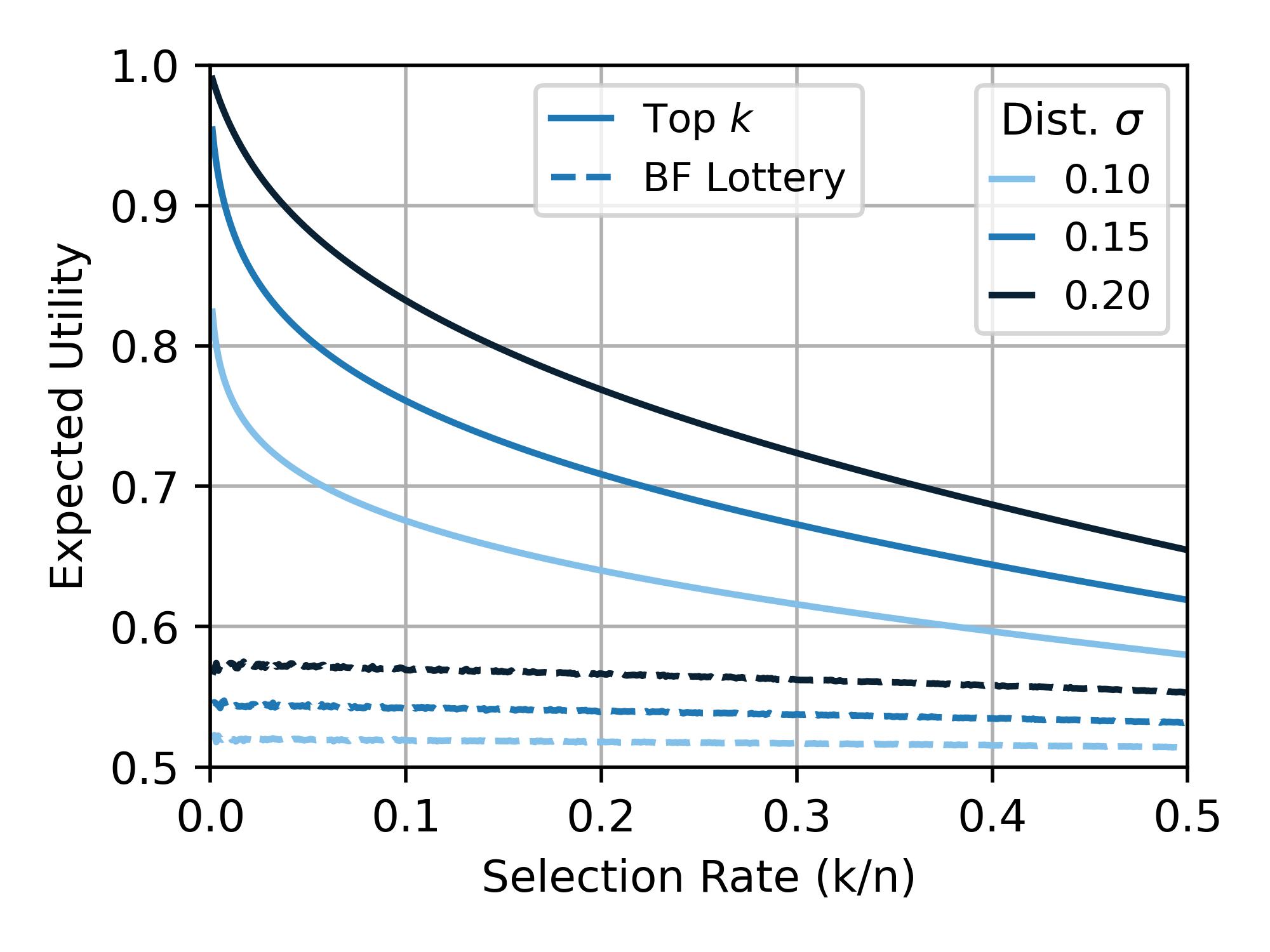}
} \\
\text{(c) - (e) Expected Utility for Varying Partial BF Randomization Rates in Ex~\ref{ex:partial_bf_lottery}} \\[2mm]
\subfloat[\centering Selection Rate = $0.1$]{
\includegraphics[width=0.32\columnwidth]{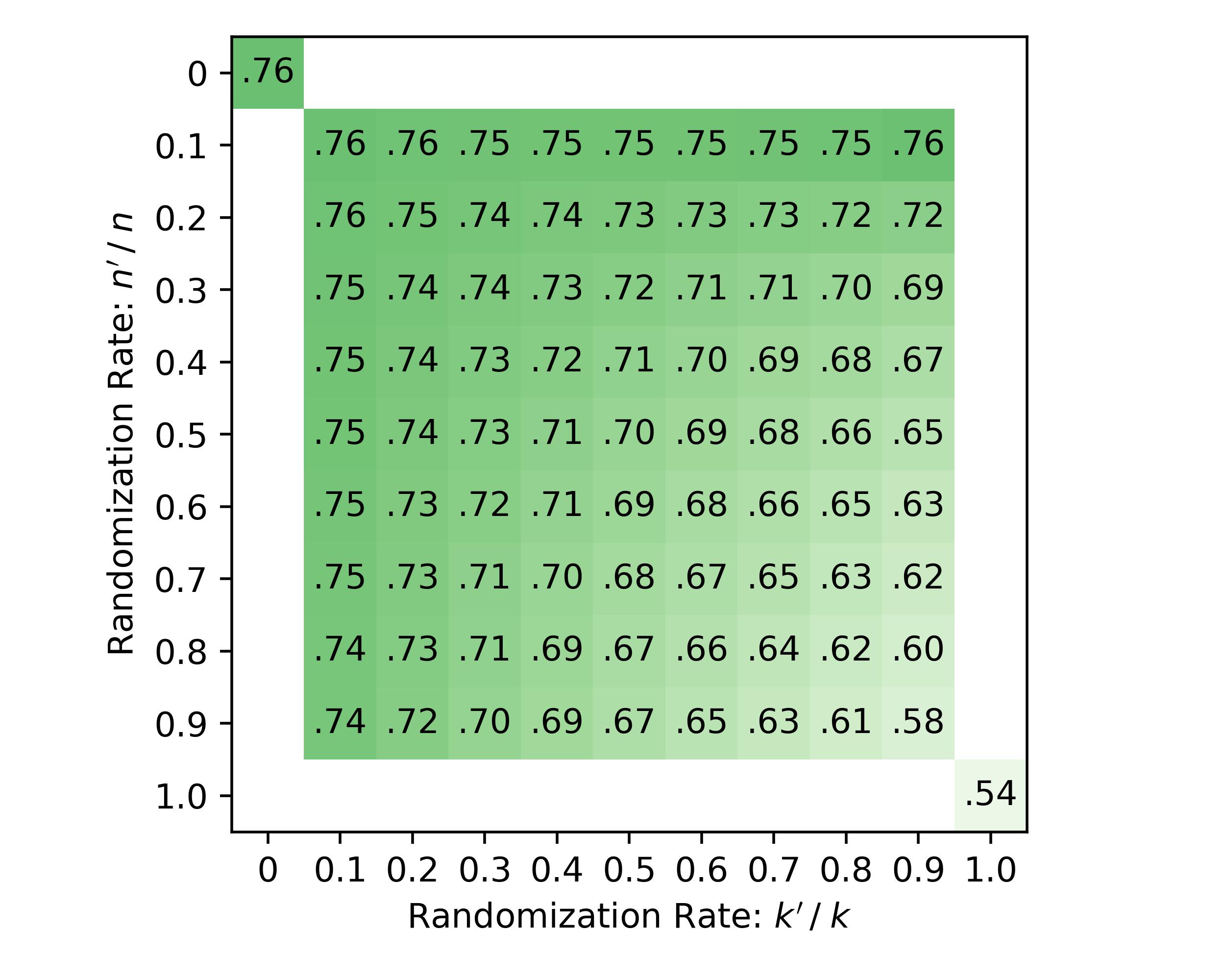}
} 
\subfloat[\centering Selection Rate = $0.25$]{
\includegraphics[width=0.32\columnwidth]{figures/normal_partial_25.jpg}
}
\subfloat[\centering Selection Rate = $0.5$]{
\includegraphics[width=0.32\columnwidth]{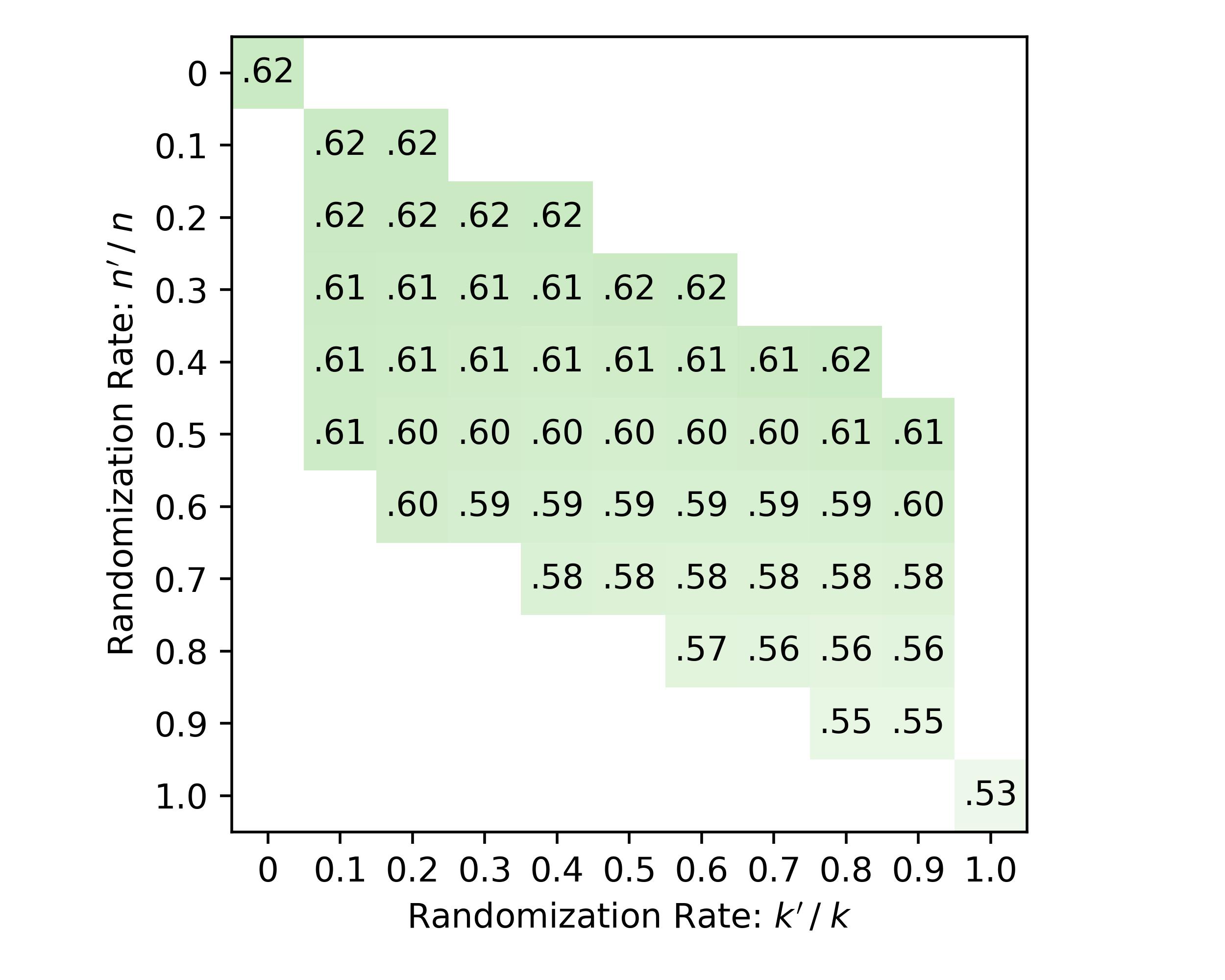}
}
\\
\text{(f) - (h) Systemic Exclusion Rate v. Expected Utility Across Varying Partial BF Randomization Rates} \\[2mm]
\subfloat[\centering Selection Rate = $0.1$]{
\includegraphics[width=0.32\columnwidth]{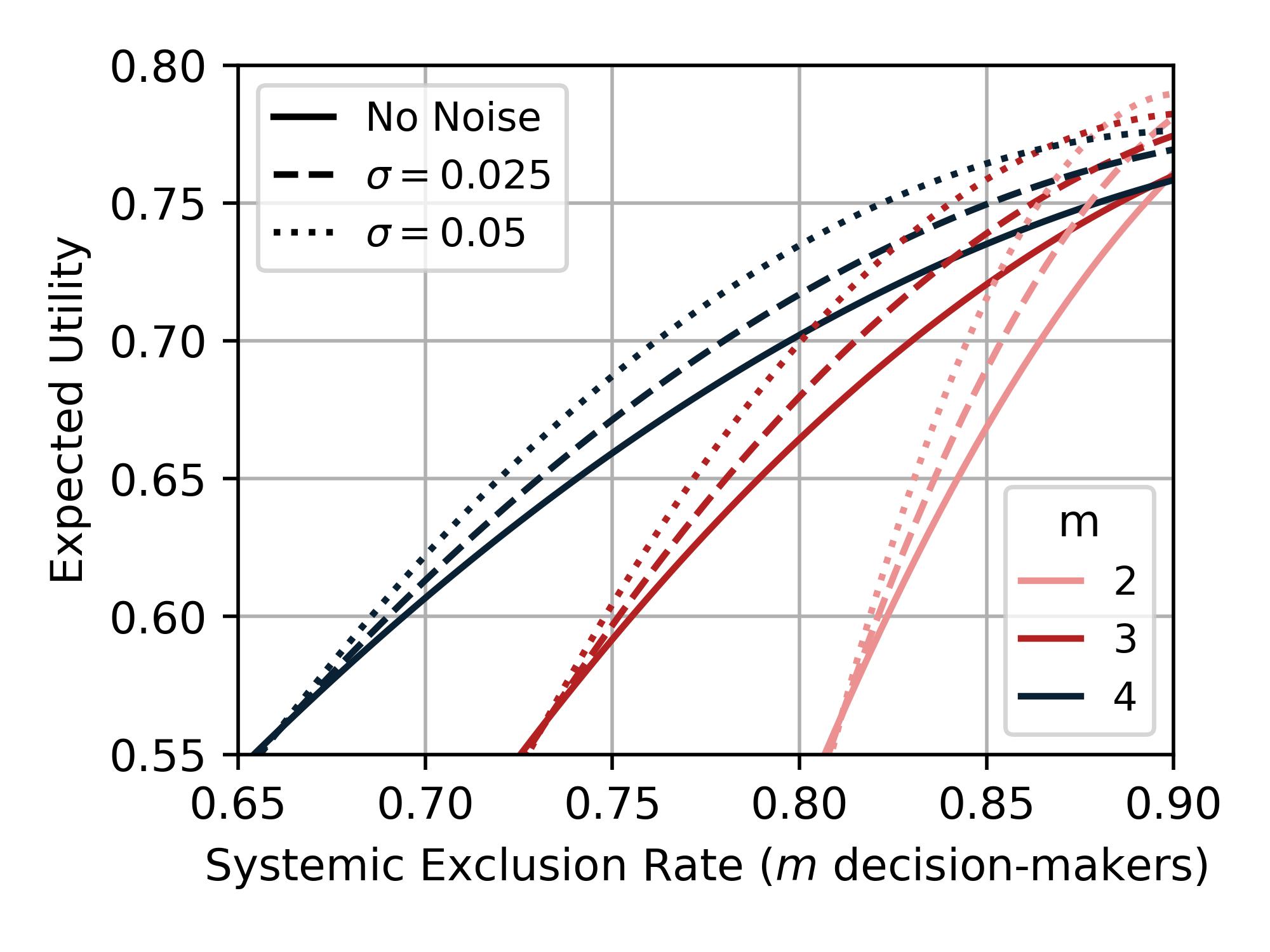}
} 
\subfloat[\centering Selection Rate = $0.25$]{
\includegraphics[width=0.32\columnwidth]{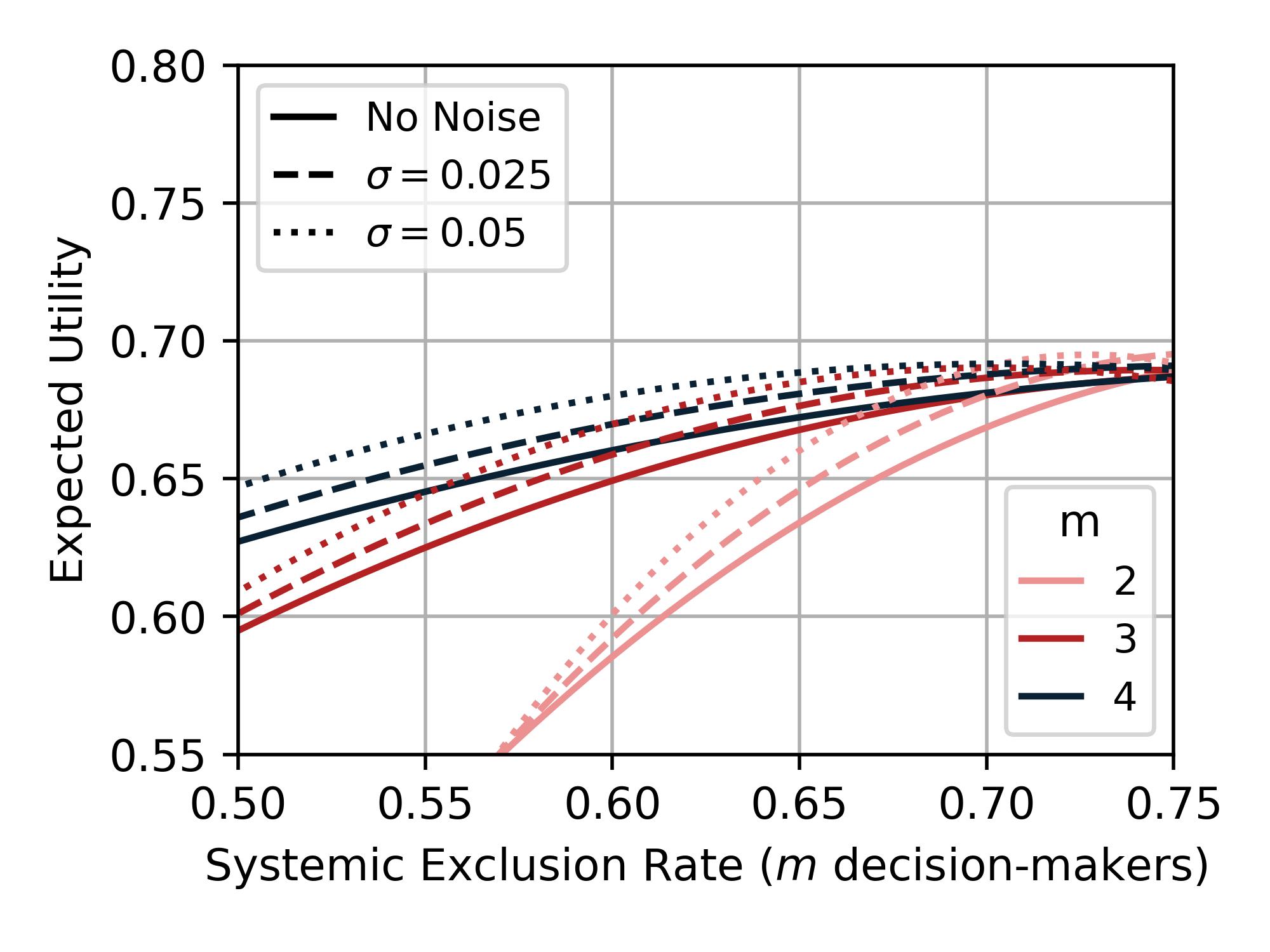}
}
\subfloat[\centering Selection Rate = $0.5$]{
\includegraphics[width=0.32\columnwidth]{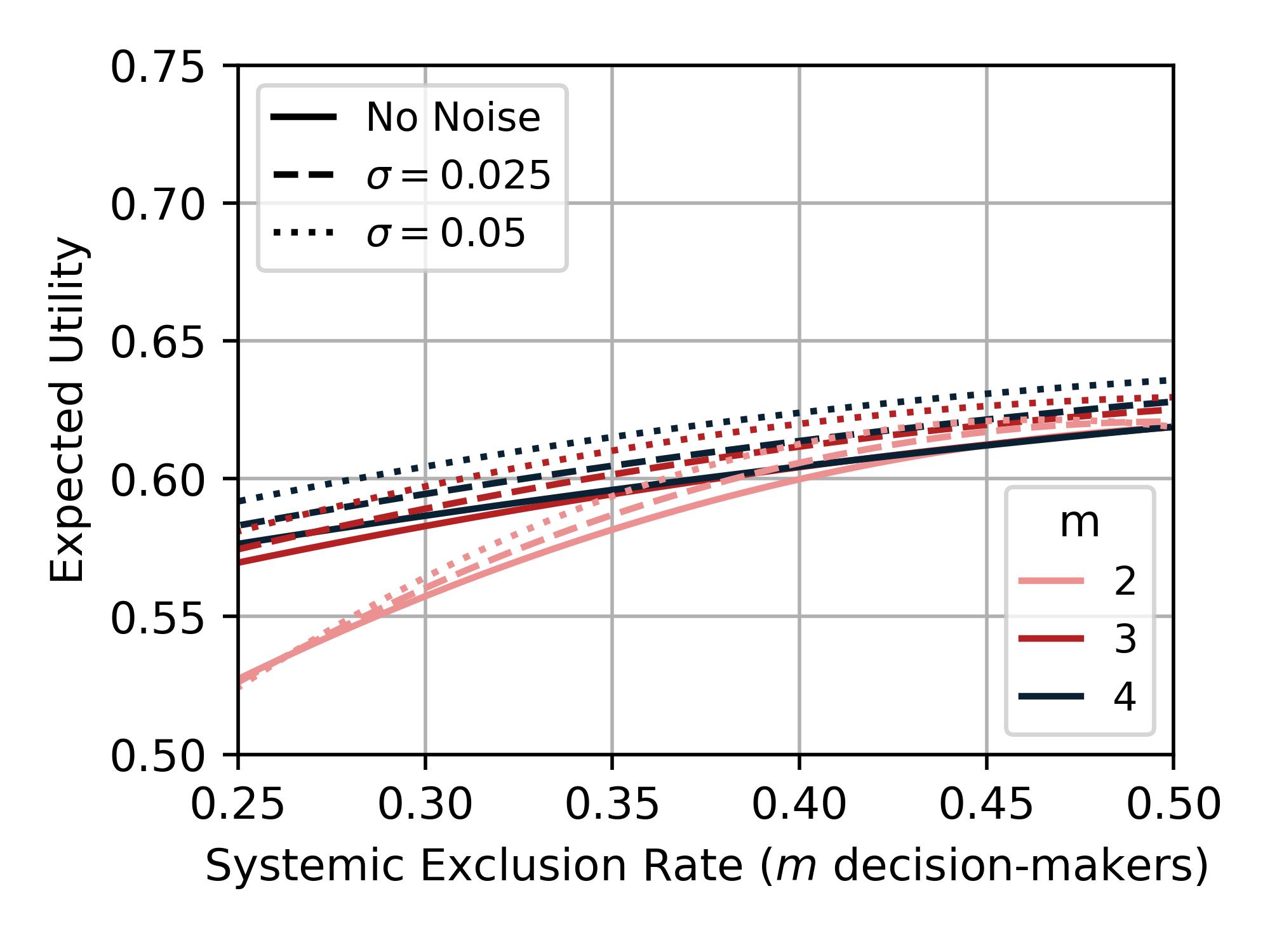}
}
\label{fig:normal}
\end{figure*}

\begin{figure*}[h!]
\centering
\caption{Inverse Normal Distribution of Claims}
\subfloat[\centering Distribution of Claims]{
\includegraphics[width=0.4\columnwidth]{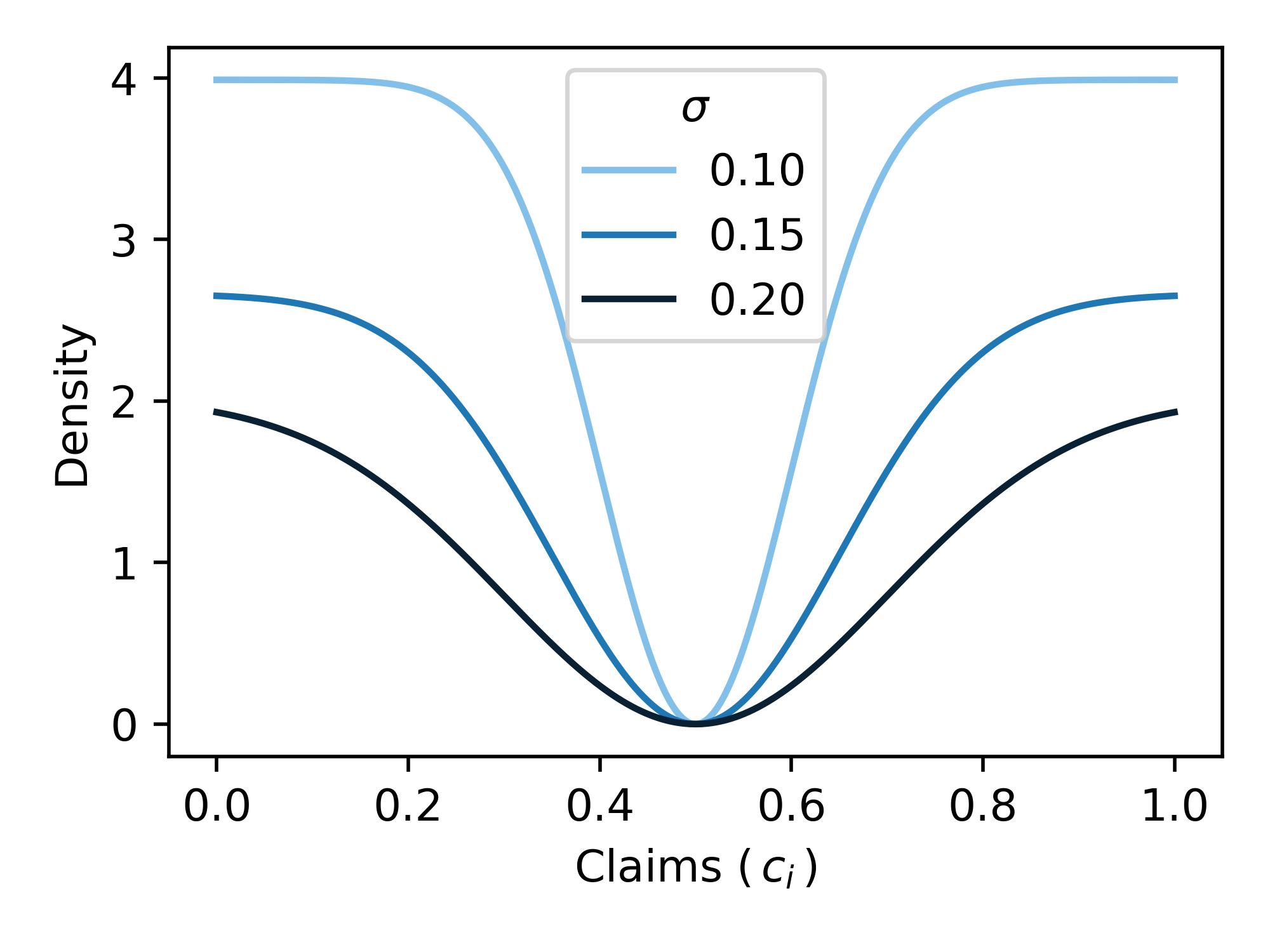}
} 
\subfloat[\centering Expected Utility for Top $k$ v. BF Lottery in Ex~\ref{ex:bf_lottery}]{
\includegraphics[width=0.4\columnwidth]{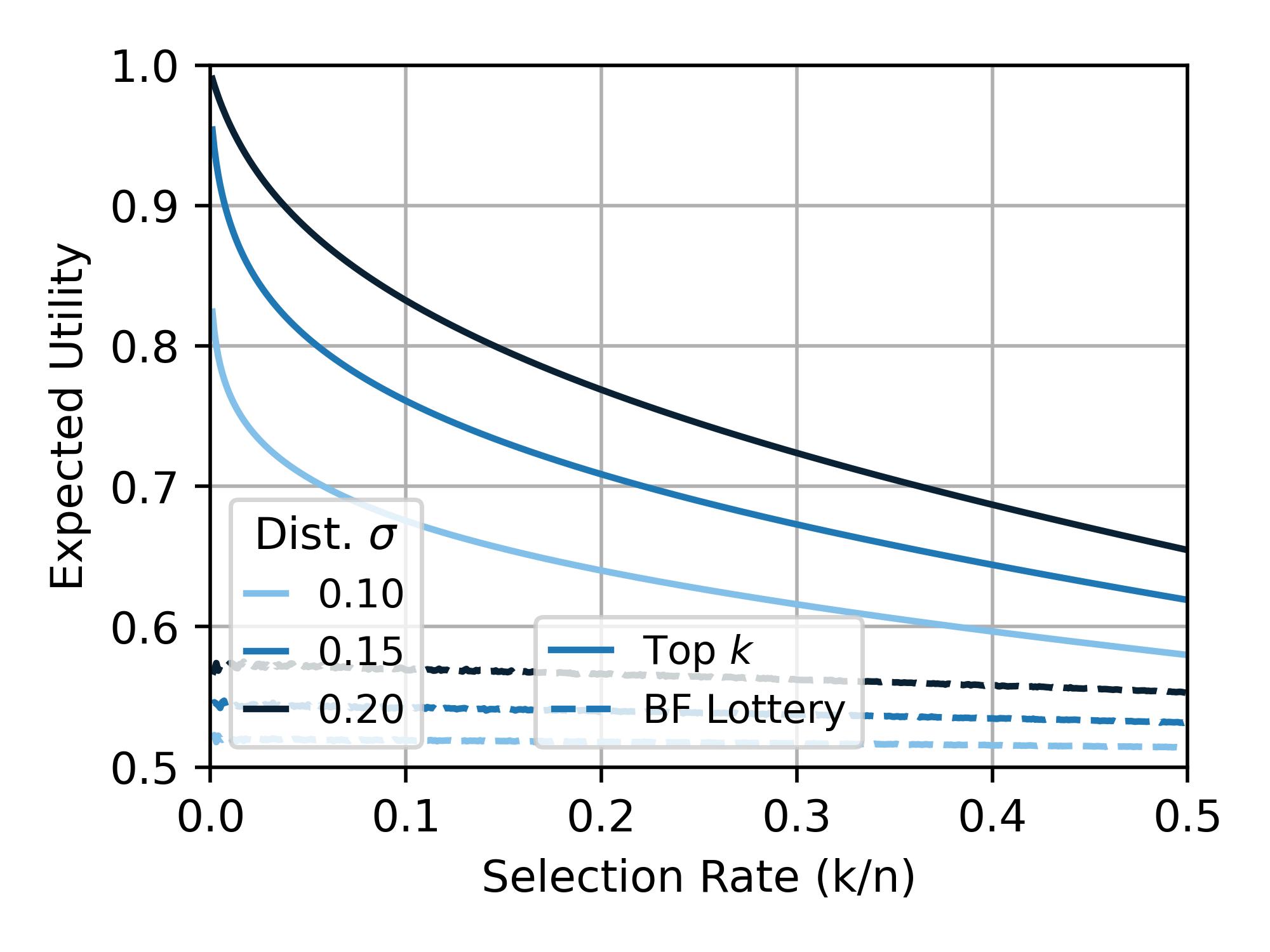}
} \\
\text{(c) - (e) Expected Utility for Varying Partial BF Randomization Rates in Ex~\ref{ex:partial_bf_lottery}} \\[2mm]
\subfloat[\centering Selection Rate = $0.1$]{
\includegraphics[width=0.32\columnwidth]{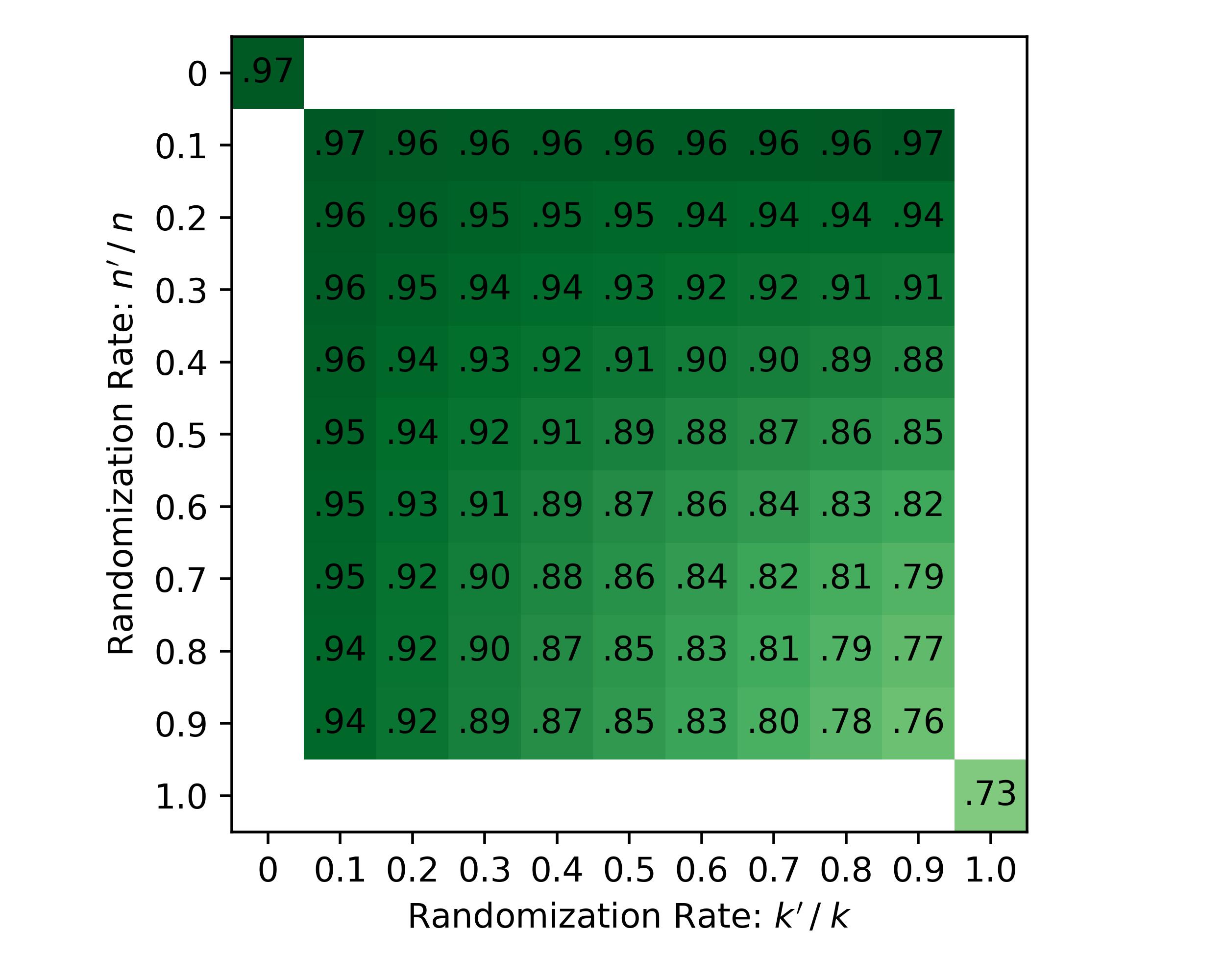}
} 
\subfloat[\centering Selection Rate = $0.25$]{
\includegraphics[width=0.32\columnwidth]{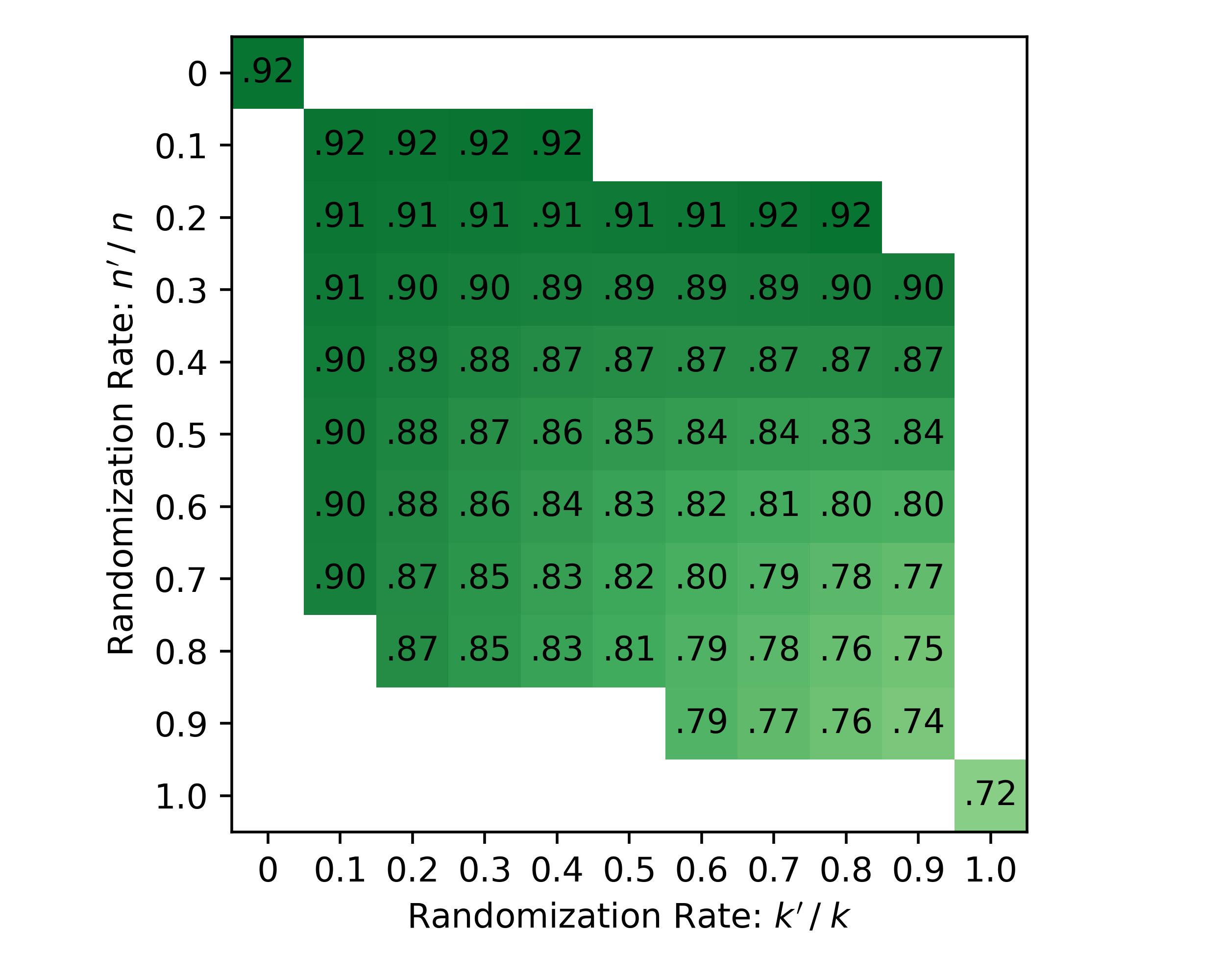}
}
\subfloat[\centering Selection Rate = $0.5$]{
\includegraphics[width=0.32\columnwidth]{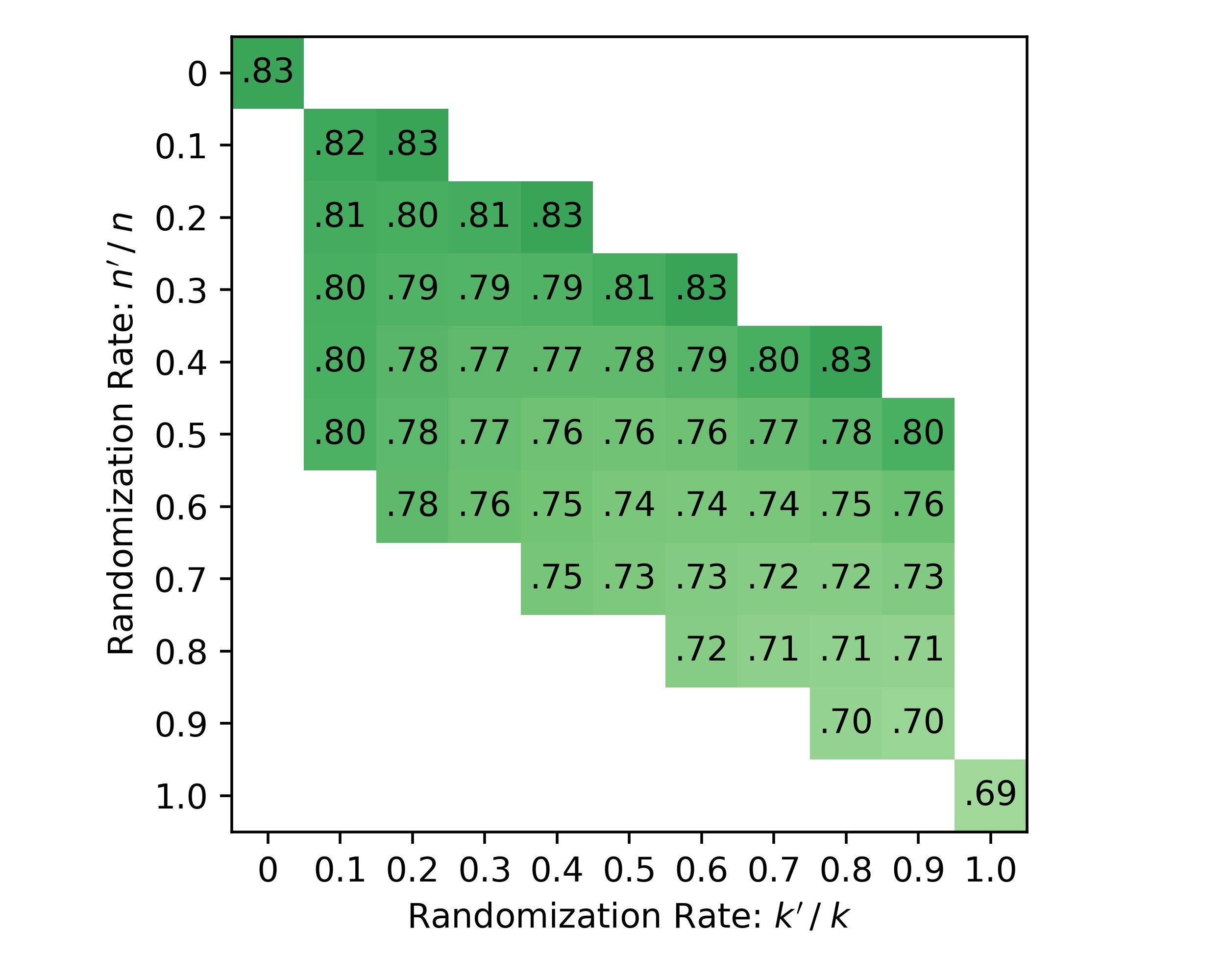}
}
\\
\text{(f) - (h) Systemic Exclusion Rate v. Expected Utility Across Varying Partial BF Randomization Rates} \\[2mm]
\subfloat[\centering Selection Rate = $0.1$]{
\includegraphics[width=0.32\columnwidth]{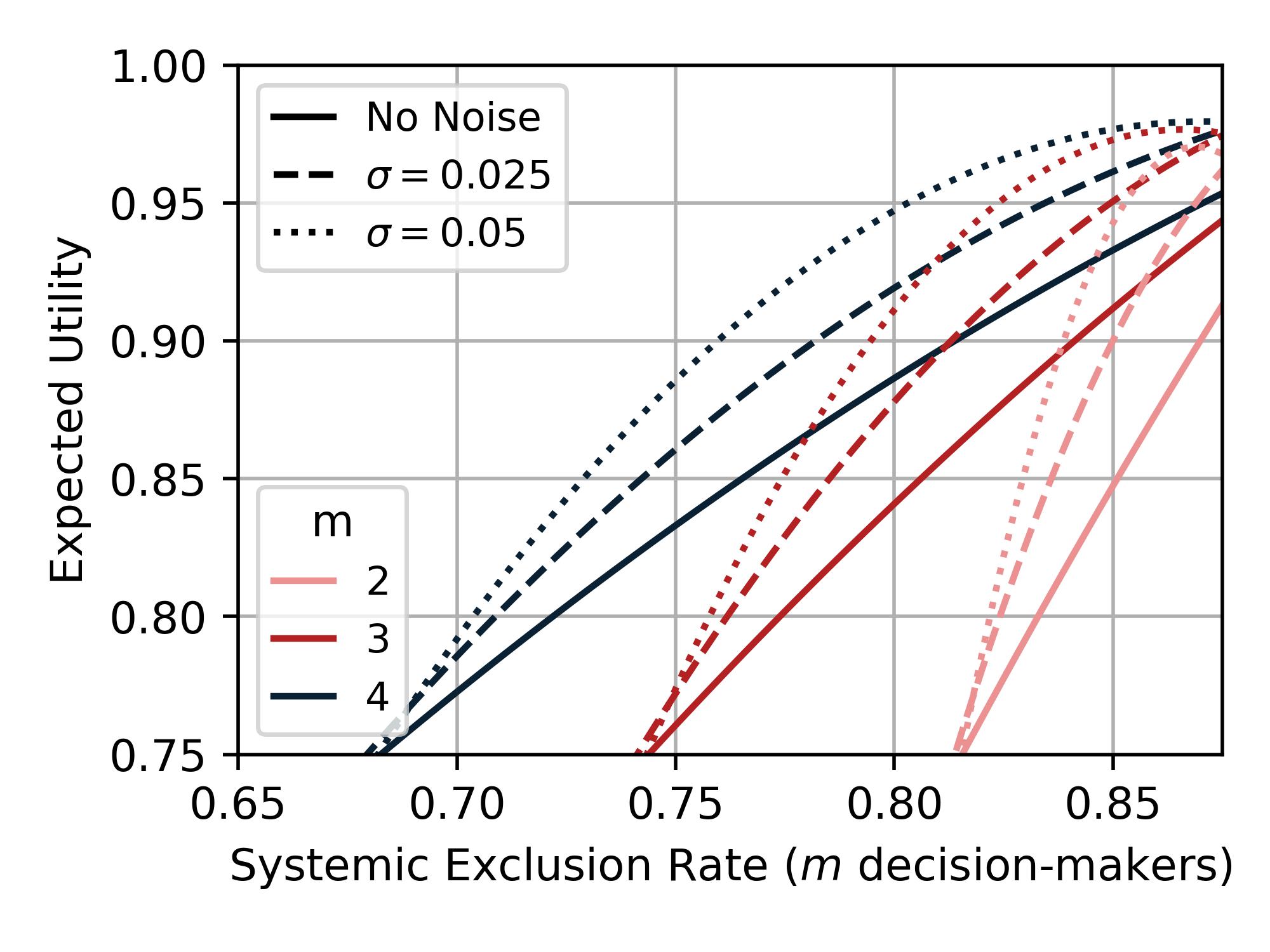}
} 
\subfloat[\centering Selection Rate = $0.25$]{
\includegraphics[width=0.32\columnwidth]{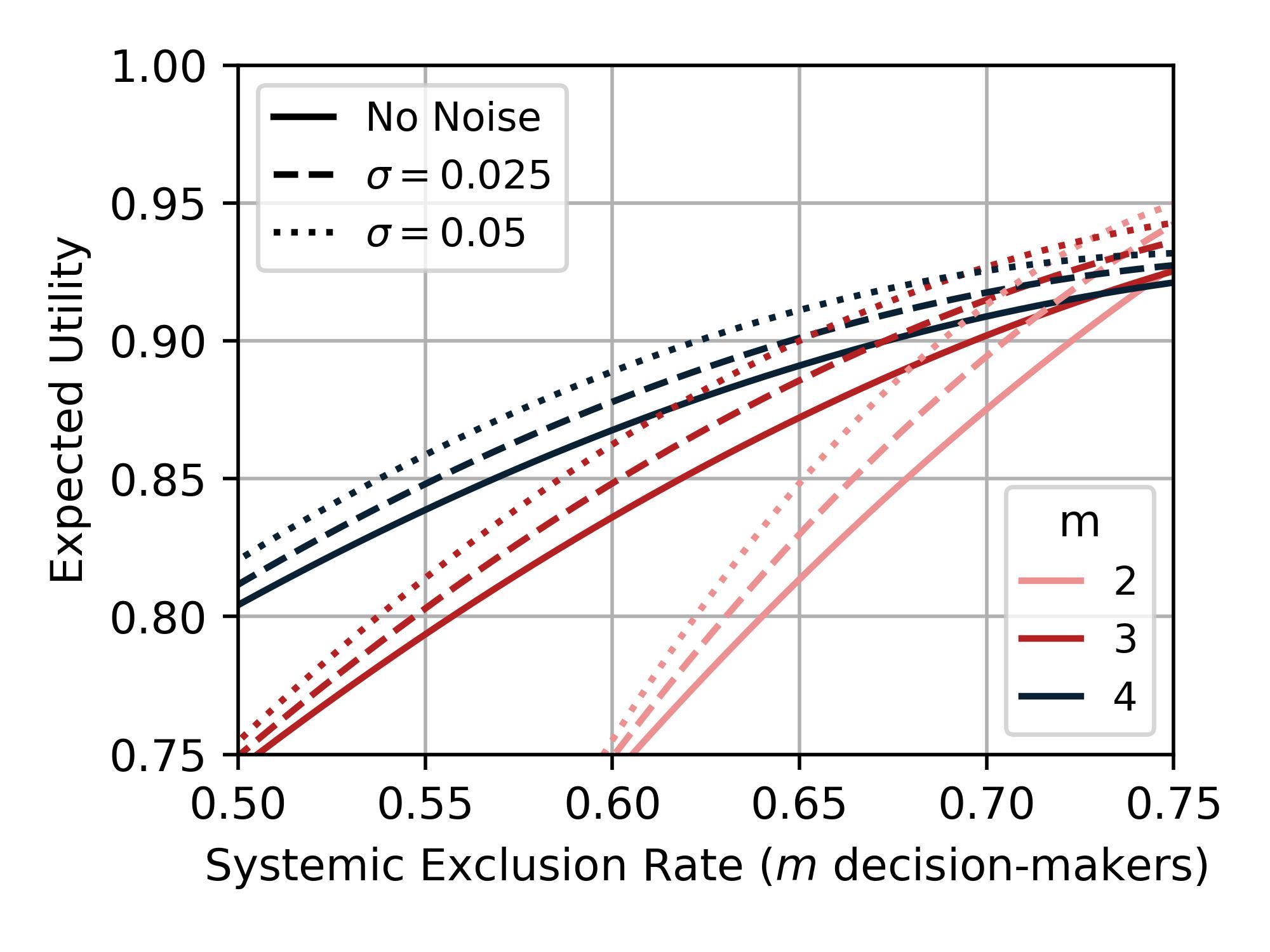}
}
\subfloat[\centering Selection Rate = $0.5$]{
\includegraphics[width=0.32\columnwidth]{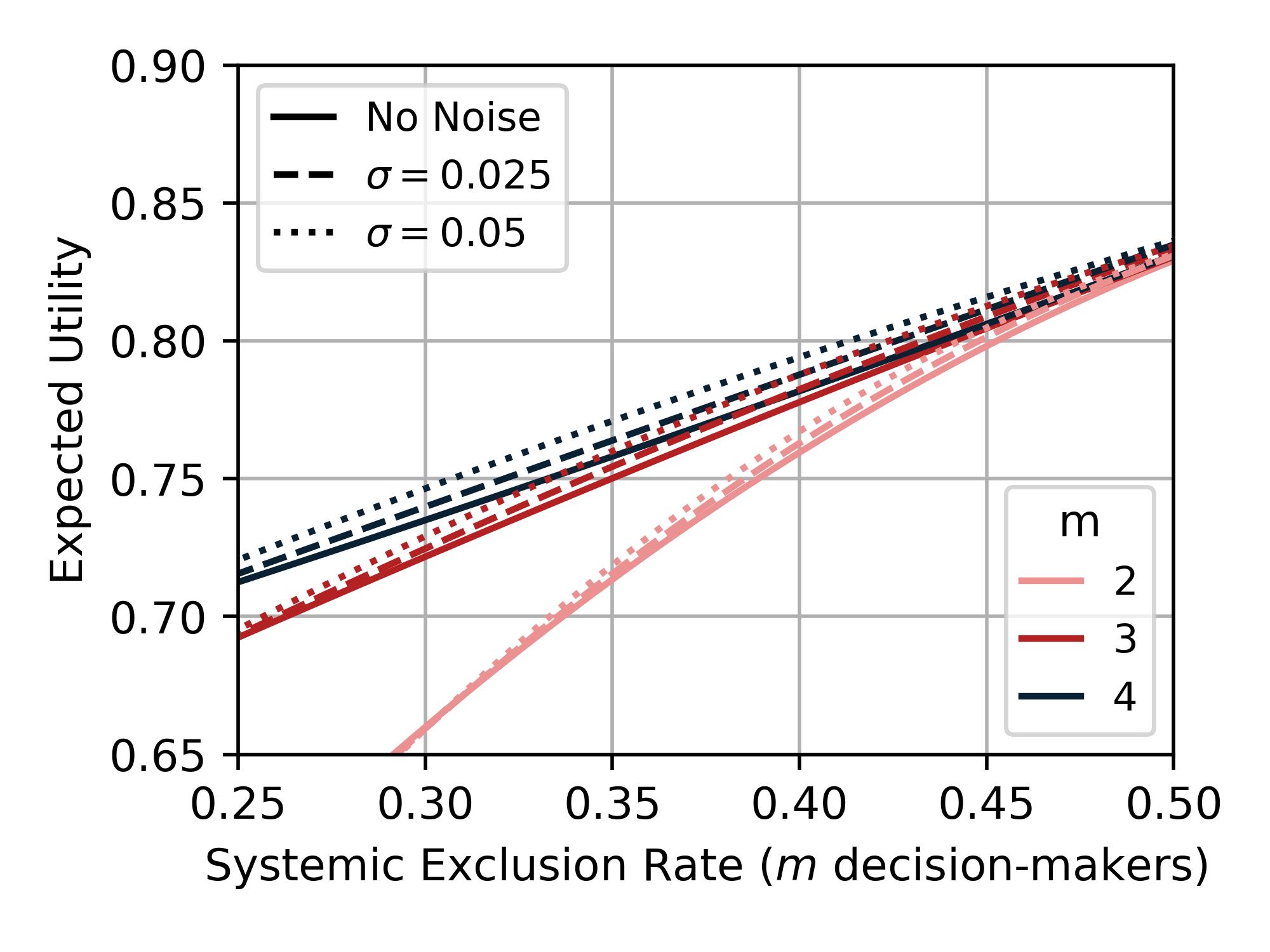}
}
\label{fig:inv_normal}
\end{figure*}

\begin{figure*}[h!]
\centering
\caption{Pareto Distribution of Claims}
\subfloat[\centering Distribution of Claims]{
\includegraphics[width=0.4\columnwidth]{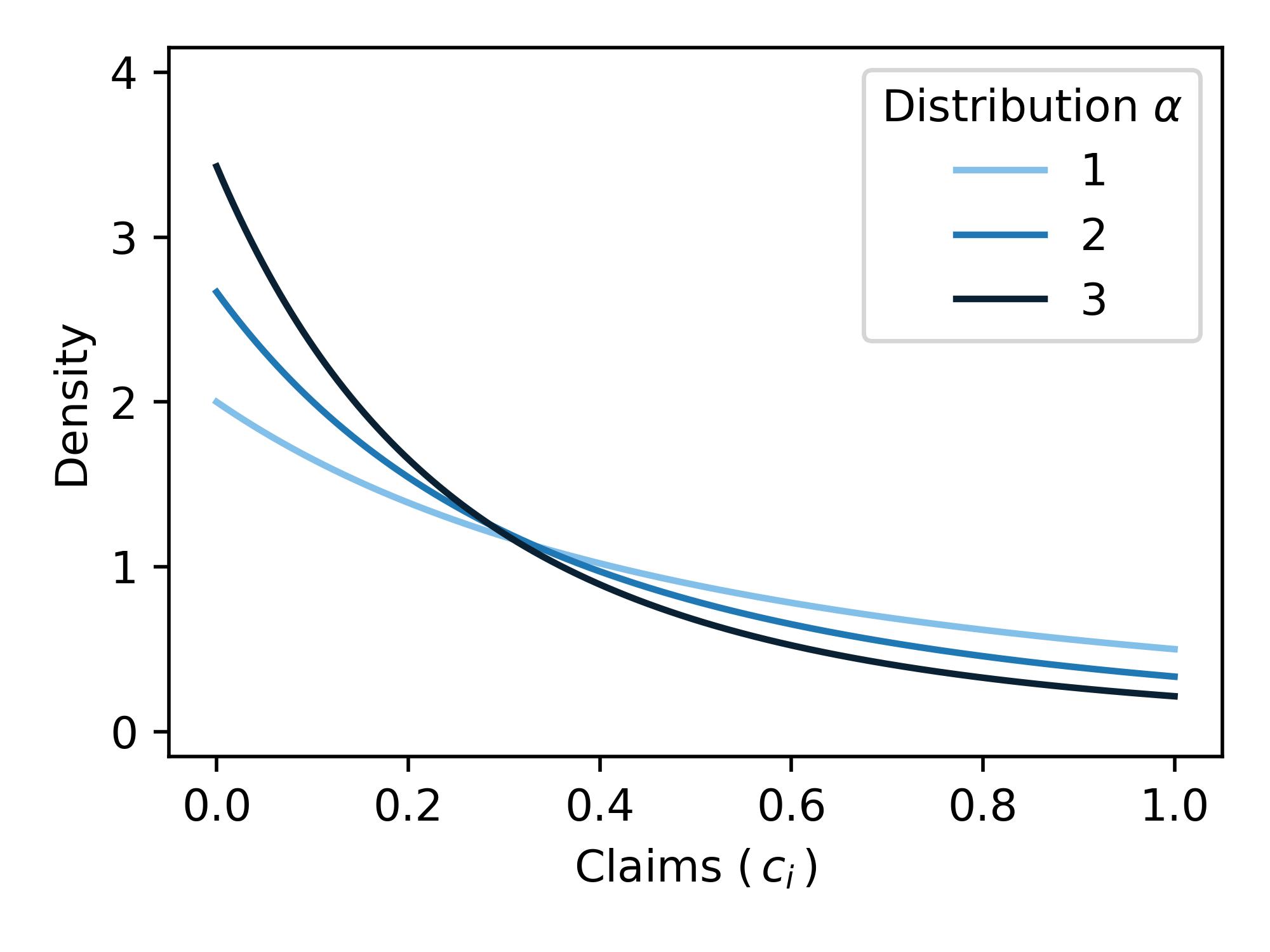}
} 
\subfloat[\centering Expected Utility for Top $k$ v. BF Lottery in Ex~\ref{ex:bf_lottery}]{
\includegraphics[width=0.4\columnwidth]{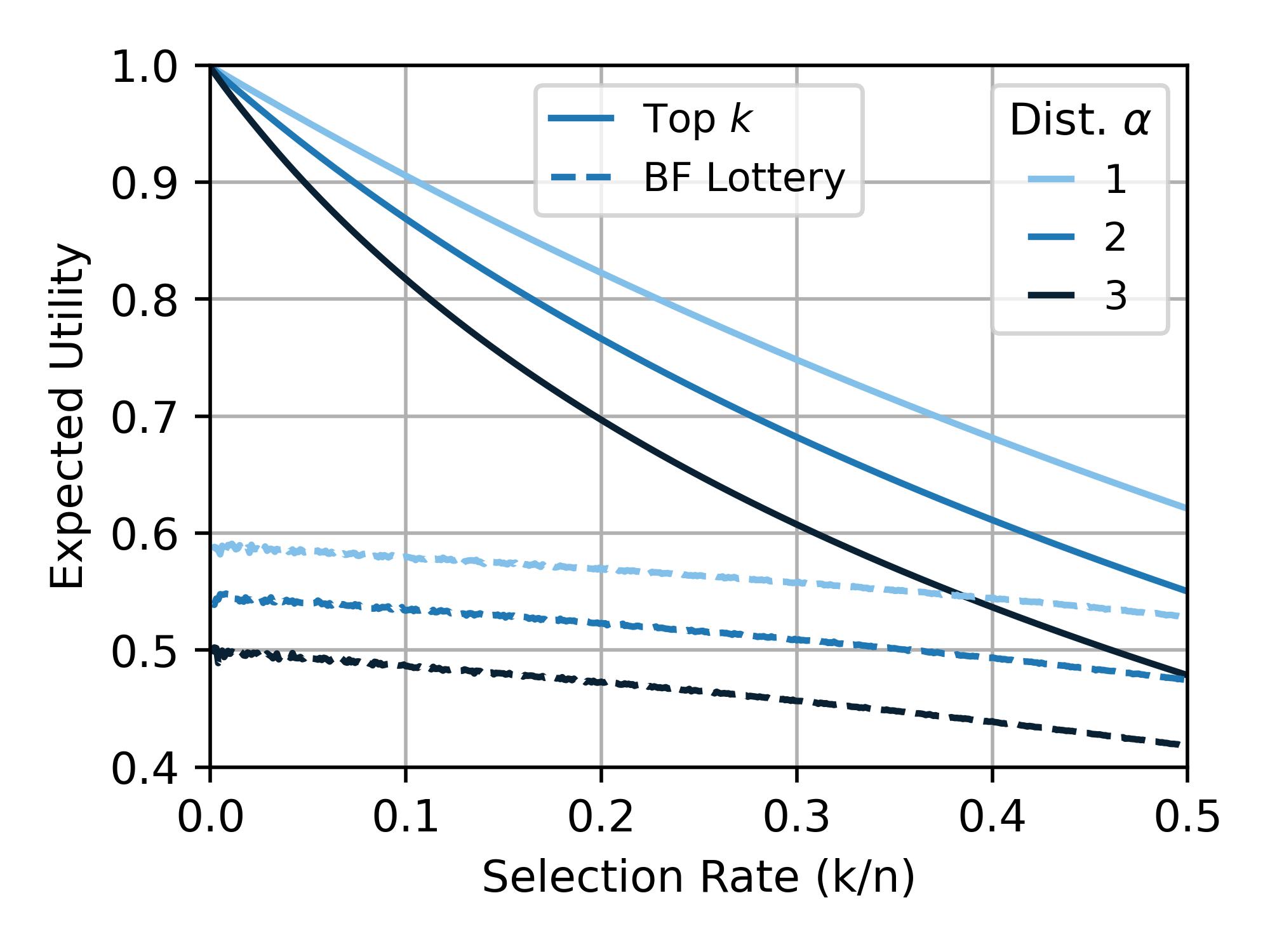}
} \\
\text{(c) - (e) Expected Utility for Varying Partial BF Randomization Rates in Ex~\ref{ex:partial_bf_lottery}} \\[2mm]
\subfloat[\centering Selection Rate = $0.1$]{
\includegraphics[width=0.32\columnwidth]{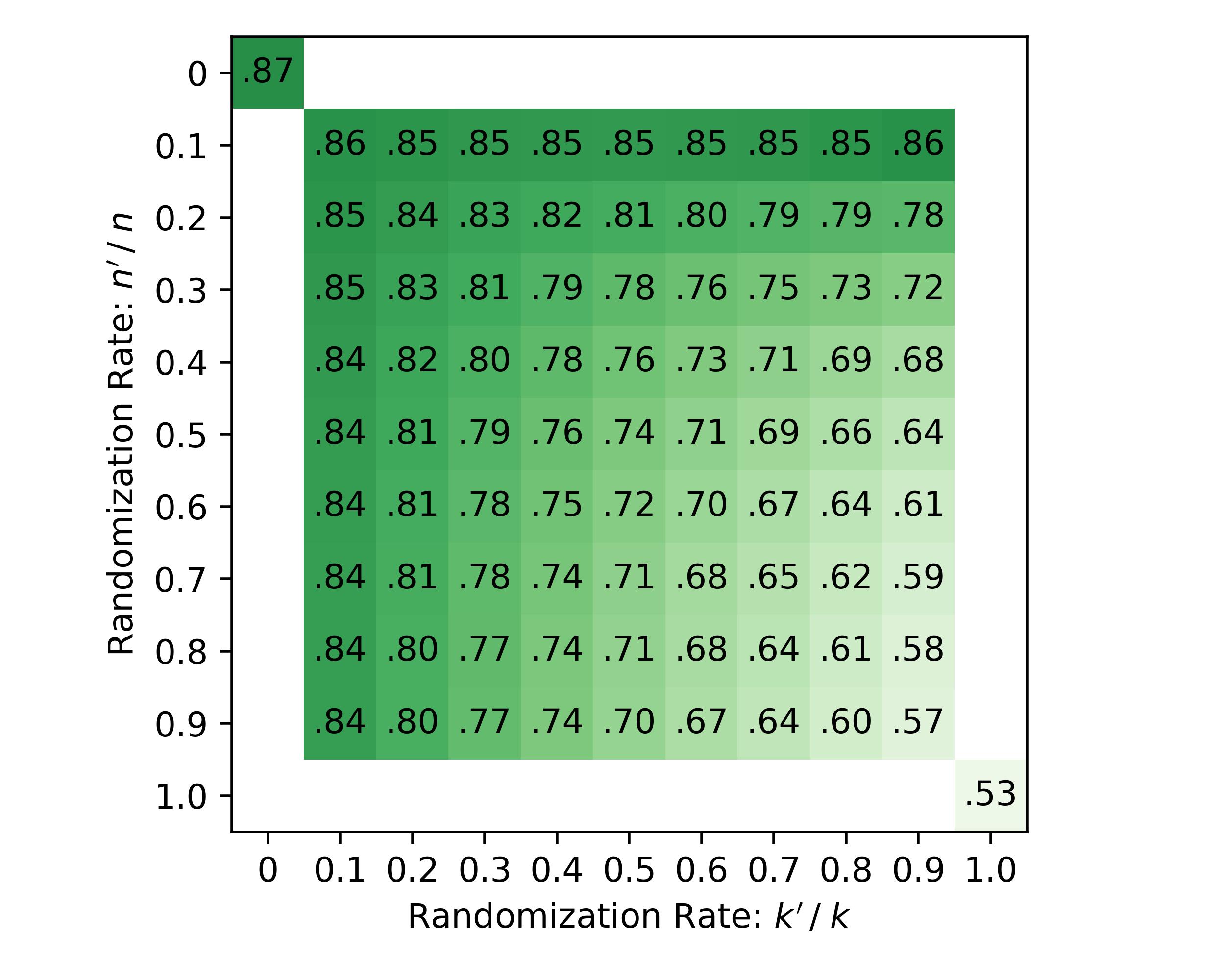}
} 
\subfloat[\centering Selection Rate = $0.25$]{
\includegraphics[width=0.32\columnwidth]{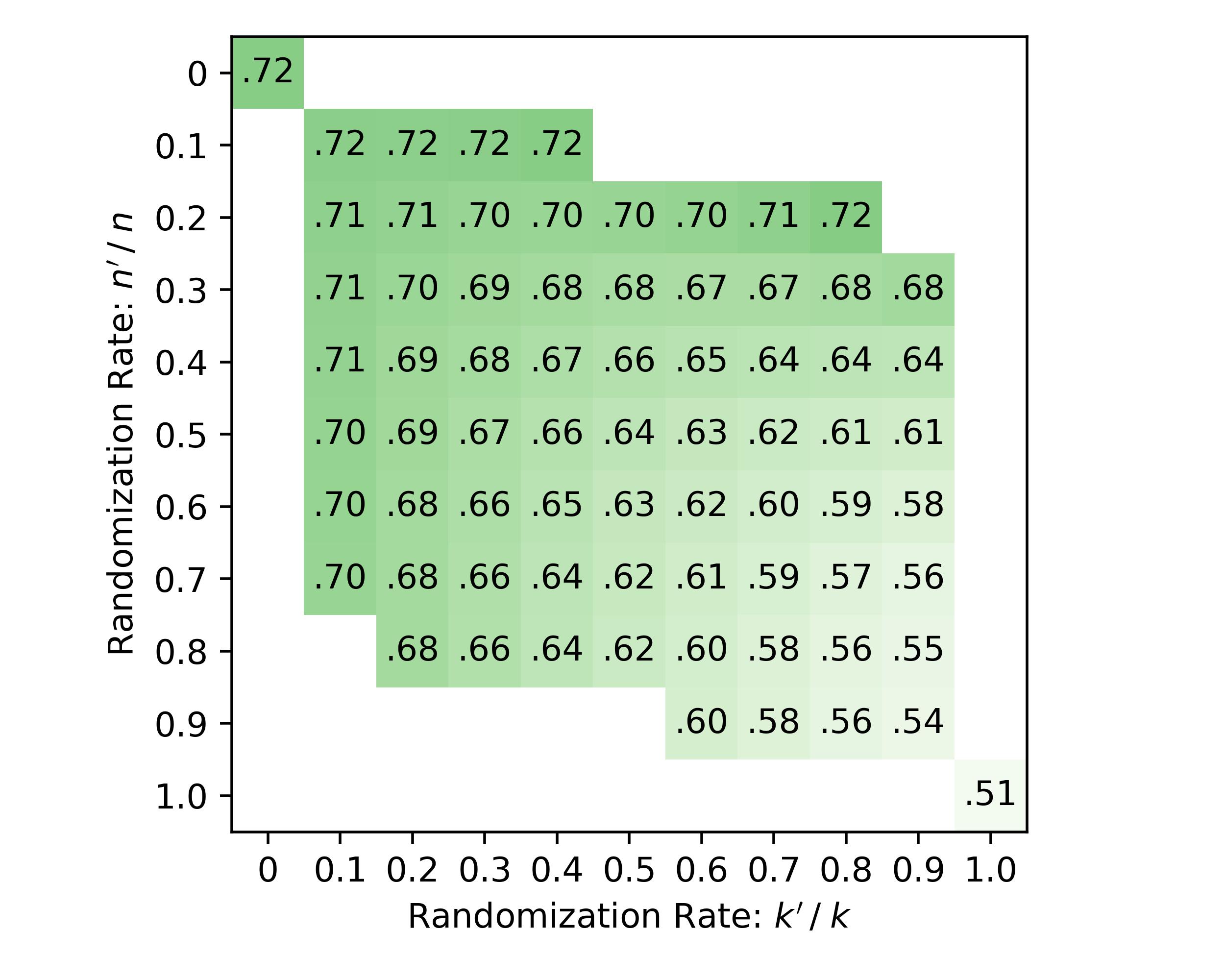}
}
\subfloat[\centering Selection Rate = $0.5$]{
\includegraphics[width=0.32\columnwidth]{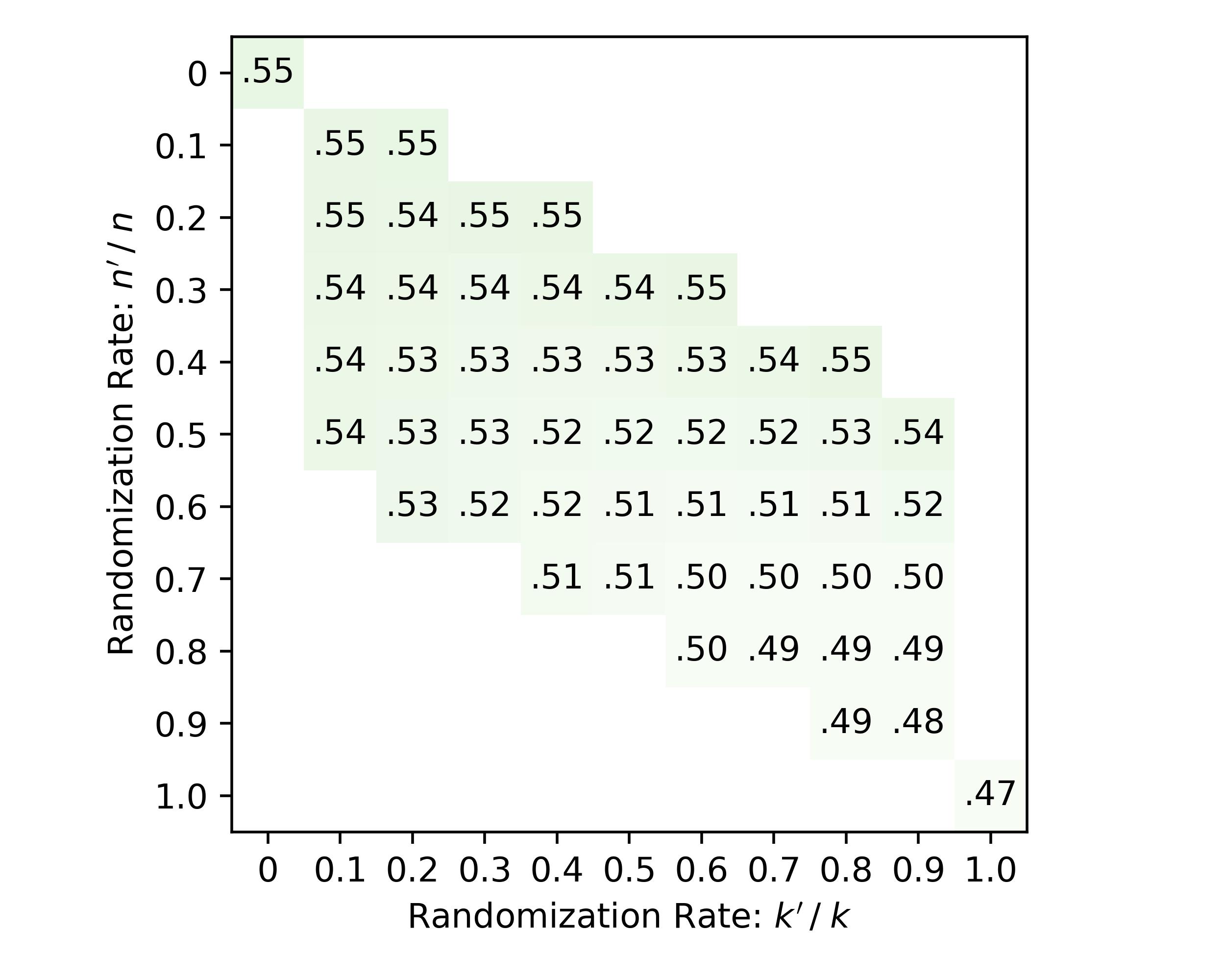}
}
\\
\text{(f) - (h) Systemic Exclusion Rate v. Expected Utility Across Varying Partial BF Randomization Rates} \\[2mm]
\subfloat[\centering Selection Rate = $0.1$]{
\includegraphics[width=0.32\columnwidth]{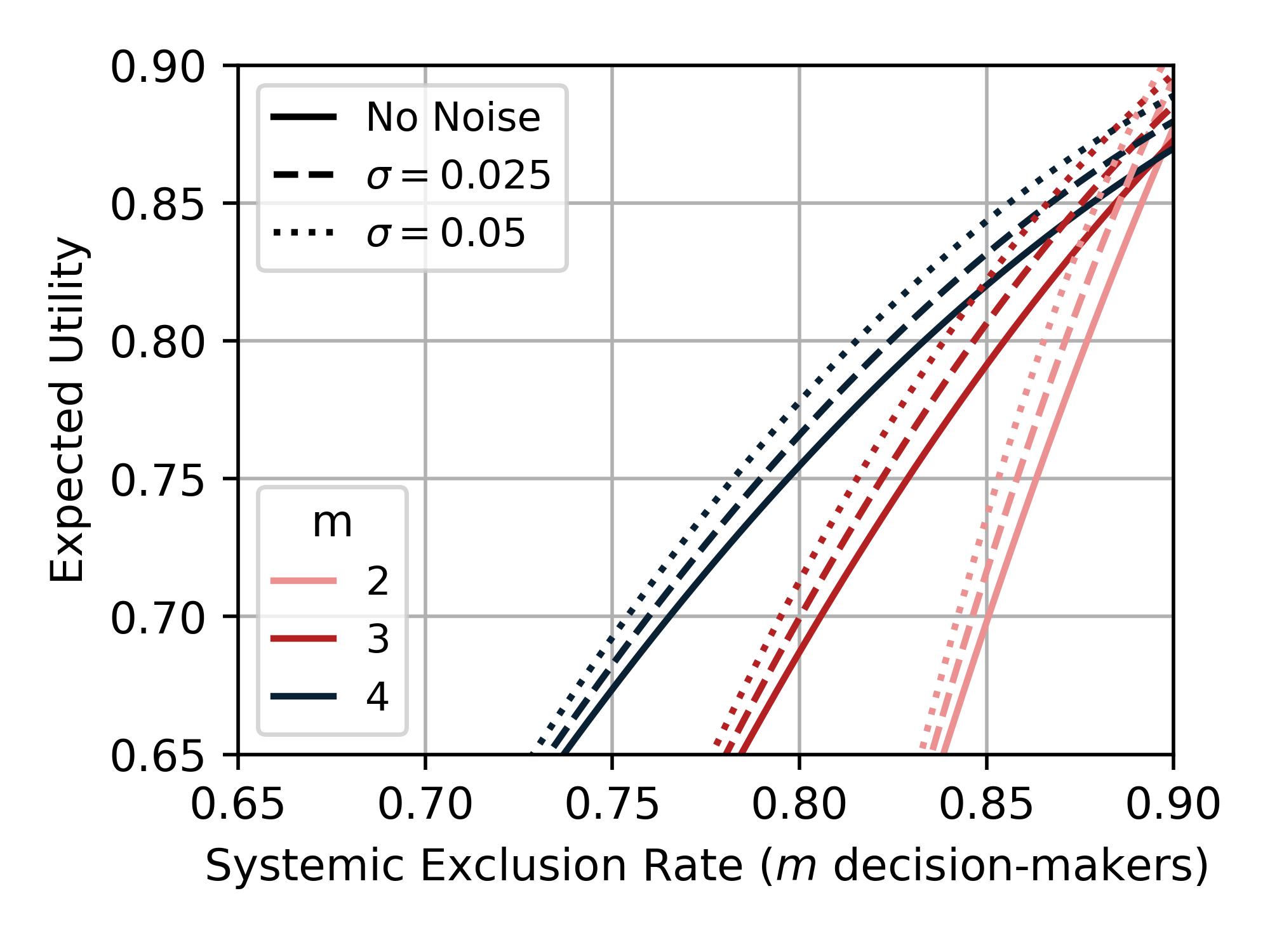}
} 
\subfloat[\centering Selection Rate = $0.25$]{
\includegraphics[width=0.32\columnwidth]{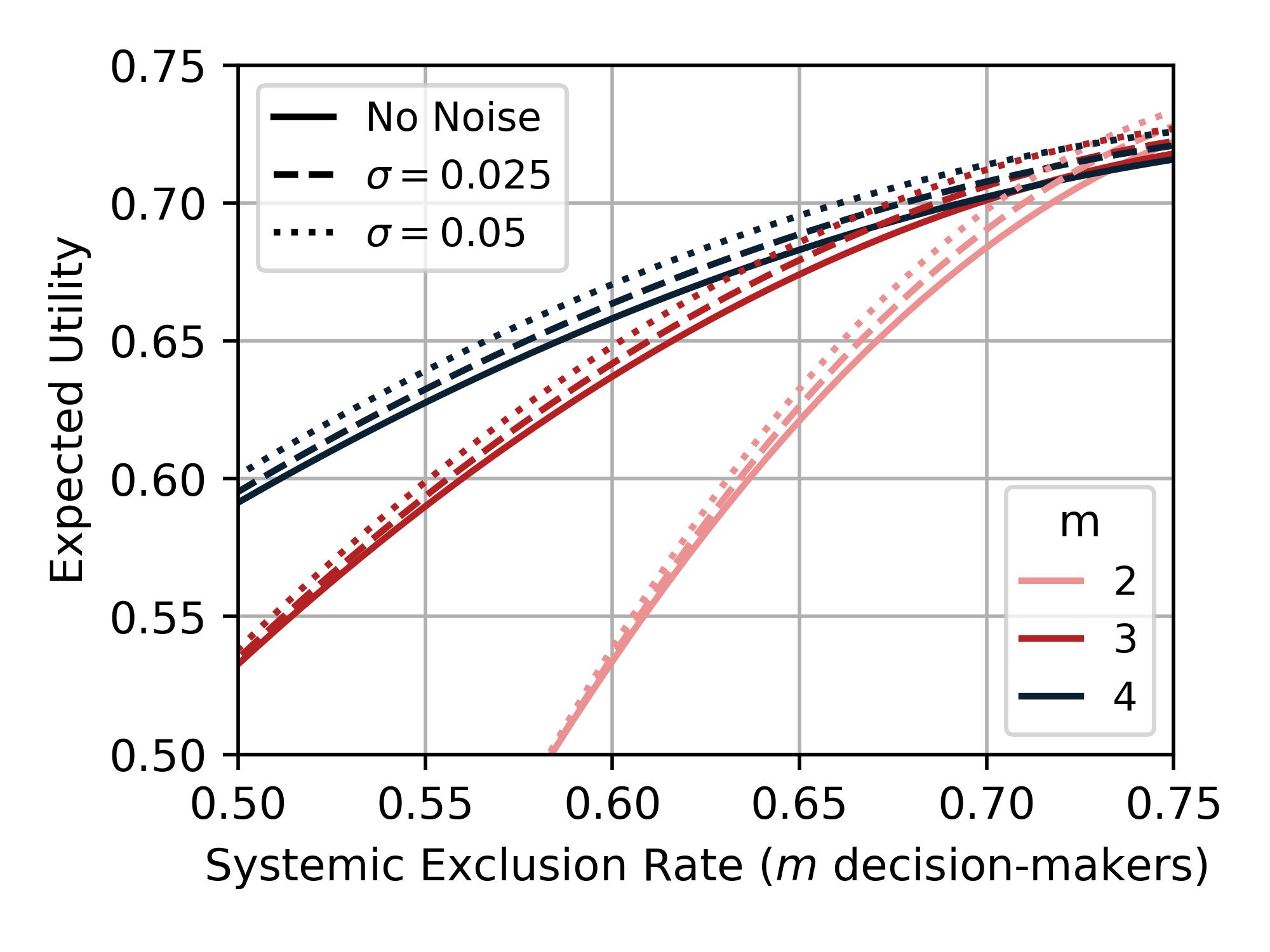}
}
\subfloat[\centering Selection Rate = $0.5$]{
\includegraphics[width=0.32\columnwidth]{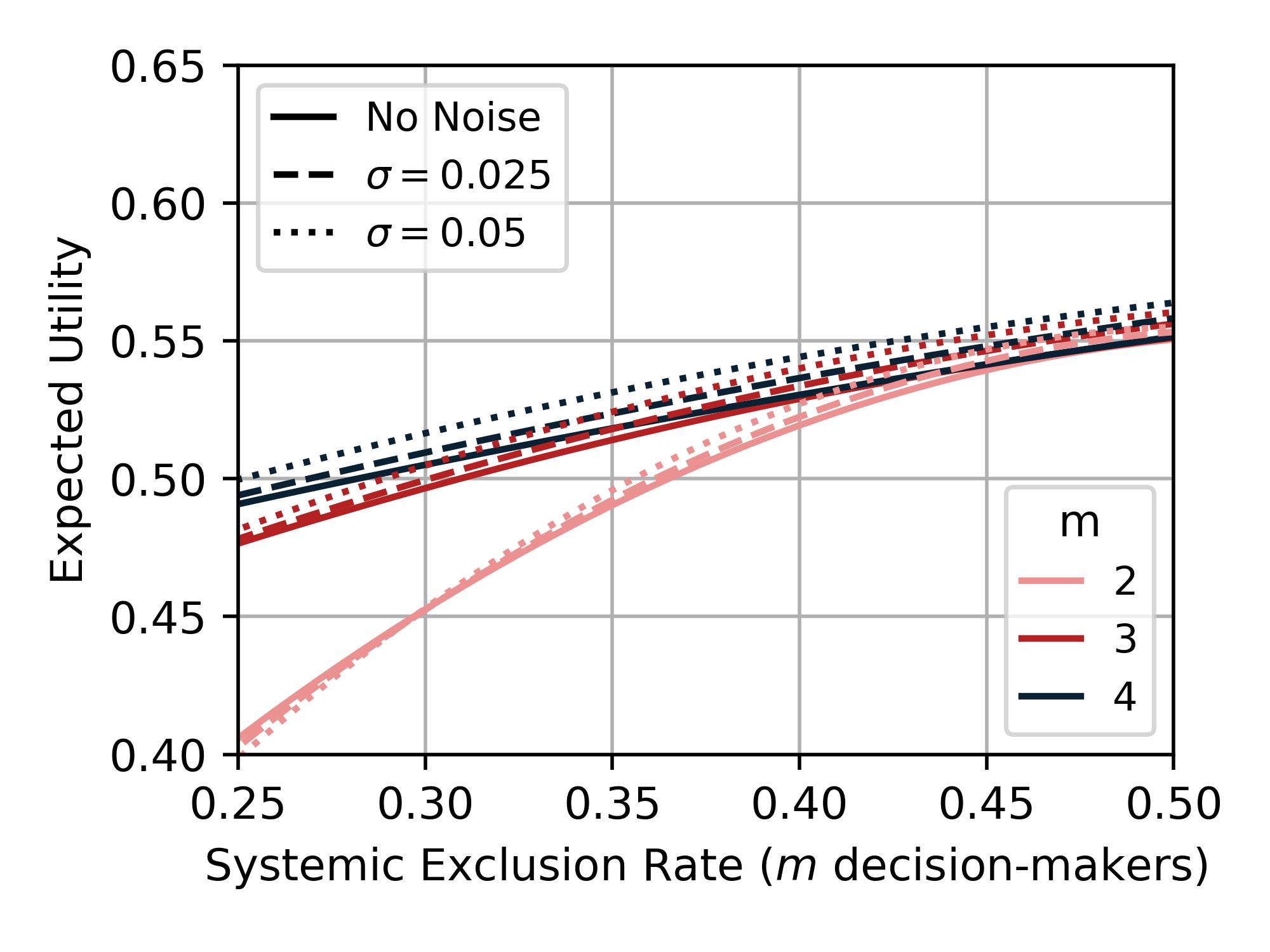}
}
\label{fig:pareto}
\end{figure*}

\begin{figure*}[h!]
\centering
\caption{Inverse Pareto Distribution of Claims}
\subfloat[\centering Distribution of Claims]{
\includegraphics[width=0.4\columnwidth]{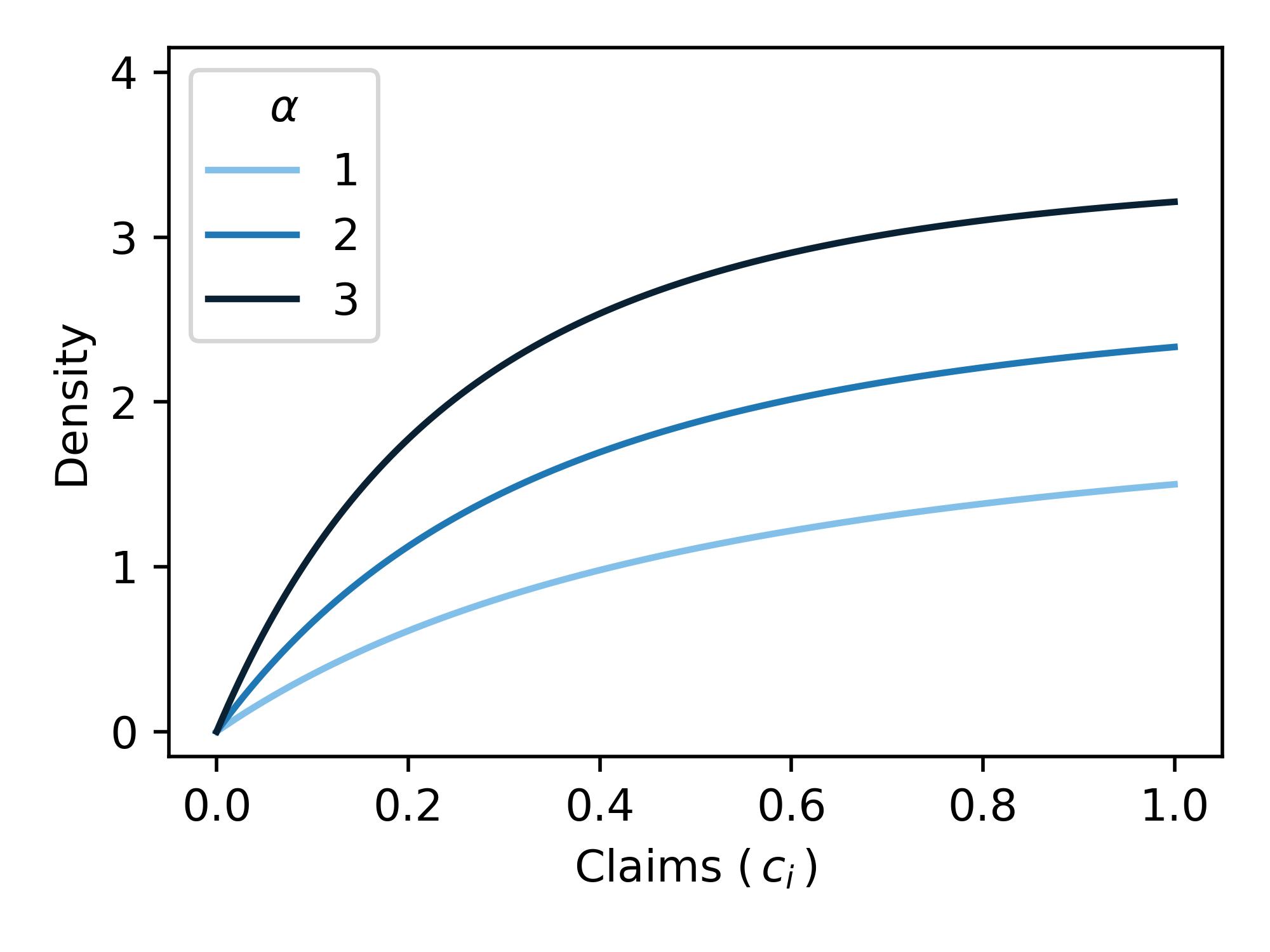}
} 
\subfloat[\centering Expected Utility for Top $k$ v. BF Lottery in Ex~\ref{ex:bf_lottery}]{
\includegraphics[width=0.4\columnwidth]{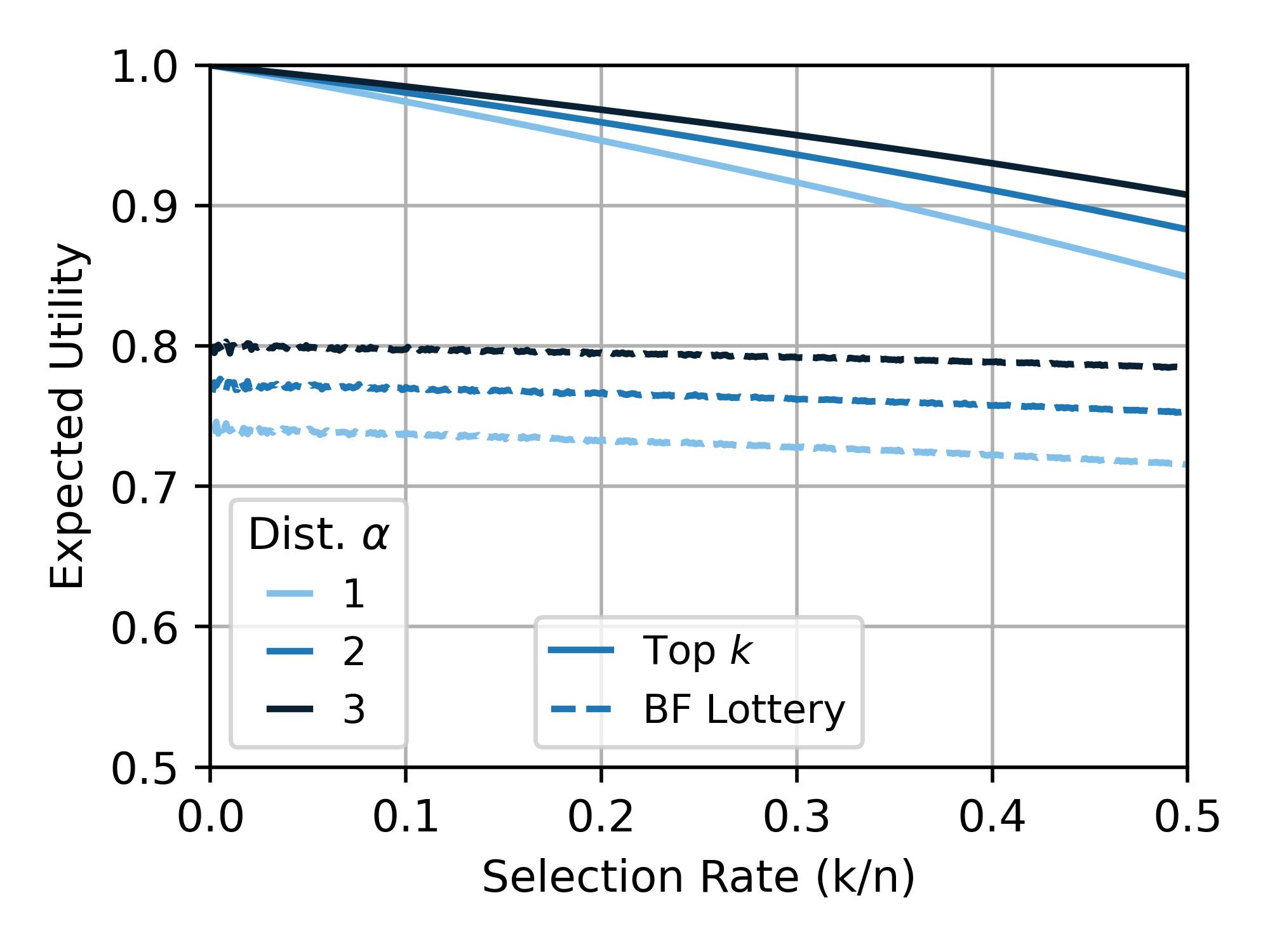}
} \\
\text{(c) - (e) Expected Utility for Varying Partial BF Randomization Rates in Ex~\ref{ex:partial_bf_lottery}} \\[2mm]
\subfloat[\centering Selection Rate = $0.1$]{
\includegraphics[width=0.32\columnwidth]{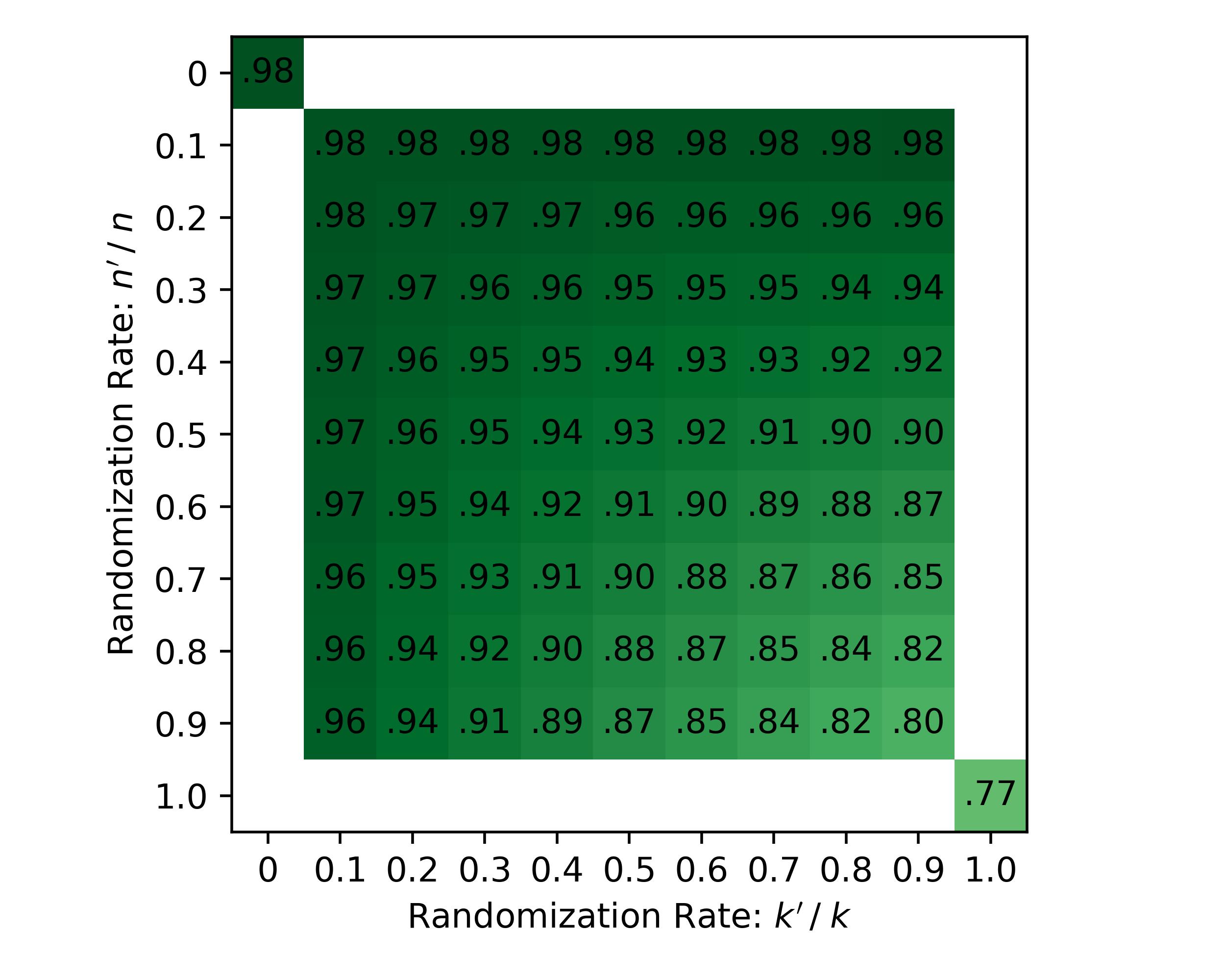}
} 
\subfloat[\centering Selection Rate = $0.25$]{
\includegraphics[width=0.32\columnwidth]{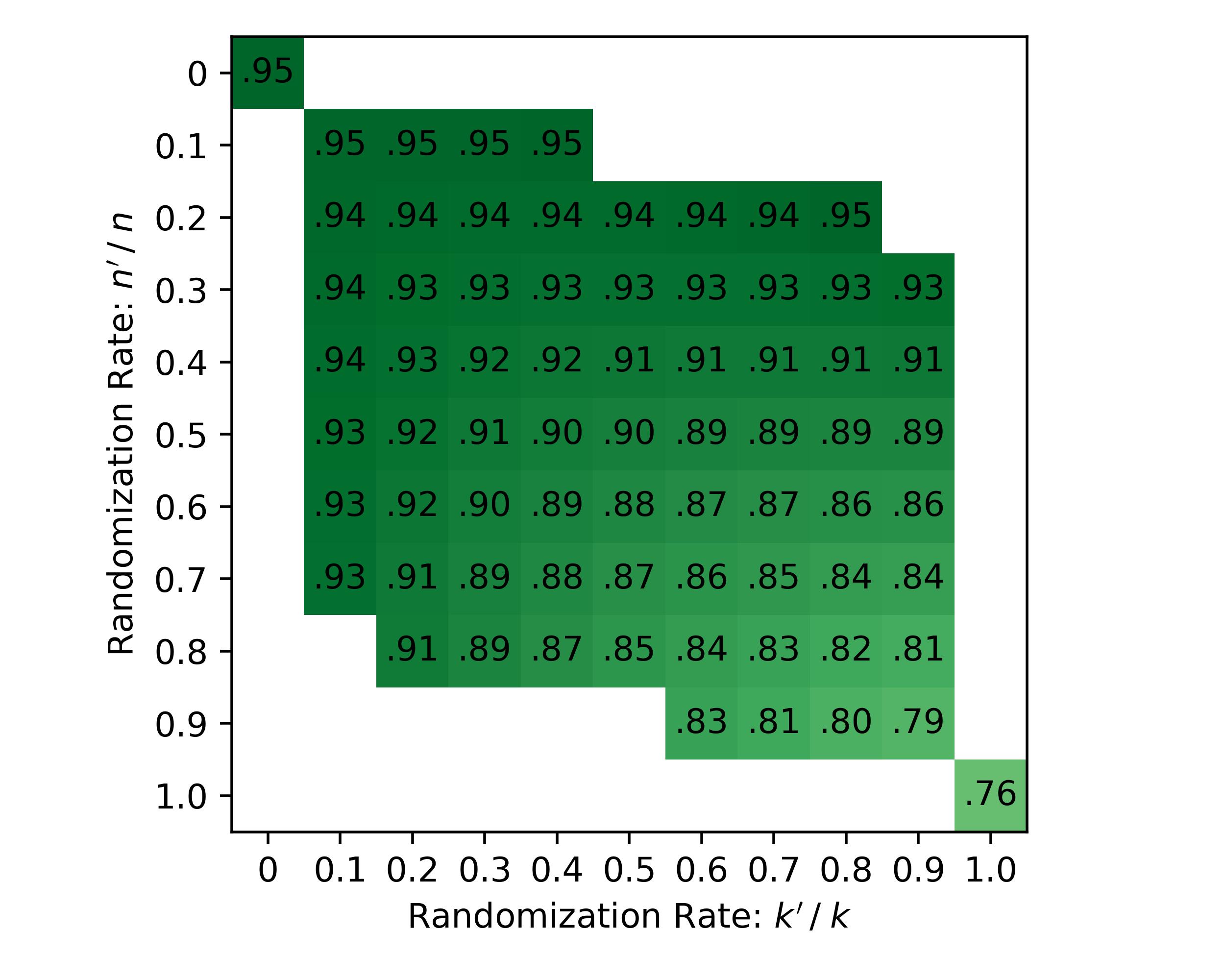}
}
\subfloat[\centering Selection Rate = $0.5$]{
\includegraphics[width=0.32\columnwidth]{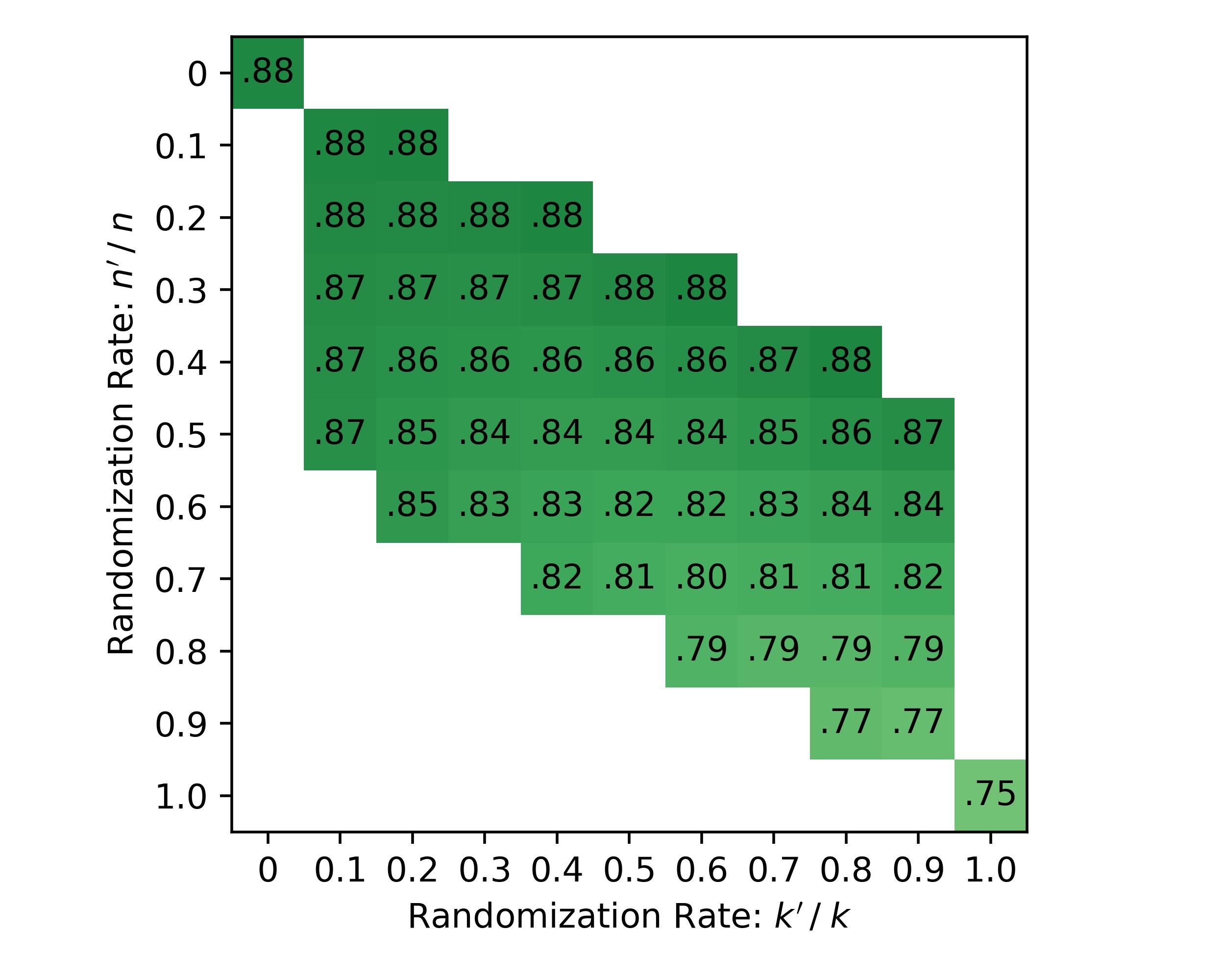}
}
\\
\text{(f) - (h) Systemic Exclusion Rate v. Expected Utility Across Varying Partial BF Randomization Rates} \\[2mm]
\subfloat[\centering Selection Rate = $0.1$]{
\includegraphics[width=0.32\columnwidth]{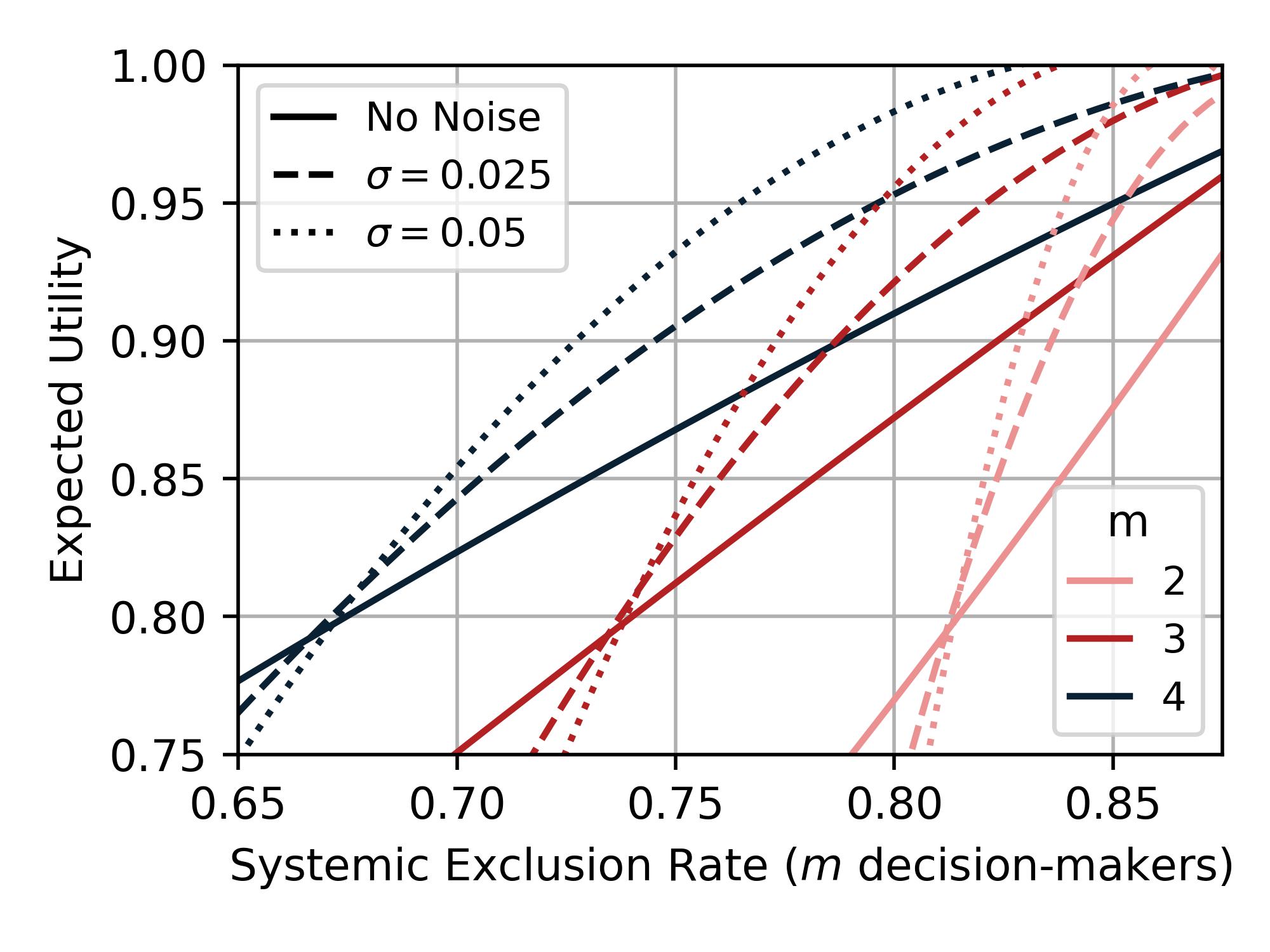}
} 
\subfloat[\centering Selection Rate = $0.25$]{
\includegraphics[width=0.32\columnwidth]{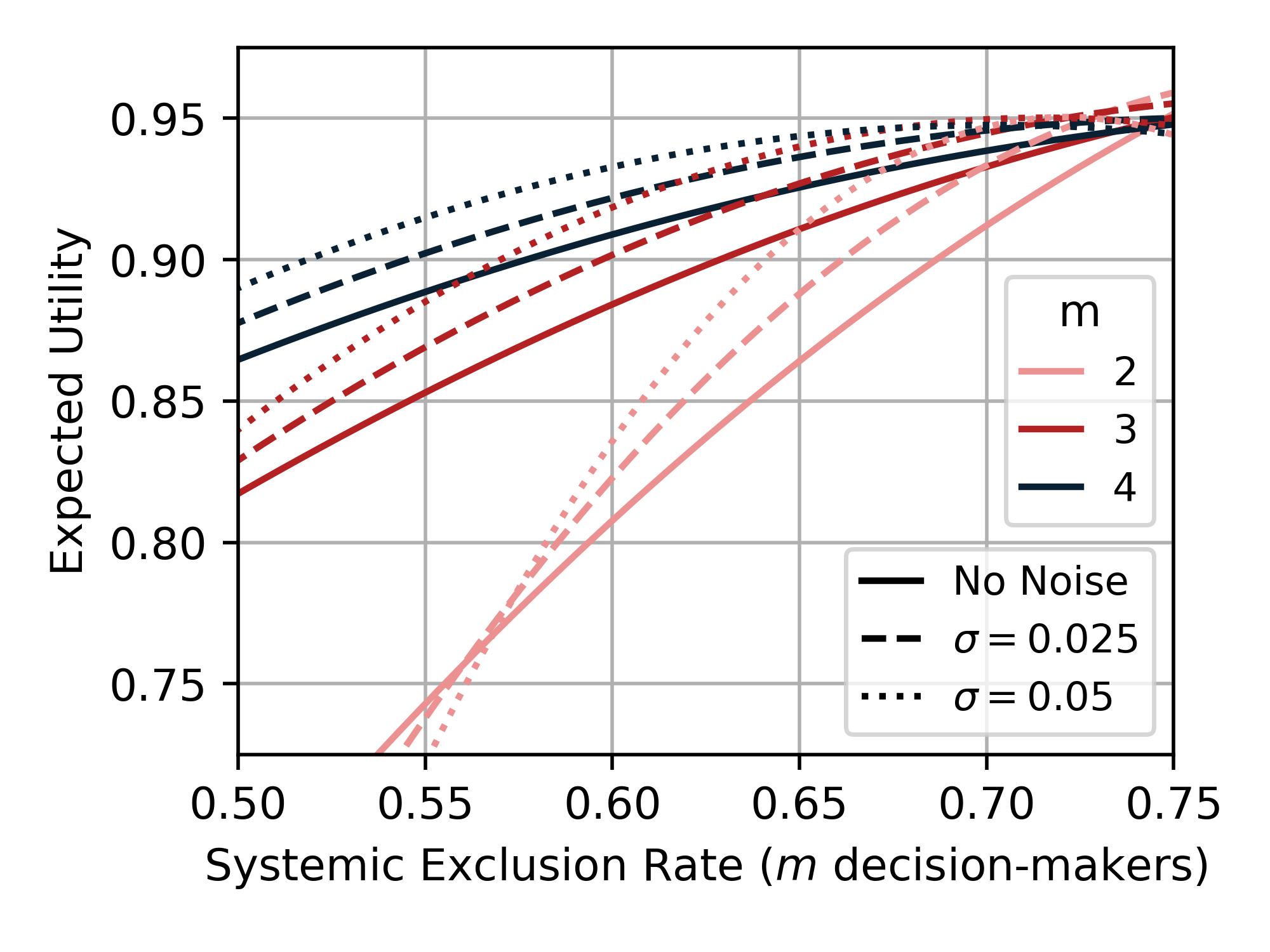}
}
\subfloat[\centering Selection Rate = $0.5$]{
\includegraphics[width=0.32\columnwidth]{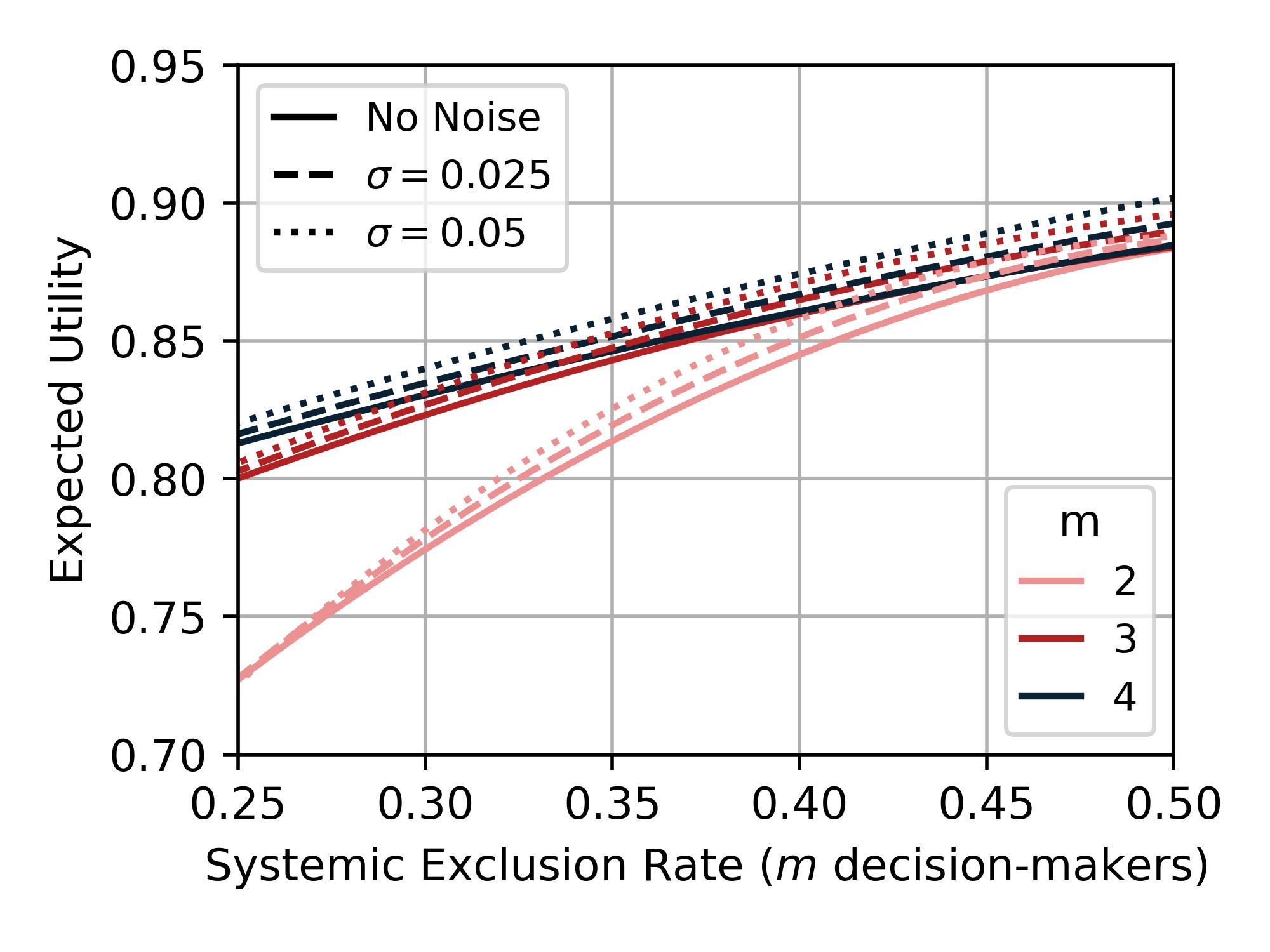}
}
\label{fig:inv_pareto}
\end{figure*}

\begin{figure*}[h!]
\centering
\caption{Uniform Distribution of Claims}
\subfloat[\centering Distribution of Claims]{
\includegraphics[width=0.4\columnwidth]{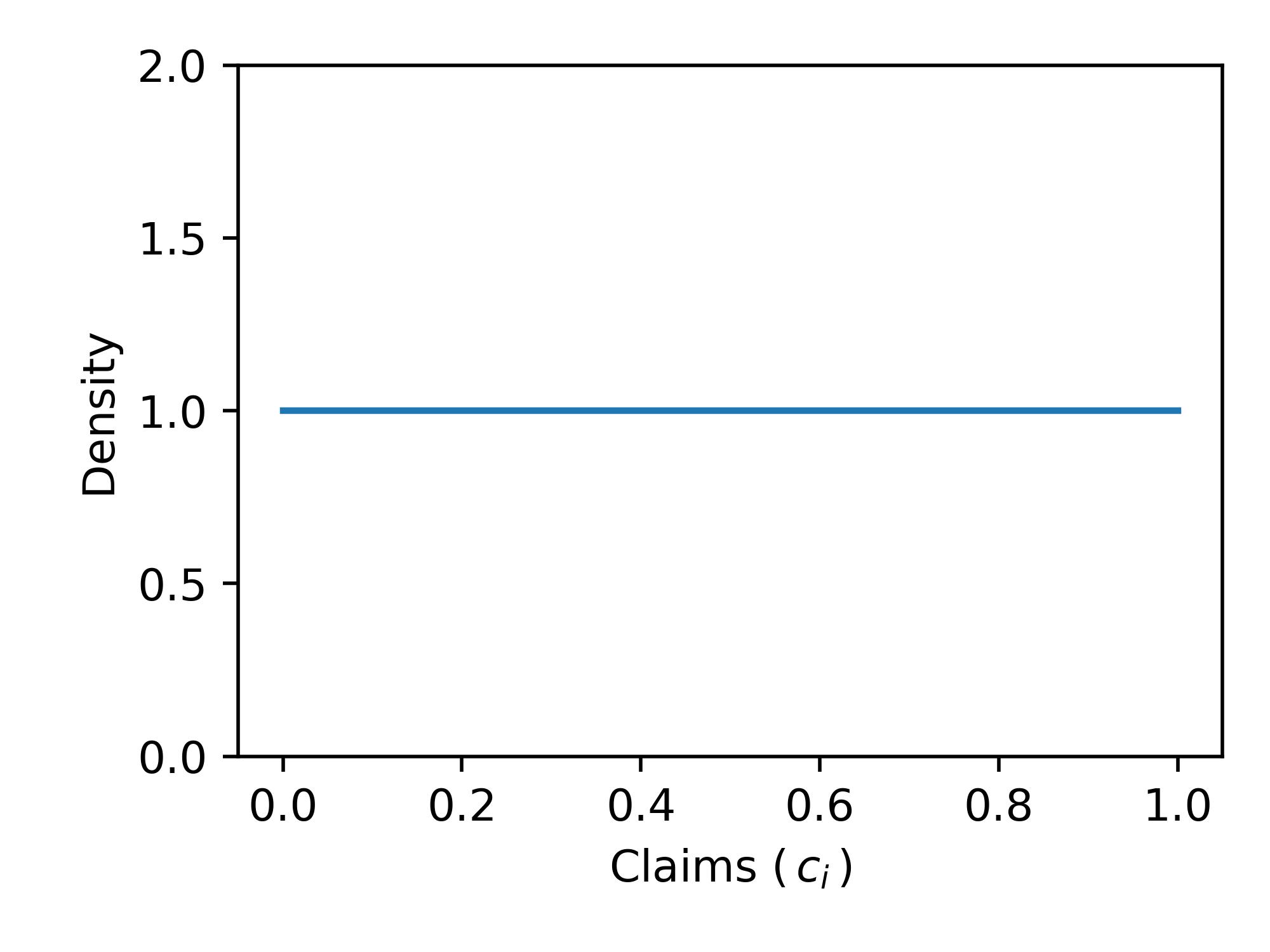}
} 
\subfloat[\centering Expected Utility for Top $k$ v. BF Lottery in Ex~\ref{ex:bf_lottery}]{
\includegraphics[width=0.4\columnwidth]{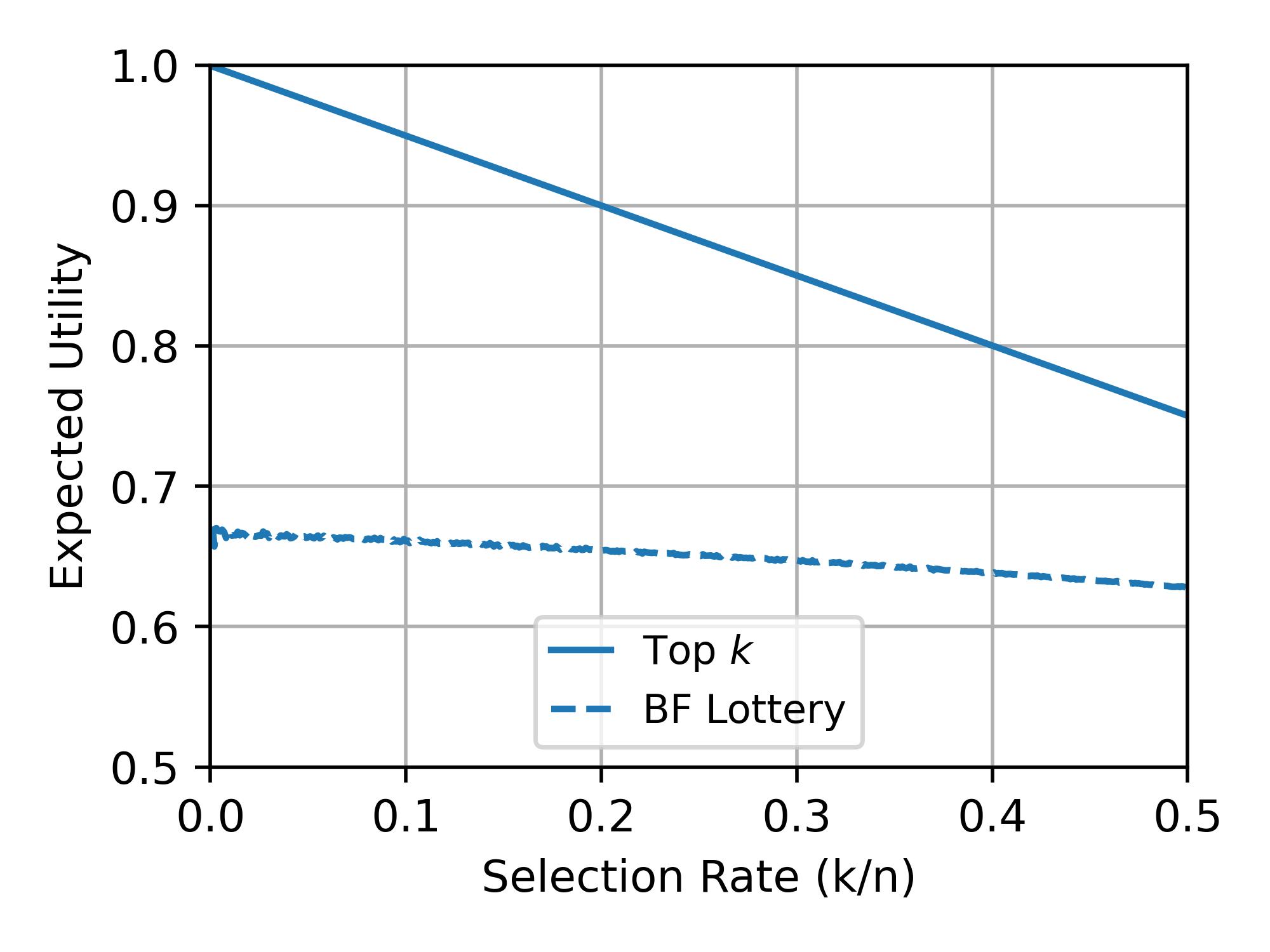}
} \\
\text{(c) - (e) Expected Utility for Varying Partial BF Randomization Rates in Ex~\ref{ex:partial_bf_lottery}} \\[2mm]
\subfloat[\centering Selection Rate = $0.1$]{
\includegraphics[width=0.32\columnwidth]{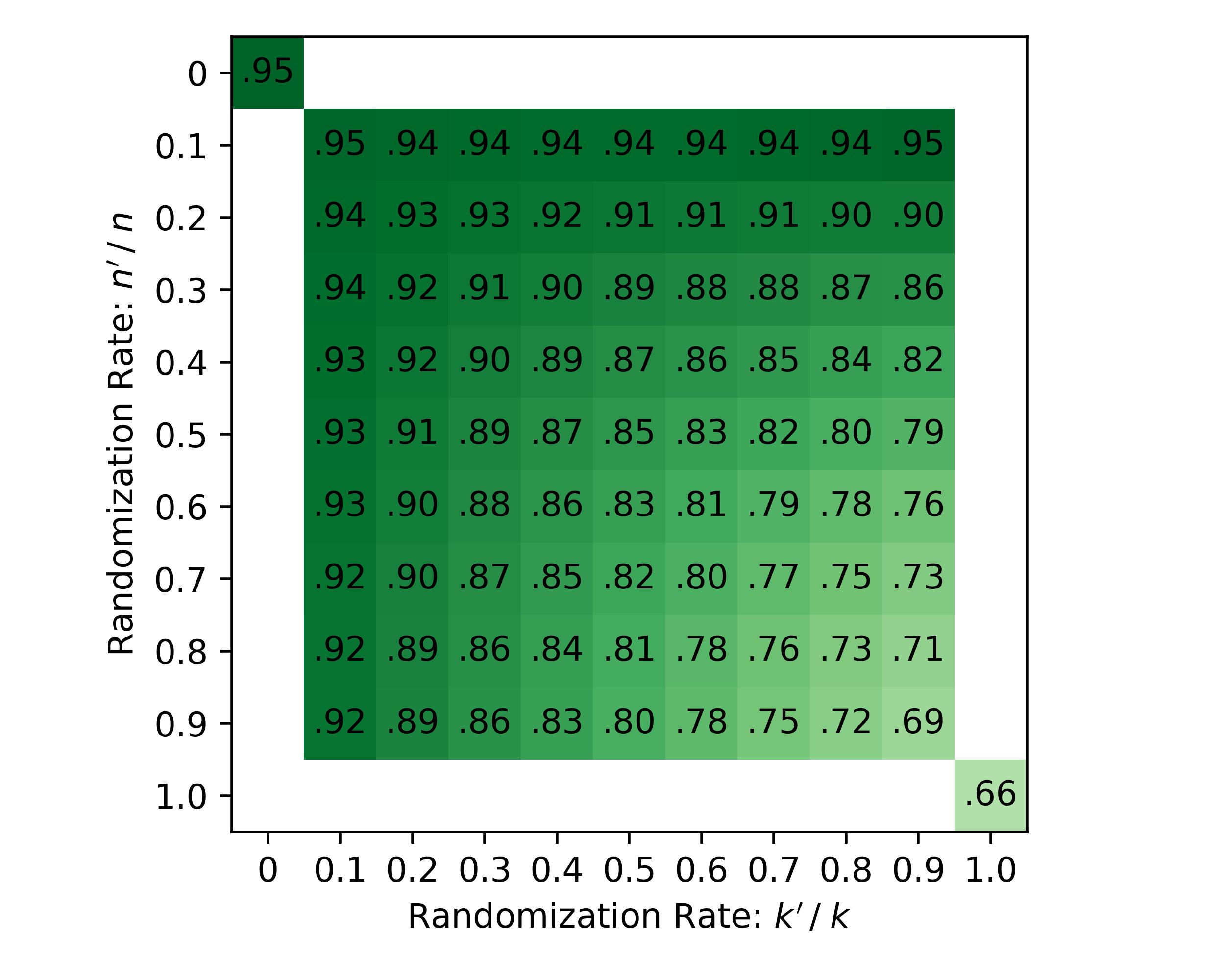}
} 
\subfloat[\centering Selection Rate = $0.25$]{
\includegraphics[width=0.32\columnwidth]{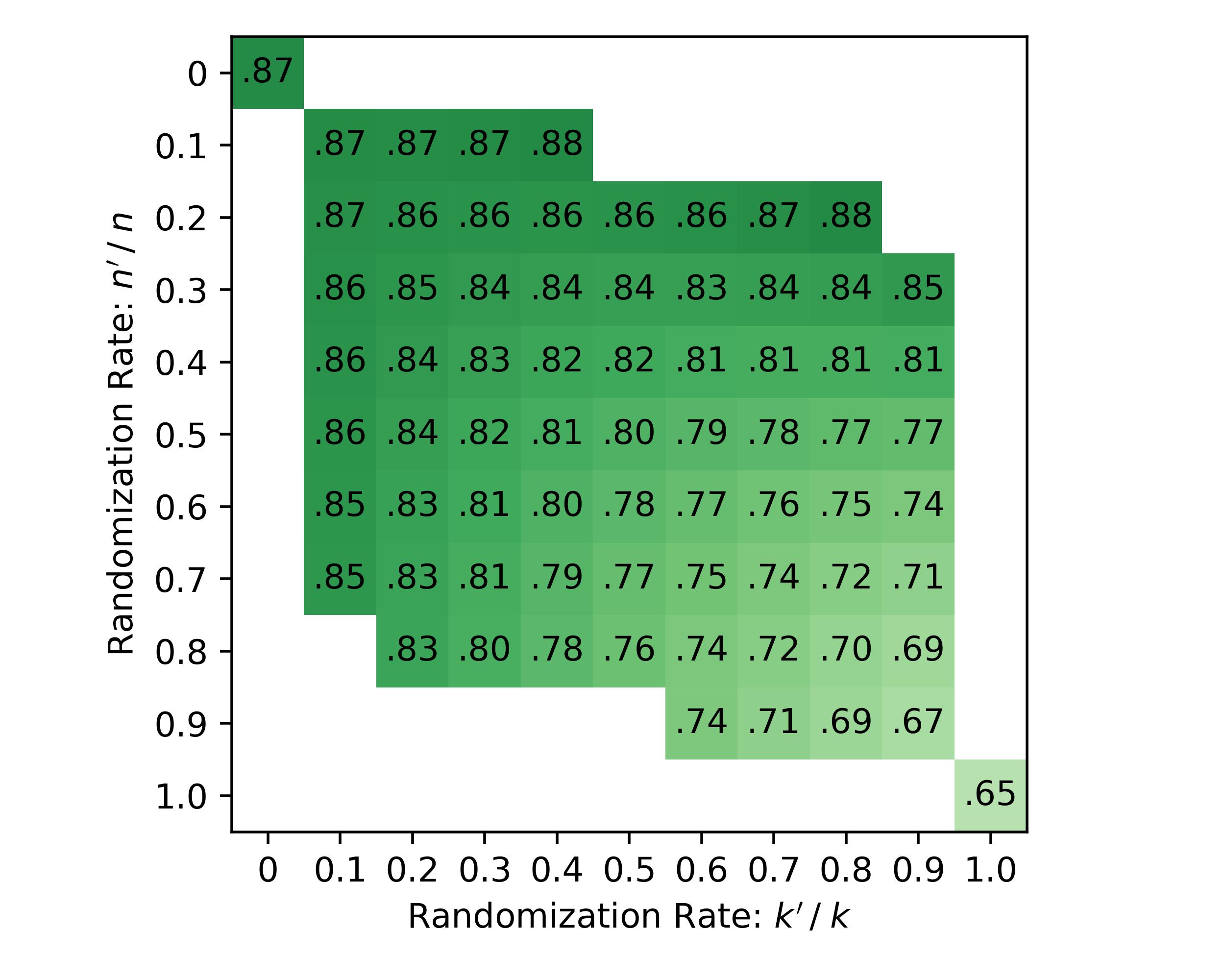}
}
\subfloat[\centering Selection Rate = $0.5$]{
\includegraphics[width=0.32\columnwidth]{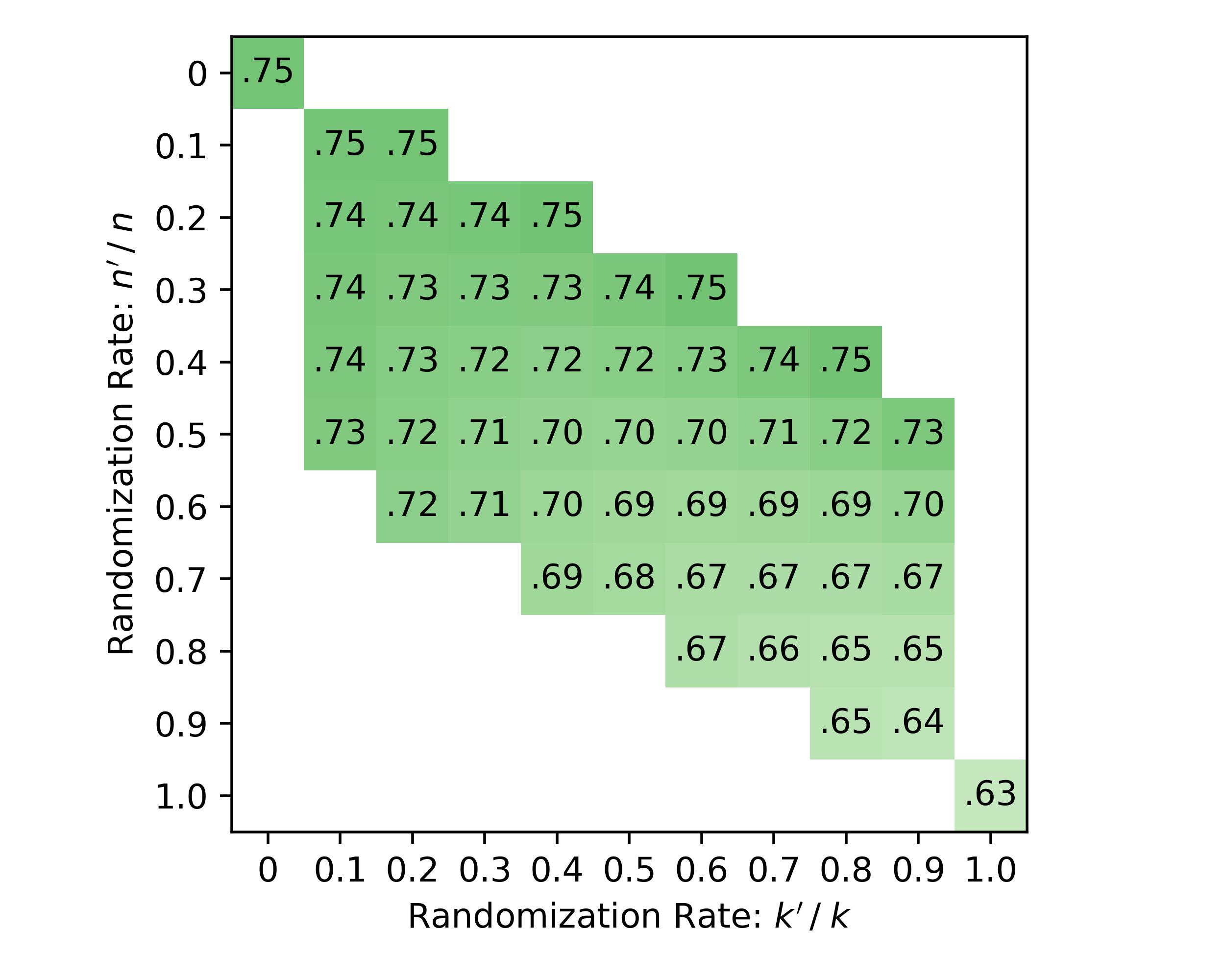}
}
\\
\text{(f) - (h) Systemic Exclusion Rate v. Expected Utility Across Varying Partial BF Randomization Rates} \\[2mm]
\subfloat[\centering Selection Rate = $0.1$]{
\includegraphics[width=0.32\columnwidth]{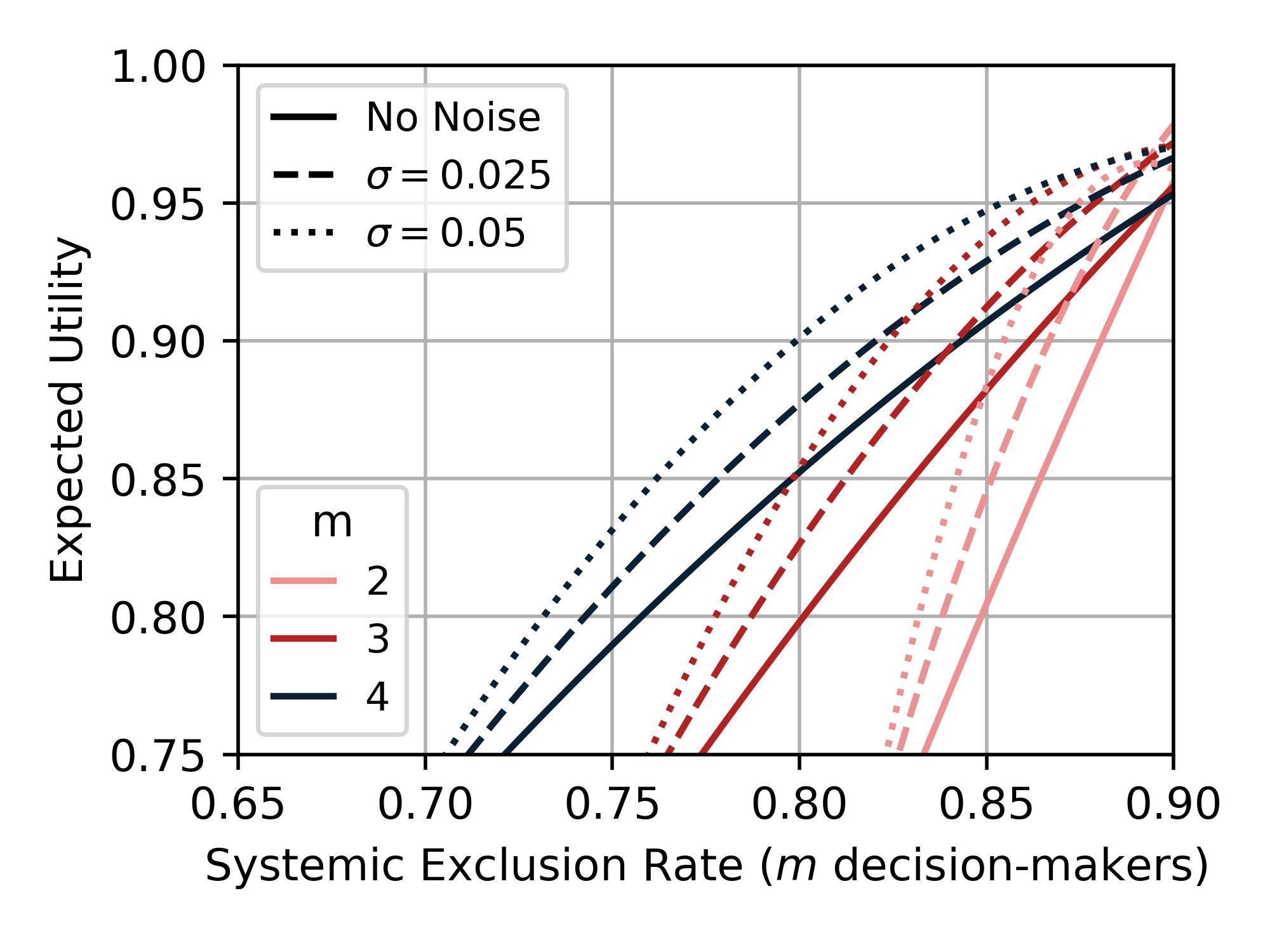}
} 
\subfloat[\centering Selection Rate = $0.25$]{
\includegraphics[width=0.32\columnwidth]{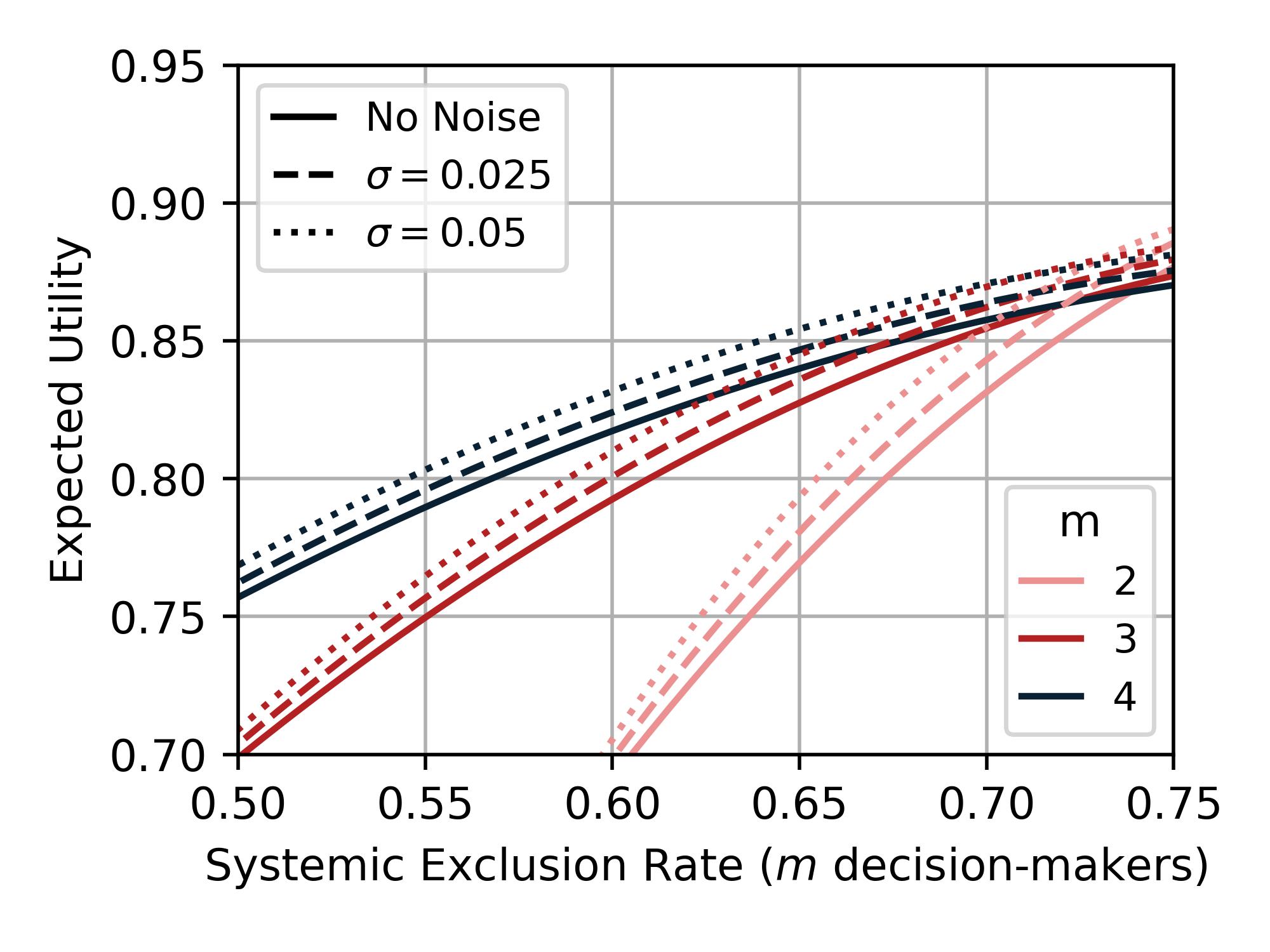}
}
\subfloat[\centering Selection Rate = $0.5$]{
\includegraphics[width=0.32\columnwidth]{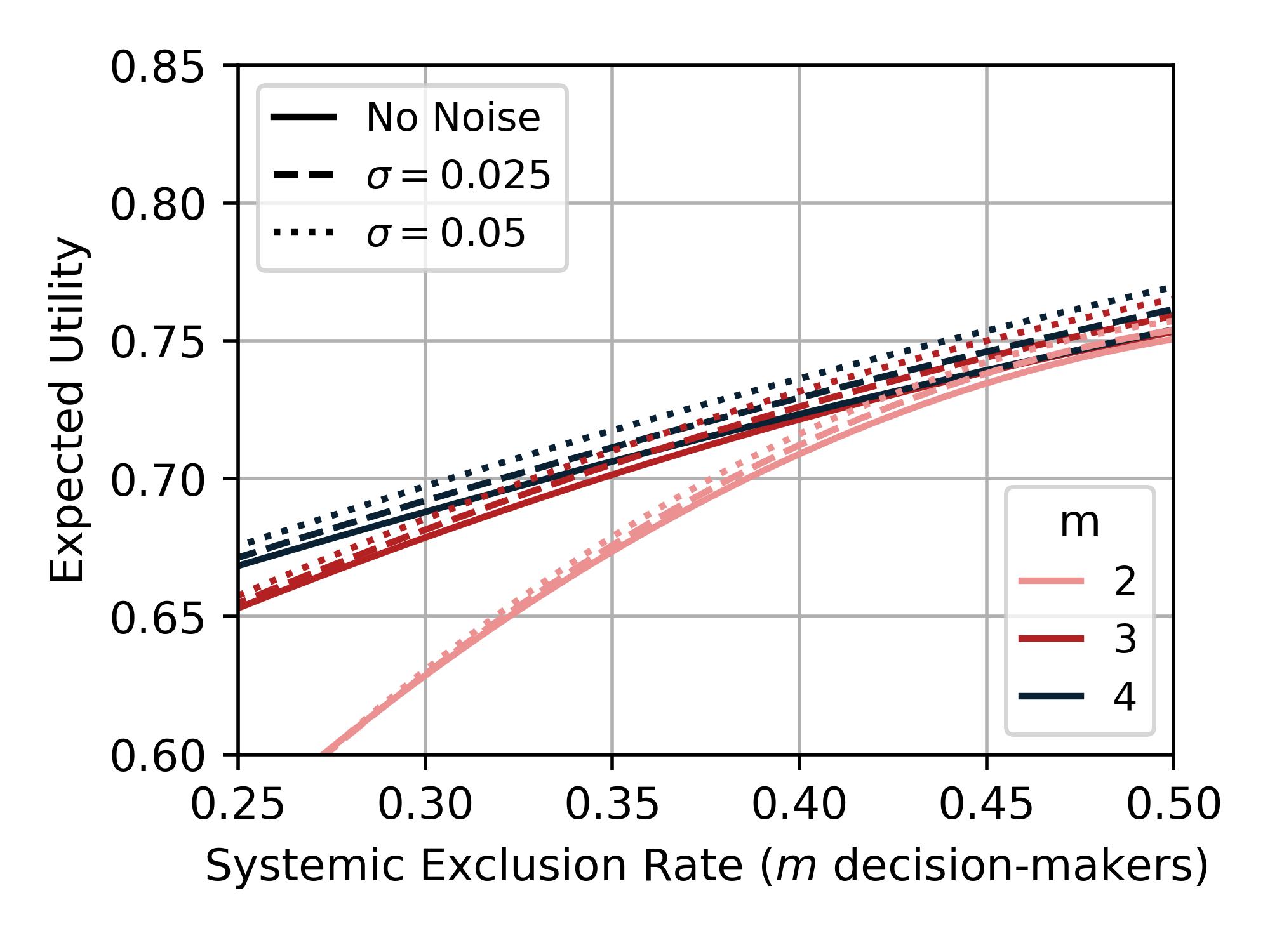}
}
\label{fig:uniform}
\end{figure*}

We simulate different distributions of claims and compare 3 different allocation types: (1) deterministic selection of the top $k$ claims, (2) the BF lottery in Example~\ref{ex:bf_lottery}, and (3) the partial BF lottery in Example~\ref{ex:partial_bf_lottery}. Specifically, we analyze the tradeoff between utility and reduction in SER under various selection rates and levels of noise. We simulate 1000 individuals and average the results over 1000 iterations for each experiment. We consider the following distributions:
\begin{itemize}[topsep=0ex,itemsep=-0.5ex]
    \item Normal: more average claims (Figure~\ref{fig:normal})
    \item Inverted Normal: more strong and weak claims (Figure~\ref{fig:inv_normal})
    \item Pareto: more weak claims (Figure~\ref{fig:pareto})
    \item Inverted Pareto: more strong claims (Figure~\ref{fig:inv_pareto})
    \item Uniform: all claims equally likely (Figure~\ref{fig:uniform})
\end{itemize}

For each distribution, we illustrate the following (letters correspond to sub-figures for each distribution): 
\begin{enumerate}[label=(\alph*),topsep=0ex,itemsep=0ex]
    \item Distribution of claims: Density under 3 different parameter choices for all distribution types except uniform (i.e. different $\sigma$ for normal or $\alpha$ for pareto)
    \item Expected utility for the top $k$ selection and BF lottery in Example~\ref{ex:bf_lottery}. We consider each parameter choice in (a).
    \item- (e) Expected utility for varying partial BF randomization rates in Example~\ref{ex:partial_bf_lottery}. We consider different choices of $k'/k$ and $n'/n$ (in increments of 0.1). We now use a fixed parameter choice ($\sigma=0.15$ for normal and $\alpha=2$ for pareto) in this sub-figure and all subsequent sub-figures. (c) uses $k/n$ = 0.1, (d) uses $k/n$ = 0.25, and (e) uses $k/n$ = 0.5.
    \setItemnumber{6}
    \item- (h) Systemic exclusion rate v. expected utility across varying partial BF randomization rates. For each of the choices of $k'/k$ and $n'/n$ in (c) - (e), we calculate the systemic exclusion rate for a given amount of decision-makers and noise added to each decision-maker's claims ($\pm\,N(0, \sigma^2)$). We then plot the tradeoff across randomization rates by showing the lowest SER possible for each percentage decrease in expected utility. We show this tradeoff for 2, 3, and 4 decision-makers, as well as for no noise, $\sigma$ = 0.025, and $\sigma$ = 0.05. (f) uses $k/n$ = 0.1, (g) uses $k/n$ = 0.25, and (h) uses $k/n$ = 0.5. 
\end{enumerate}

\begin{figure*}[h!]
\centering
\caption{\centering Reduction in SER using the BF Lottery in Example~\ref{ex:bf_lottery} (c.f. Figure~\subref{fig:claims_ser} in Main Text);\newline Each decision-maker has a noisy estimation of claims ($\pm\,N(0, \sigma^2)$), (a) - (f) show different $\sigma$ and selection rates $k/n$}
\subfloat[\centering $k/n$ = 0.1, $\sigma$ = 0.025]{
\includegraphics[width=0.4\columnwidth]{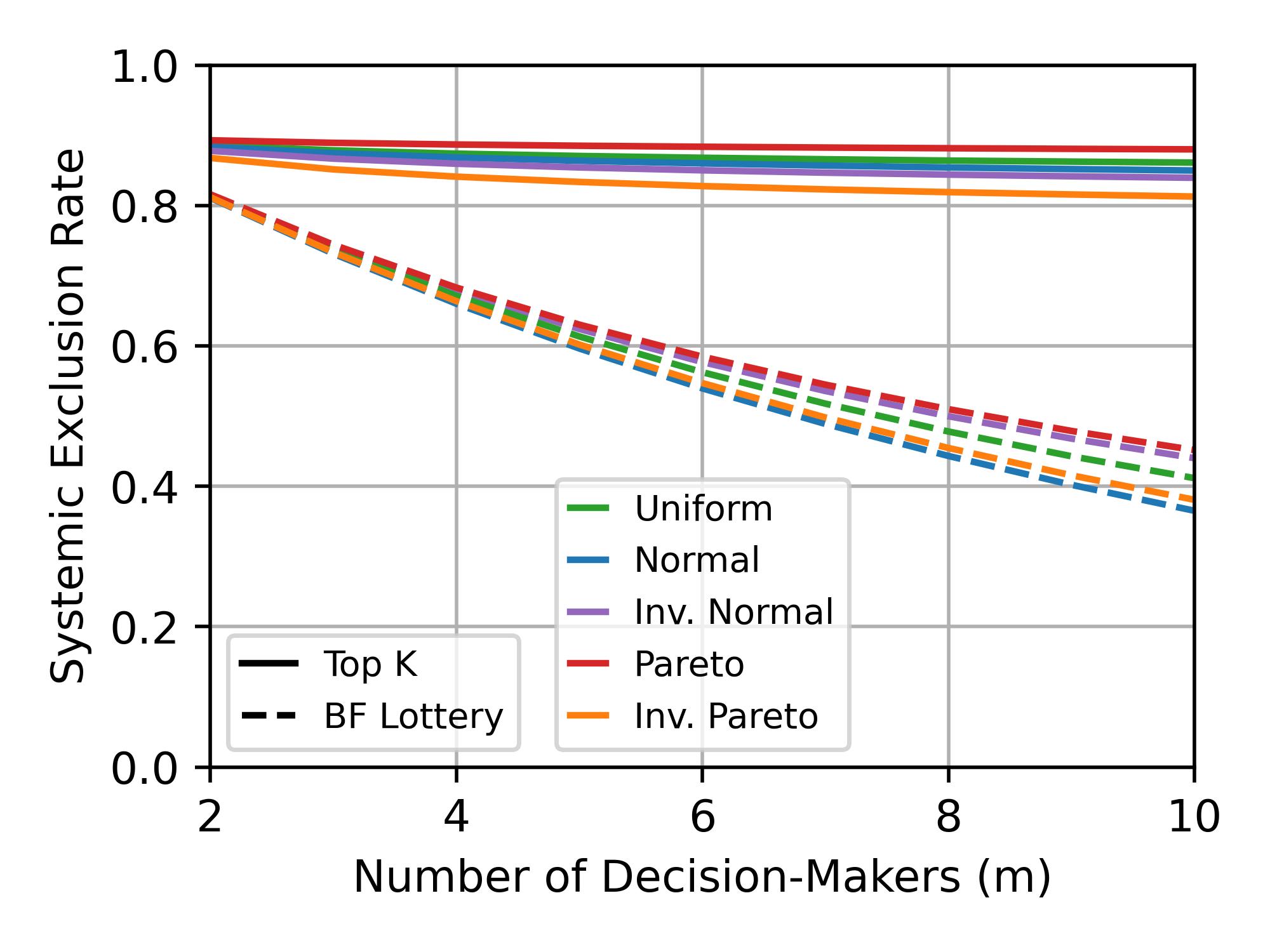}
} 
\subfloat[\centering $k/n$ = 0.1, $\sigma$ = 0.05]{
\includegraphics[width=0.4\columnwidth]{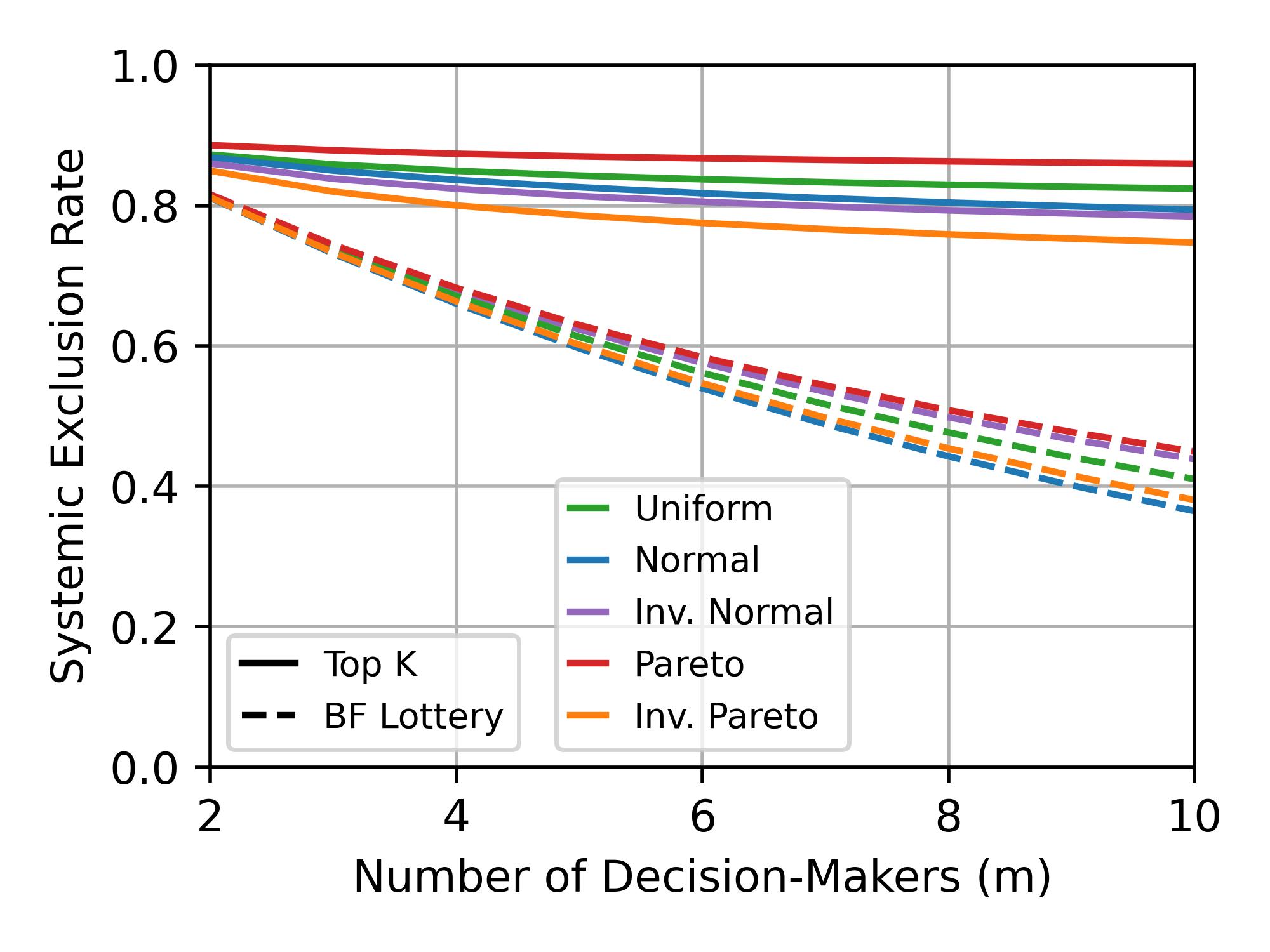}
} \\
\subfloat[\centering $k/n$ = 0.25, $\sigma$ = 0.025]{
\includegraphics[width=0.4\columnwidth]{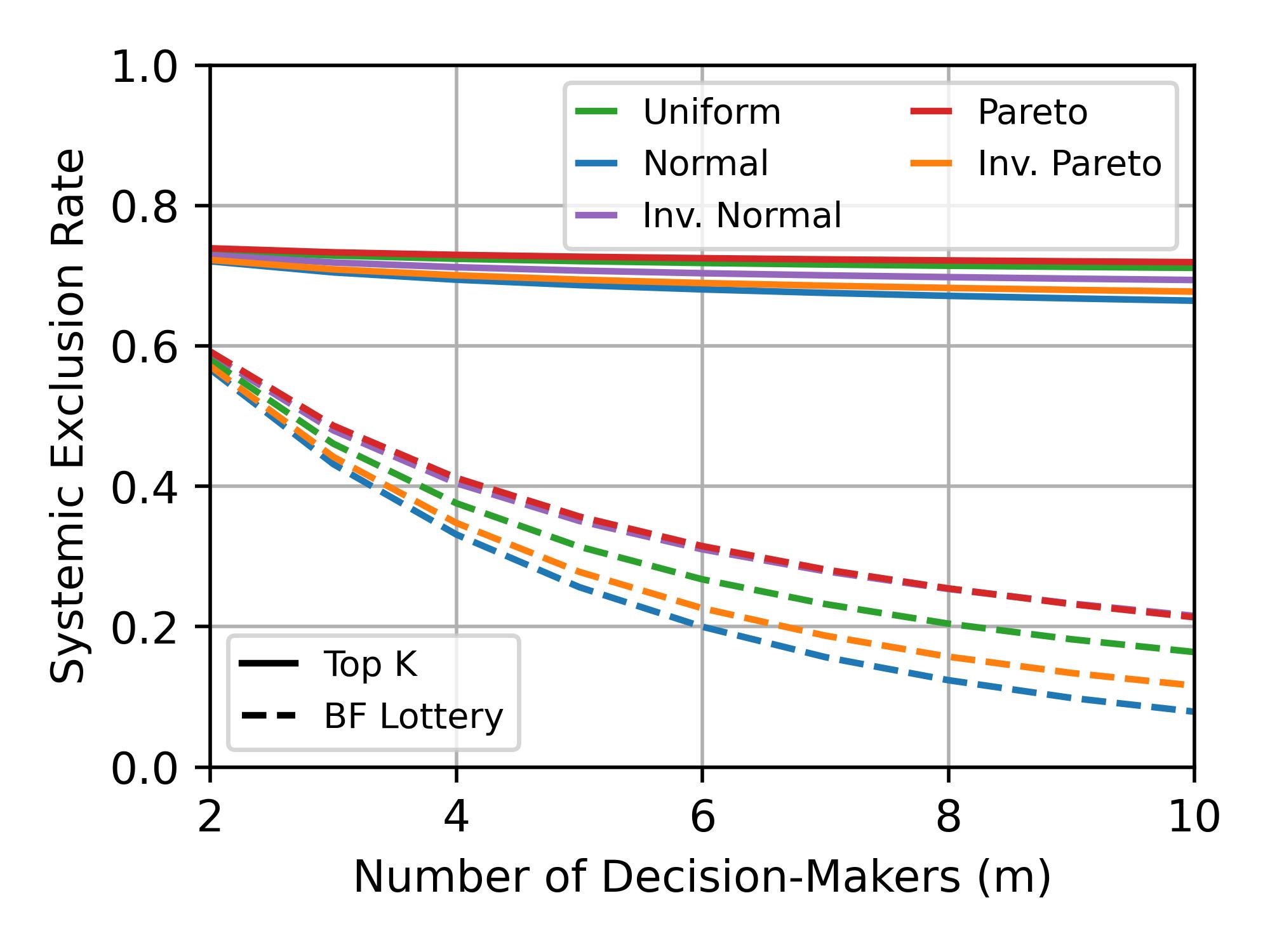}
} 
\subfloat[\centering $k/n$ = 0.25, $\sigma$ = 0.05]{
\includegraphics[width=0.4\columnwidth]{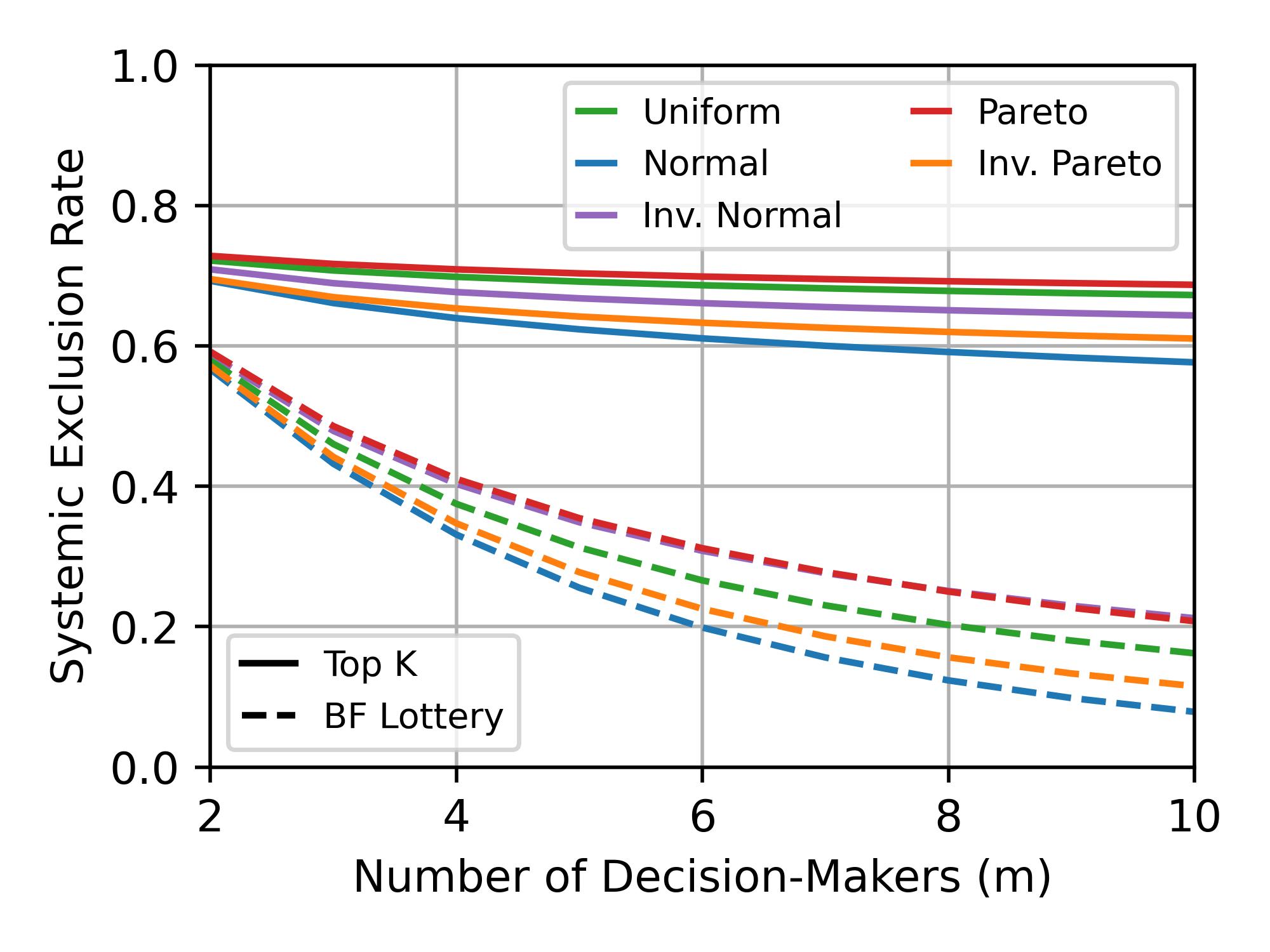}
} \\
\subfloat[\centering $k/n$ = 0.5, $\sigma$ = 0.025]{
\includegraphics[width=0.4\columnwidth]{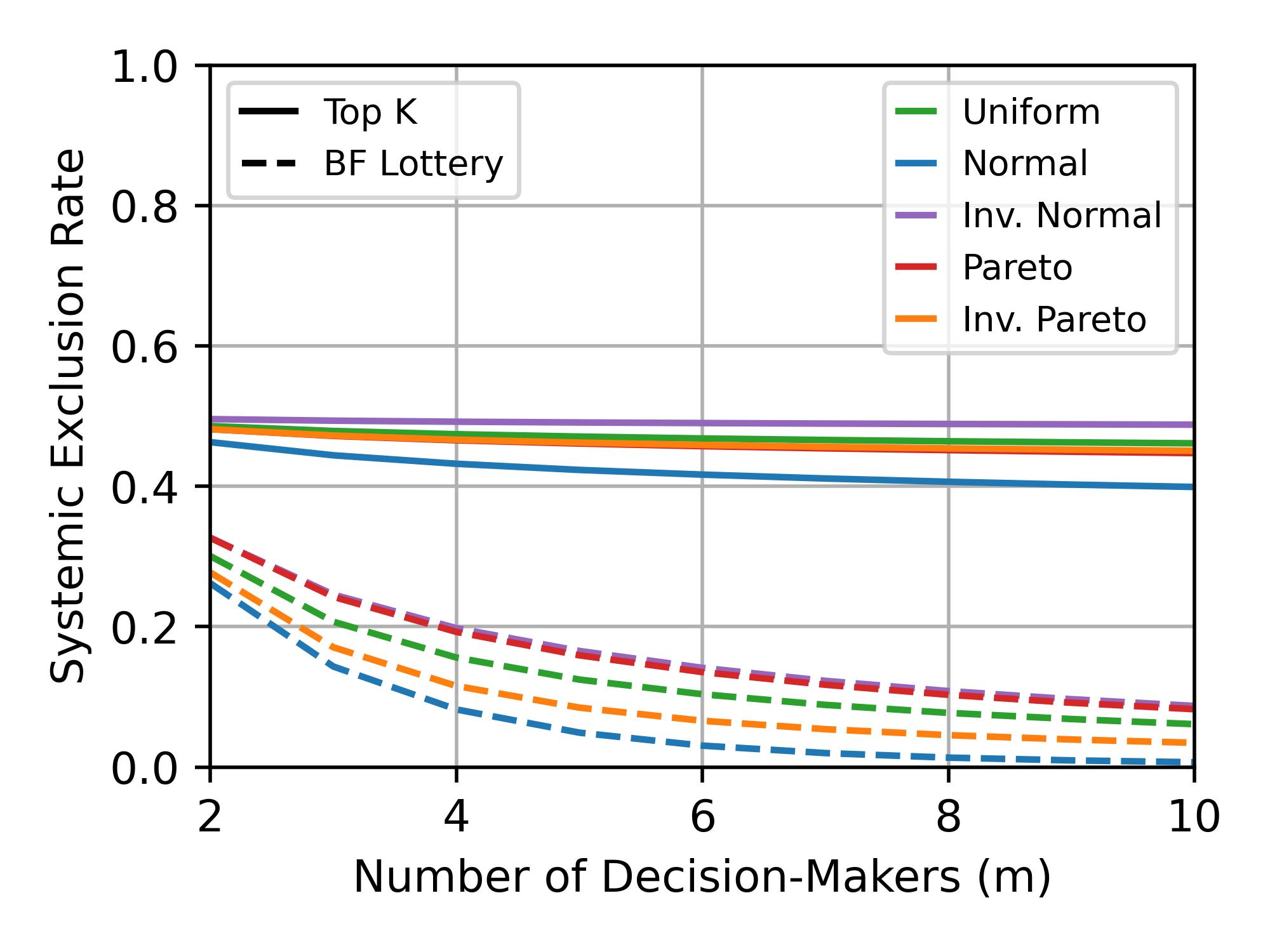}
} 
\subfloat[\centering $k/n$ = 0.5, $\sigma$ = 0.05]{
\includegraphics[width=0.4\columnwidth]{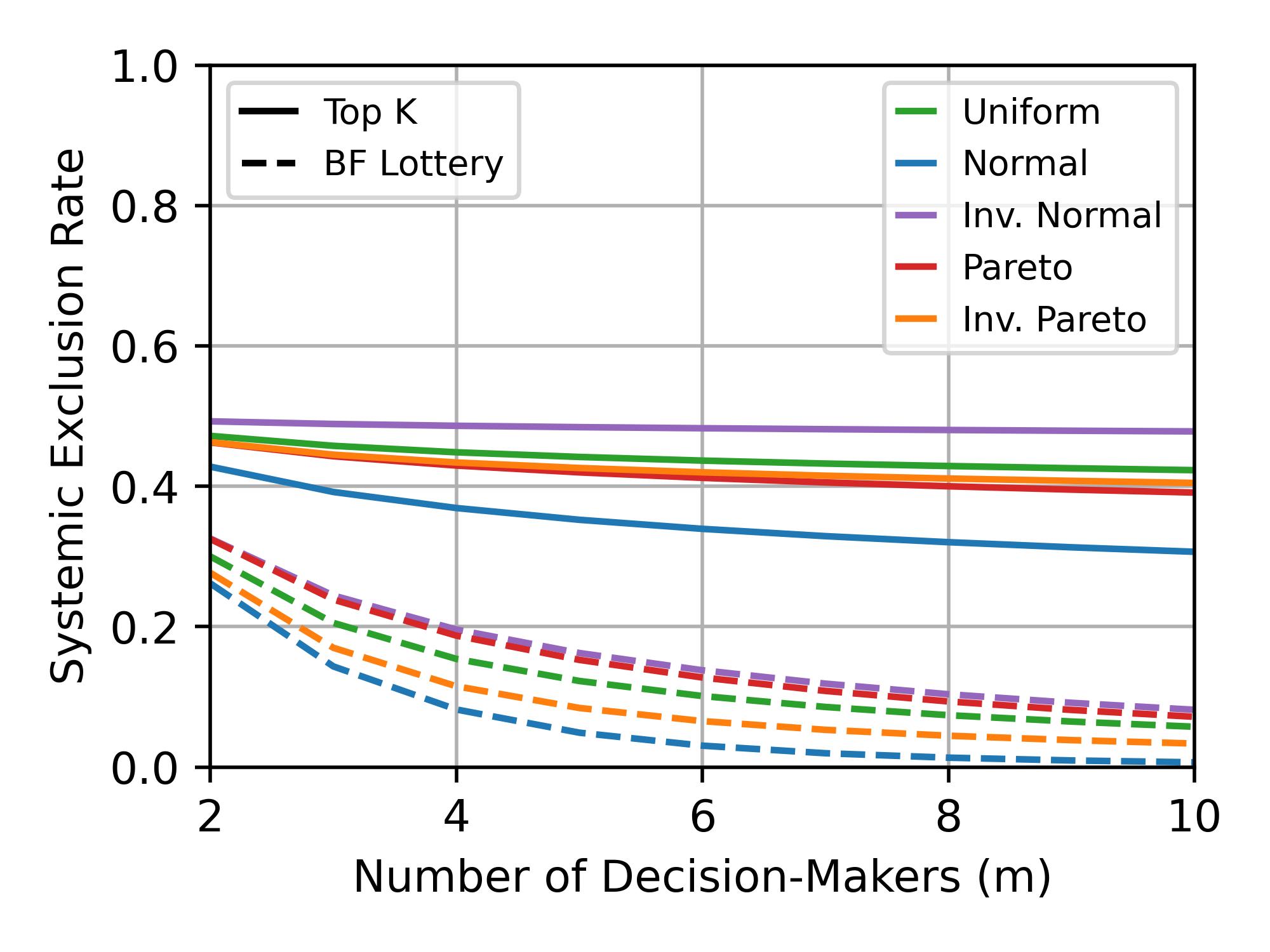}
} \\
\label{fig:ser_full_bf}
\end{figure*}

In Figure~\ref{fig:ser_full_bf}, we show the reduction in systemic exclusion rate using the full BF lottery in Example~\ref{ex:bf_lottery} for all distributions of claims together. This replicates Figure~\subref{fig:claims_ser} in the main text, but for different amounts of noise added to each decision-maker's claims, as well as different selection rates. 

\subsection{When Claims Are Unknown Experiments}
\label{app:claims_unknown}

\begin{figure*}[h!]
\centering
\caption{Swiss Unemployment Data Experiments}
\subfloat[\centering Distribution of Claims]{
\includegraphics[width=0.4\columnwidth]{figures/2_swiss_distribution.jpg}
} \\
\text{(b) - (d) Utility From Randomizing Near the Decision-Boundary} \\[2mm]
\subfloat[\centering Selection Rate = $0.1$]{
\includegraphics[width=0.32\columnwidth]{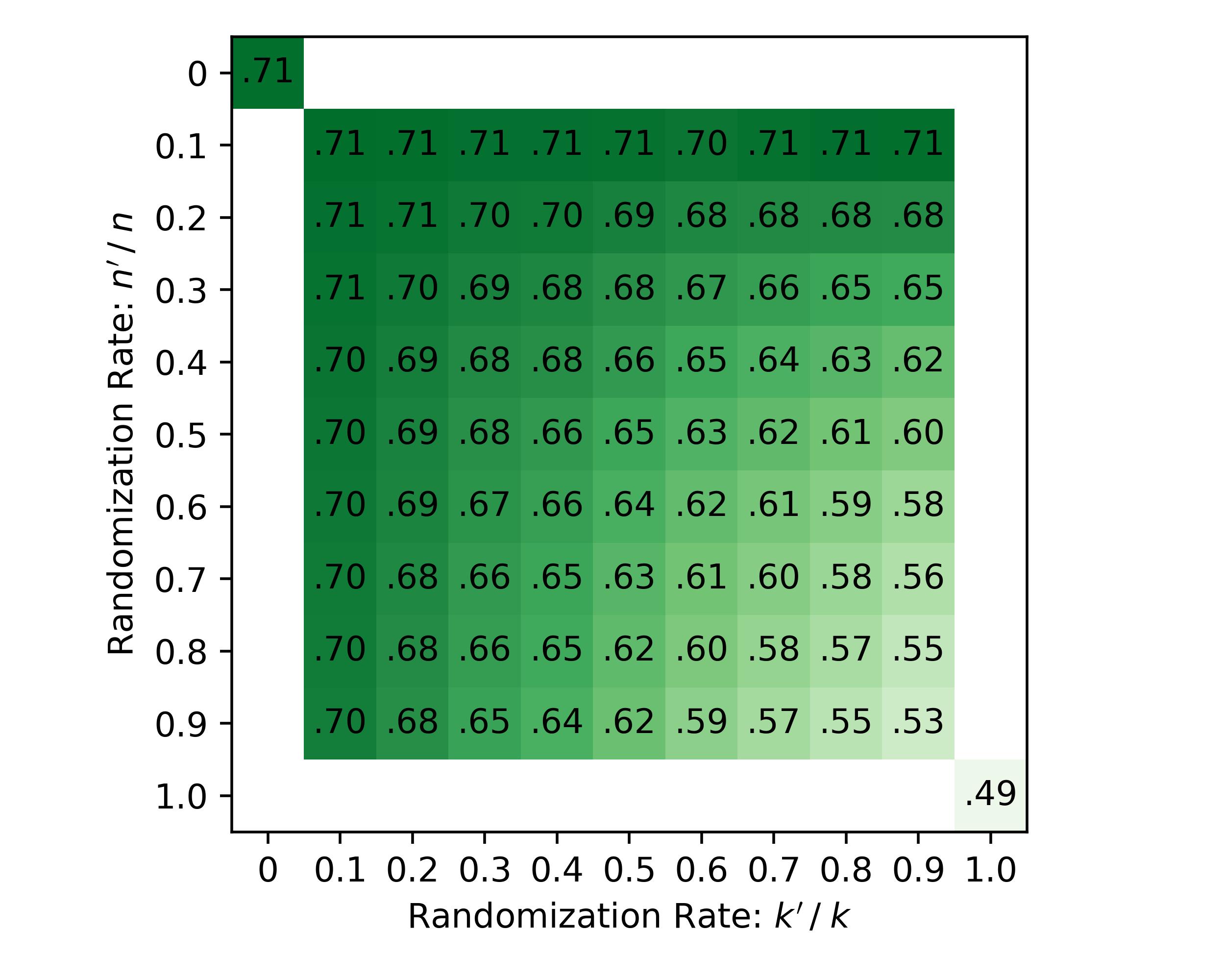}
} 
\subfloat[\centering Selection Rate = $0.25$]{
\includegraphics[width=0.32\columnwidth]{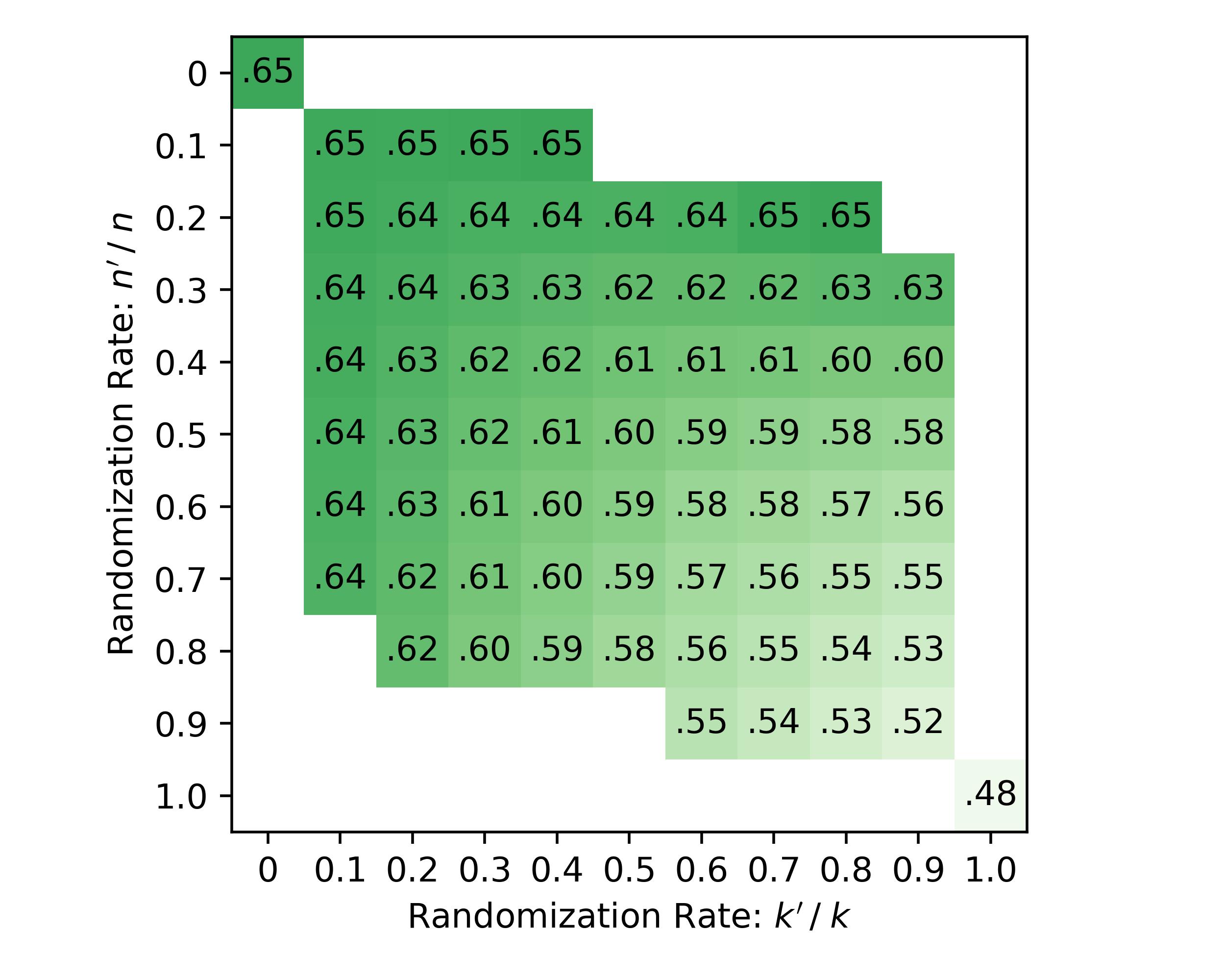}
}
\subfloat[\centering Selection Rate = $0.5$]{
\includegraphics[width=0.32\columnwidth]{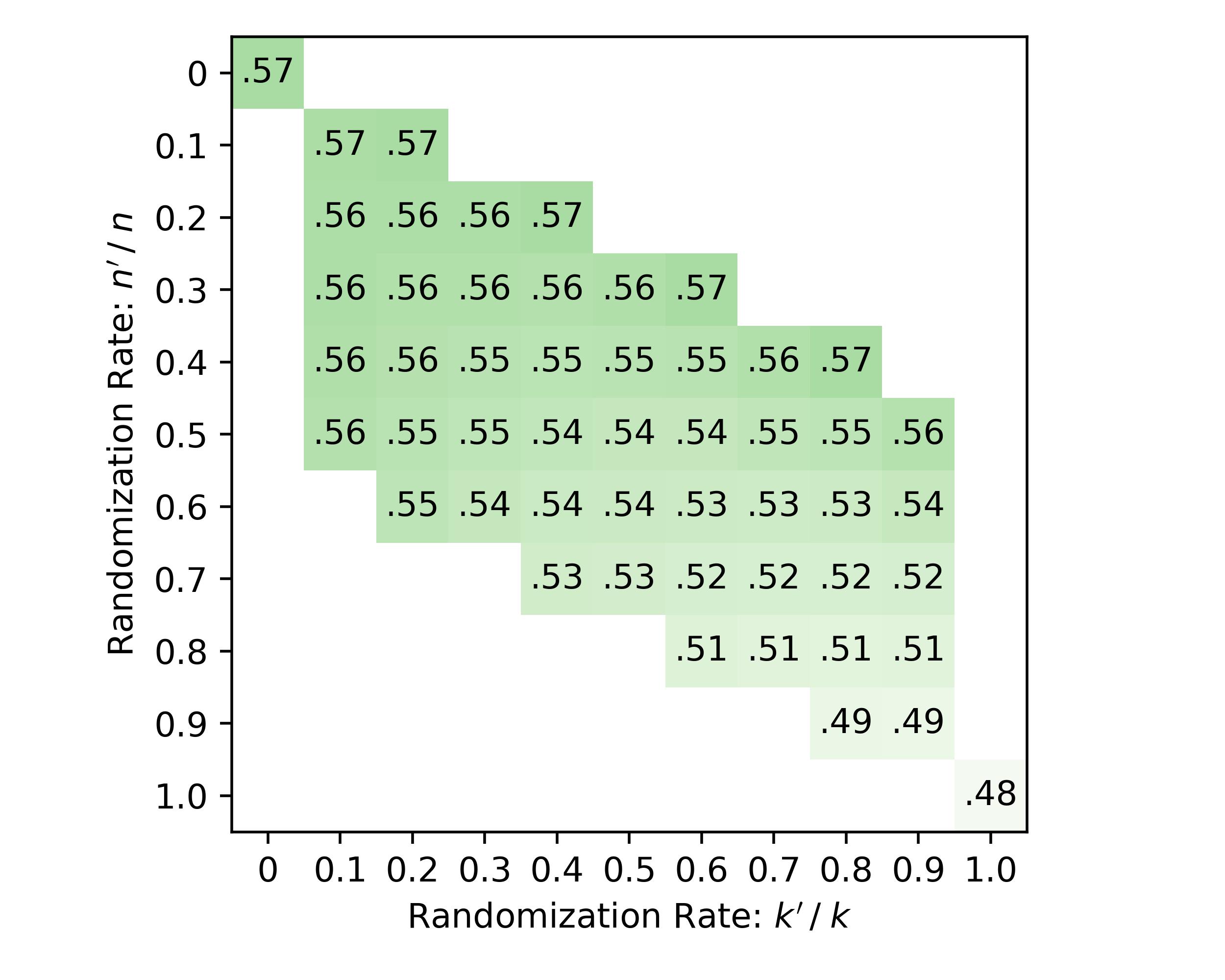}
}
\\
\text{(e) - (g) Systemic Exclusion Rate v. Utility Tradeoff for Each Randomization Method} \\[2mm]
\subfloat[\centering Selection Rate = $0.1$]{
\includegraphics[width=0.32\columnwidth]{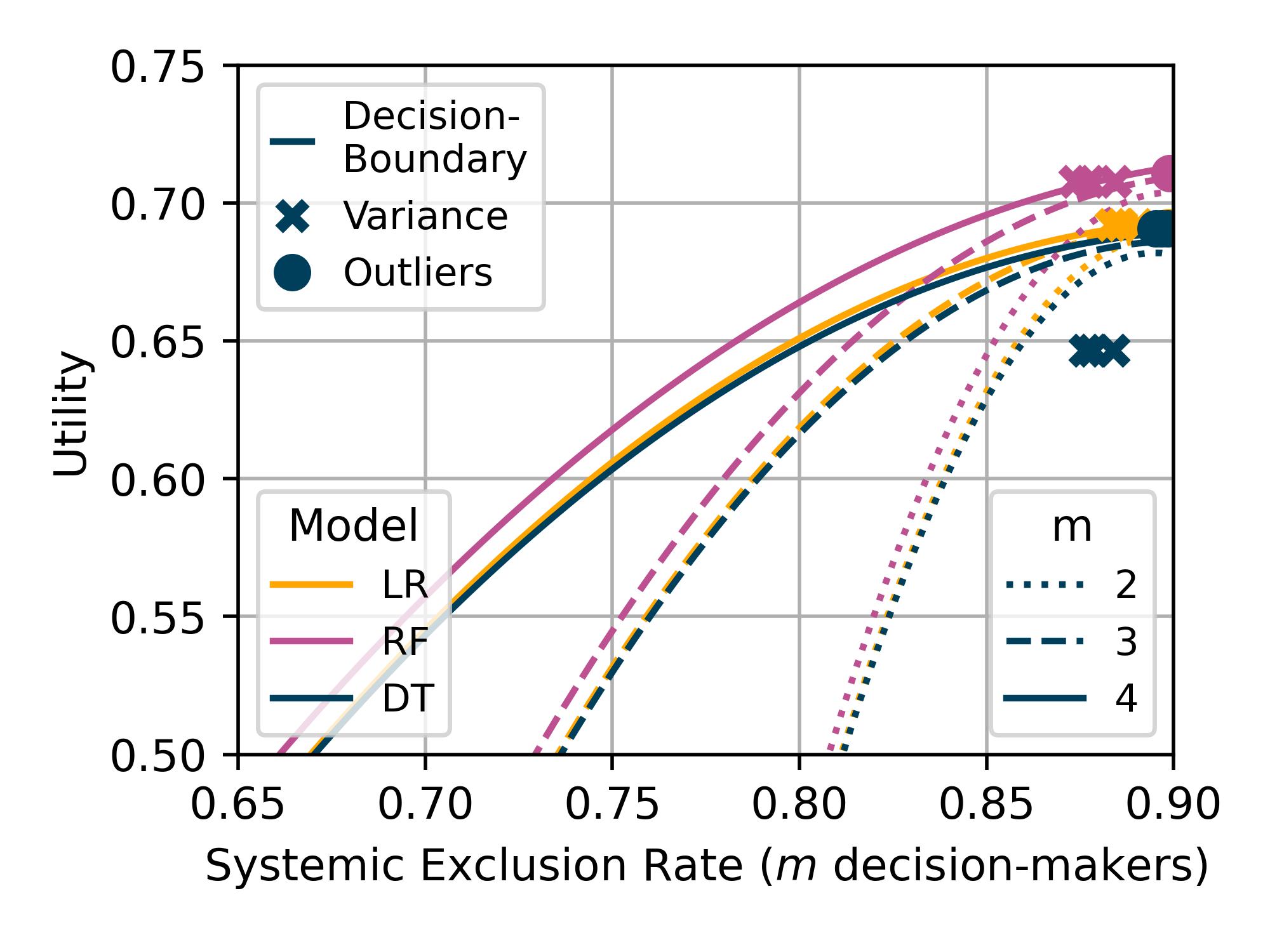}
} 
\subfloat[\centering Selection Rate = $0.25$]{
\includegraphics[width=0.32\columnwidth]{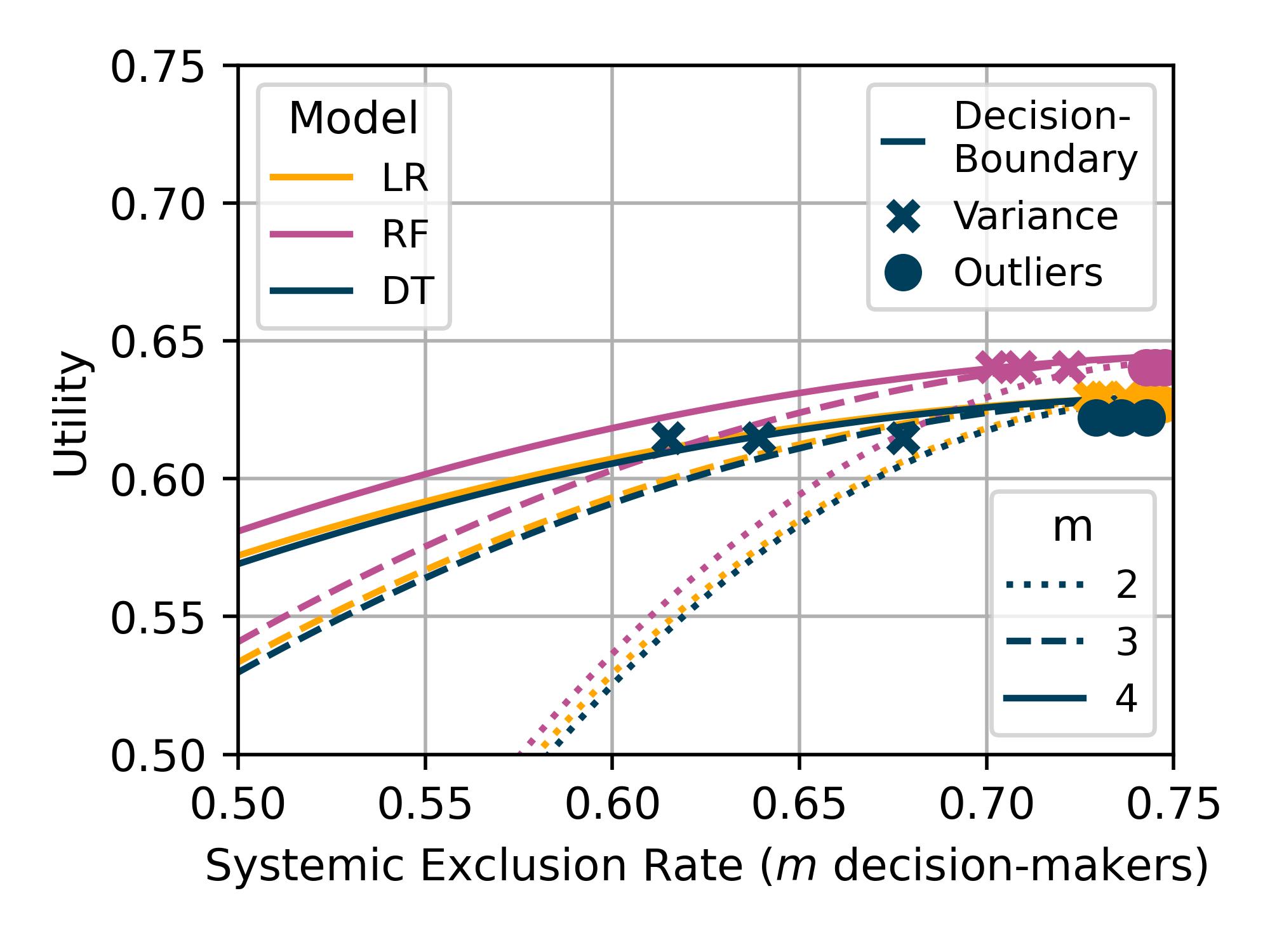}
}
\subfloat[\centering Selection Rate = $0.5$]{
\includegraphics[width=0.32\columnwidth]{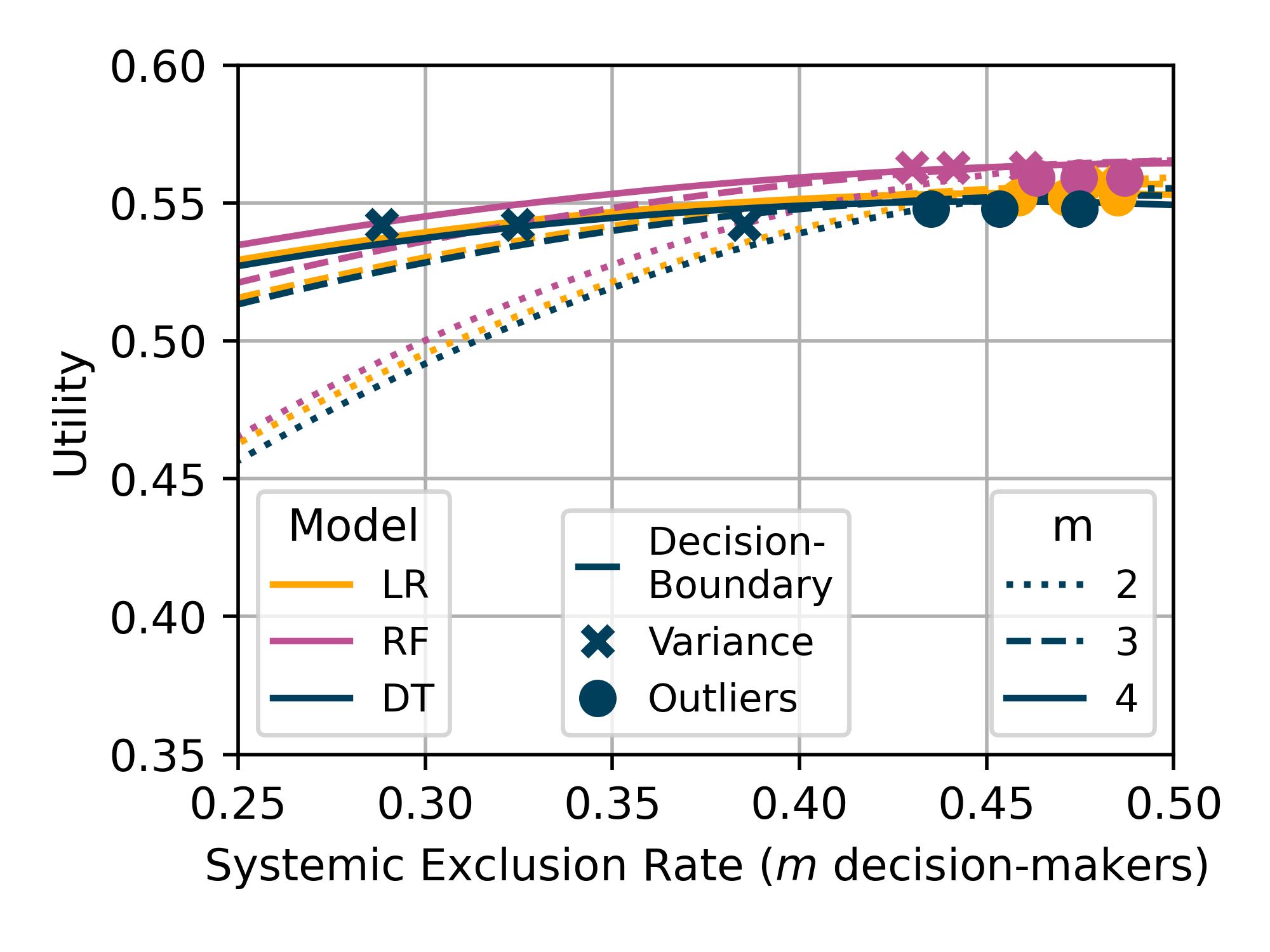}
}
\\

\label{fig:swiss}
\end{figure*}

\begin{figure*}[t!]
\centering
\caption{Census Income Data Experiments}
\subfloat[\centering Distribution of Claims]{
\includegraphics[width=0.4\columnwidth]{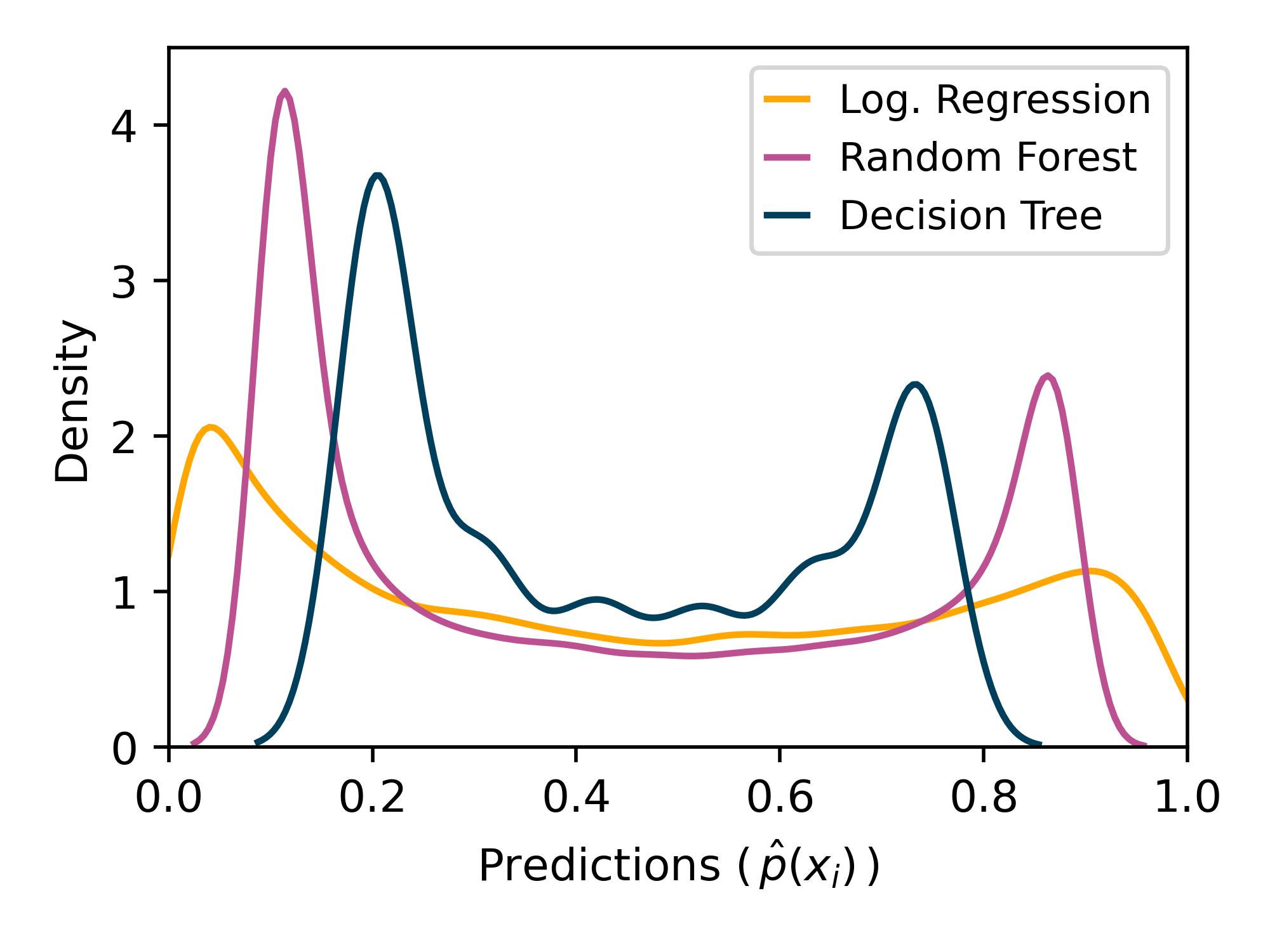}
} \\
\text{(b) - (d) Utility From Randomizing Near the Decision-Boundary} \\[2mm]
\subfloat[\centering Selection Rate = $0.1$]{
\includegraphics[width=0.32\columnwidth]{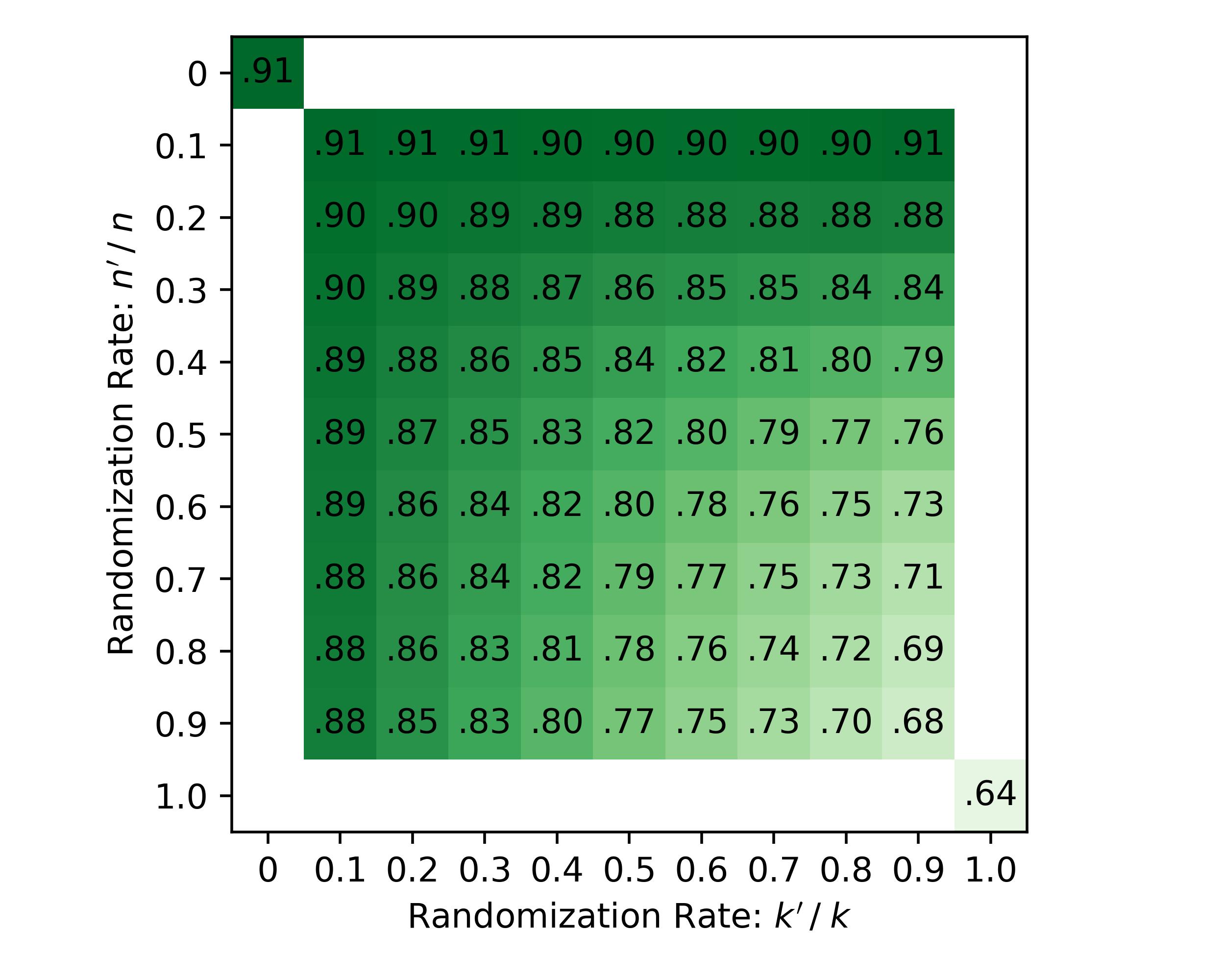}
} 
\subfloat[\centering Selection Rate = $0.25$]{
\includegraphics[width=0.32\columnwidth]{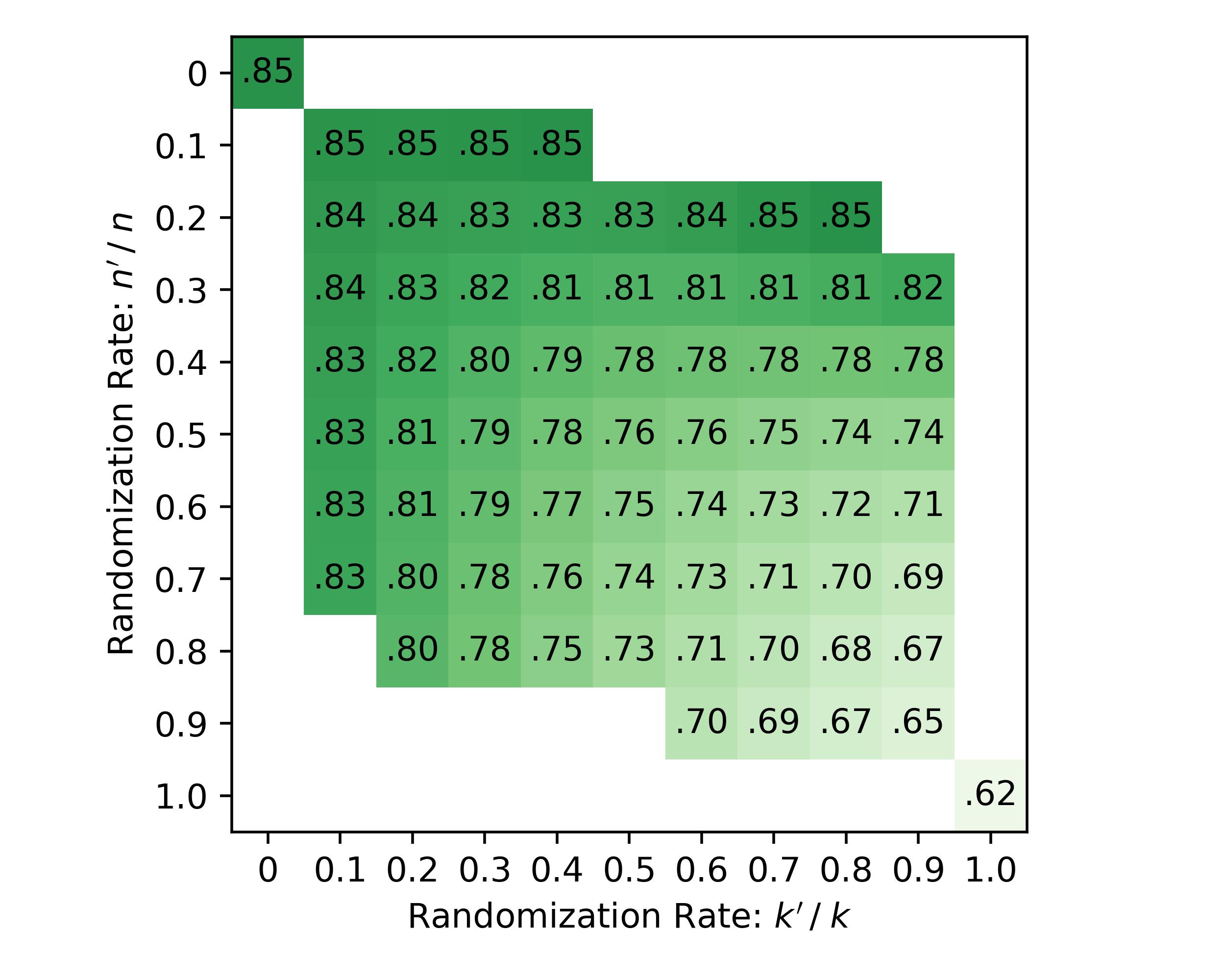}
}
\subfloat[\centering Selection Rate = $0.5$]{
\includegraphics[width=0.32\columnwidth]{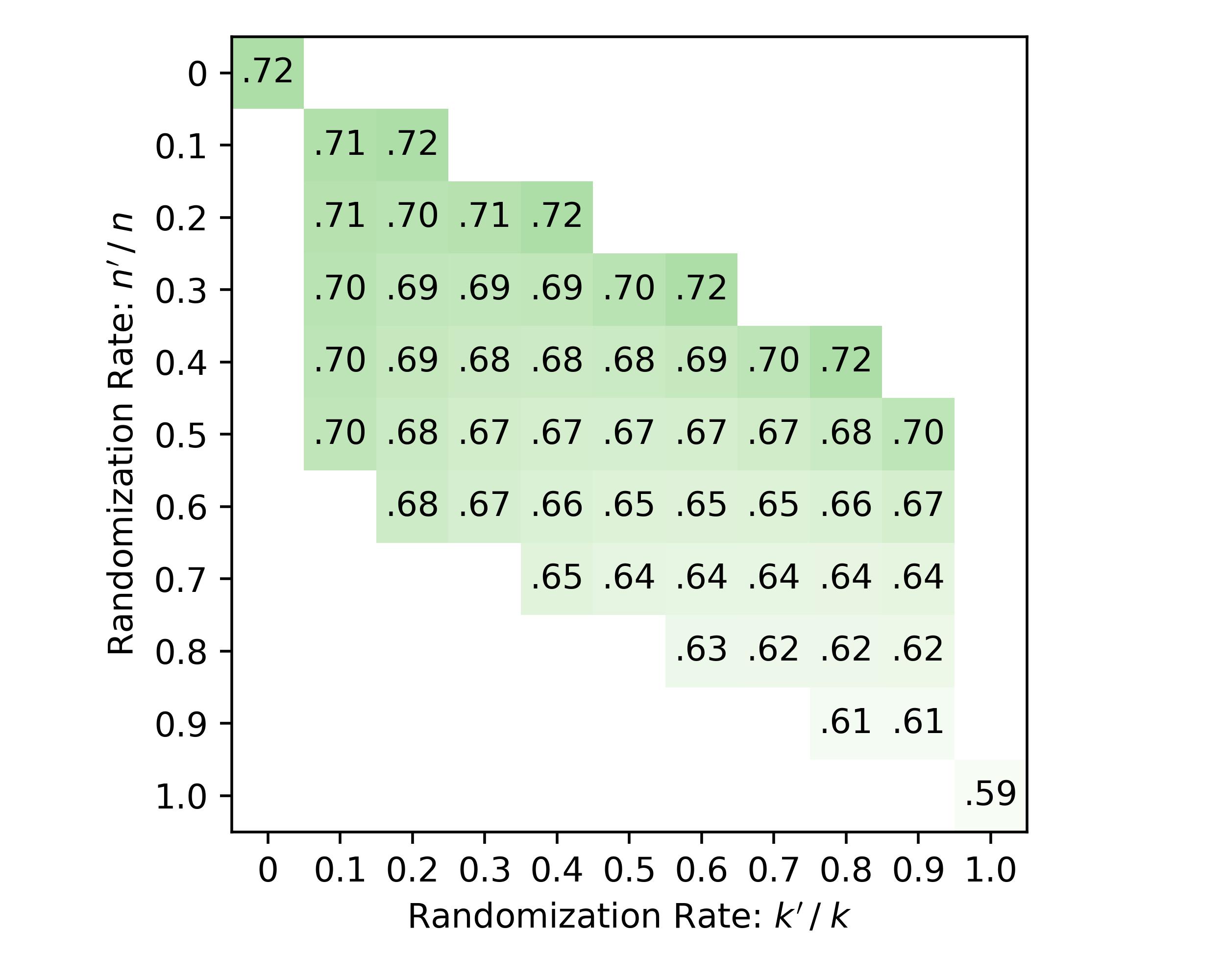}
}
\\
\text{(e) - (g) Systemic Exclusion Rate v. Utility Tradeoff for Each Randomization Method} \\[2mm]
\subfloat[\centering Selection Rate = $0.1$]{
\includegraphics[width=0.32\columnwidth]{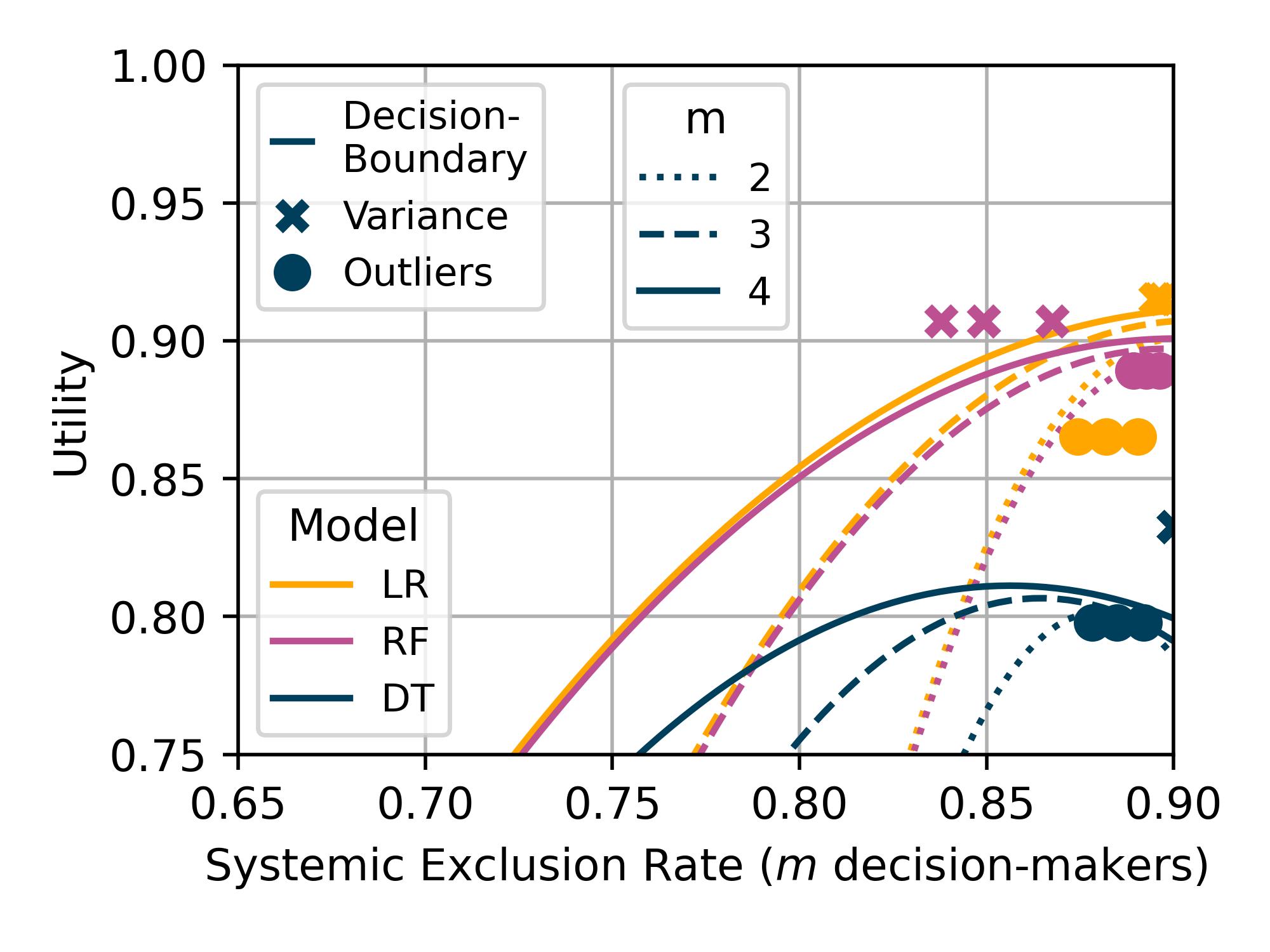}
} 
\subfloat[\centering Selection Rate = $0.25$]{
\includegraphics[width=0.32\columnwidth]{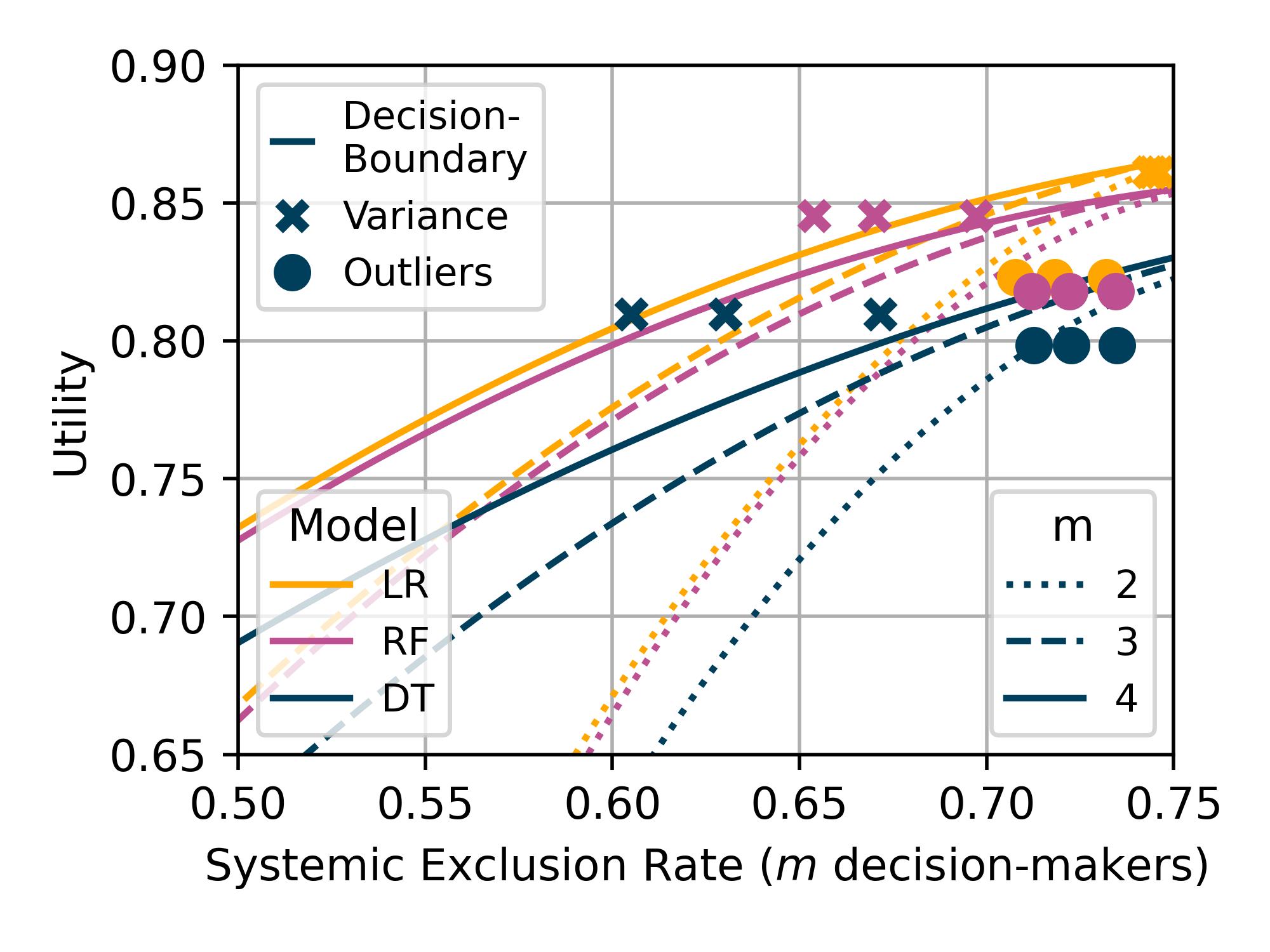}
}
\subfloat[\centering Selection Rate = $0.5$]{
\includegraphics[width=0.32\columnwidth]{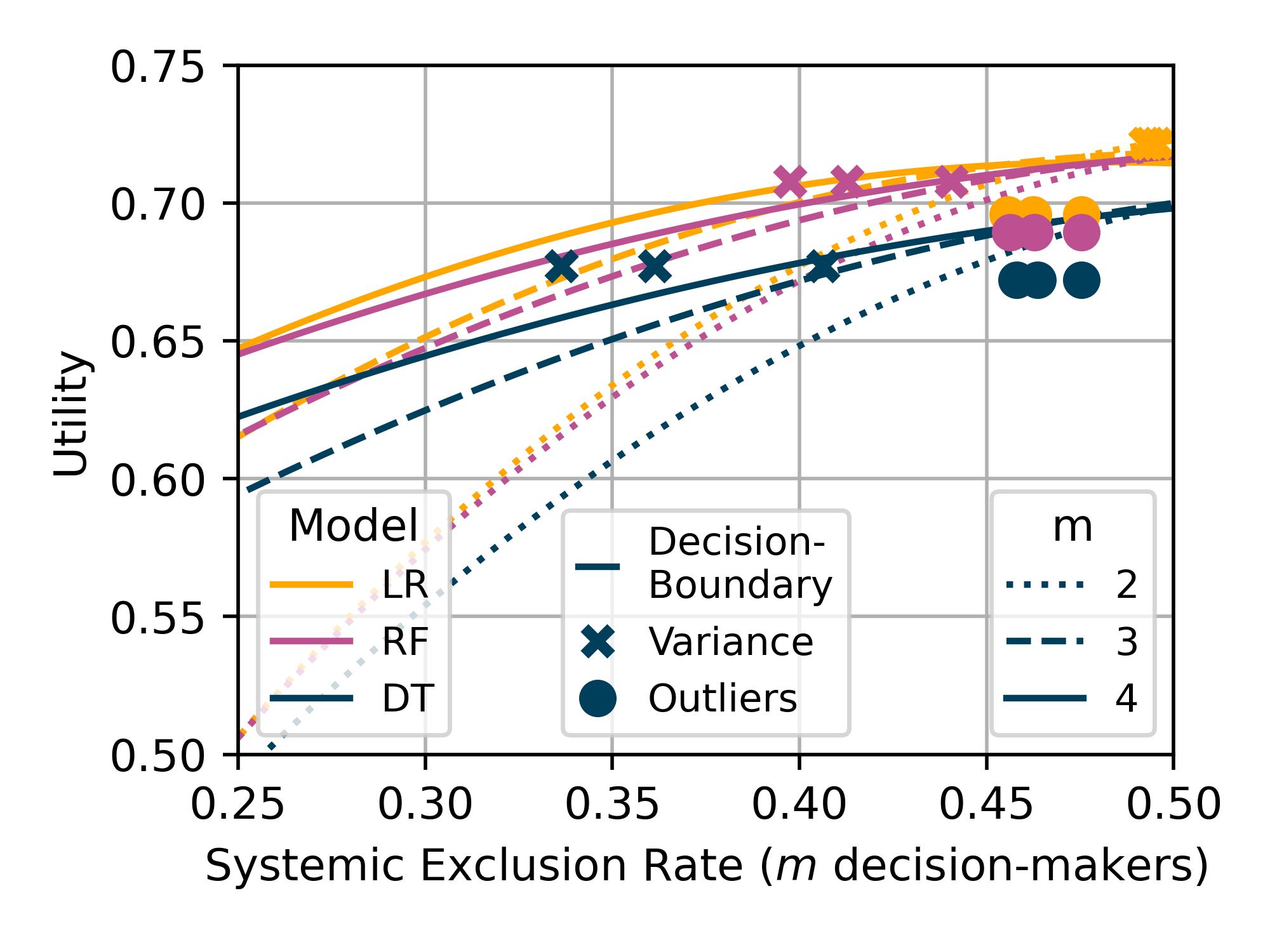}
}
\\
\label{fig:census}
\end{figure*}

\begin{figure*}[h!]
\centering
\caption{\centering Density of Predictions by Point Estimate $\hat{p}(x_i)$ \& Uncertainty Metric (Darker Colors = Higher Density)}
\text{(a) - (c) Swiss Unemployment Data -- Predictions x Std Dev Across Bootstrapped Predictions} \\[2mm]
\subfloat[\centering Logistic Regression]{
\includegraphics[width=0.30\columnwidth]{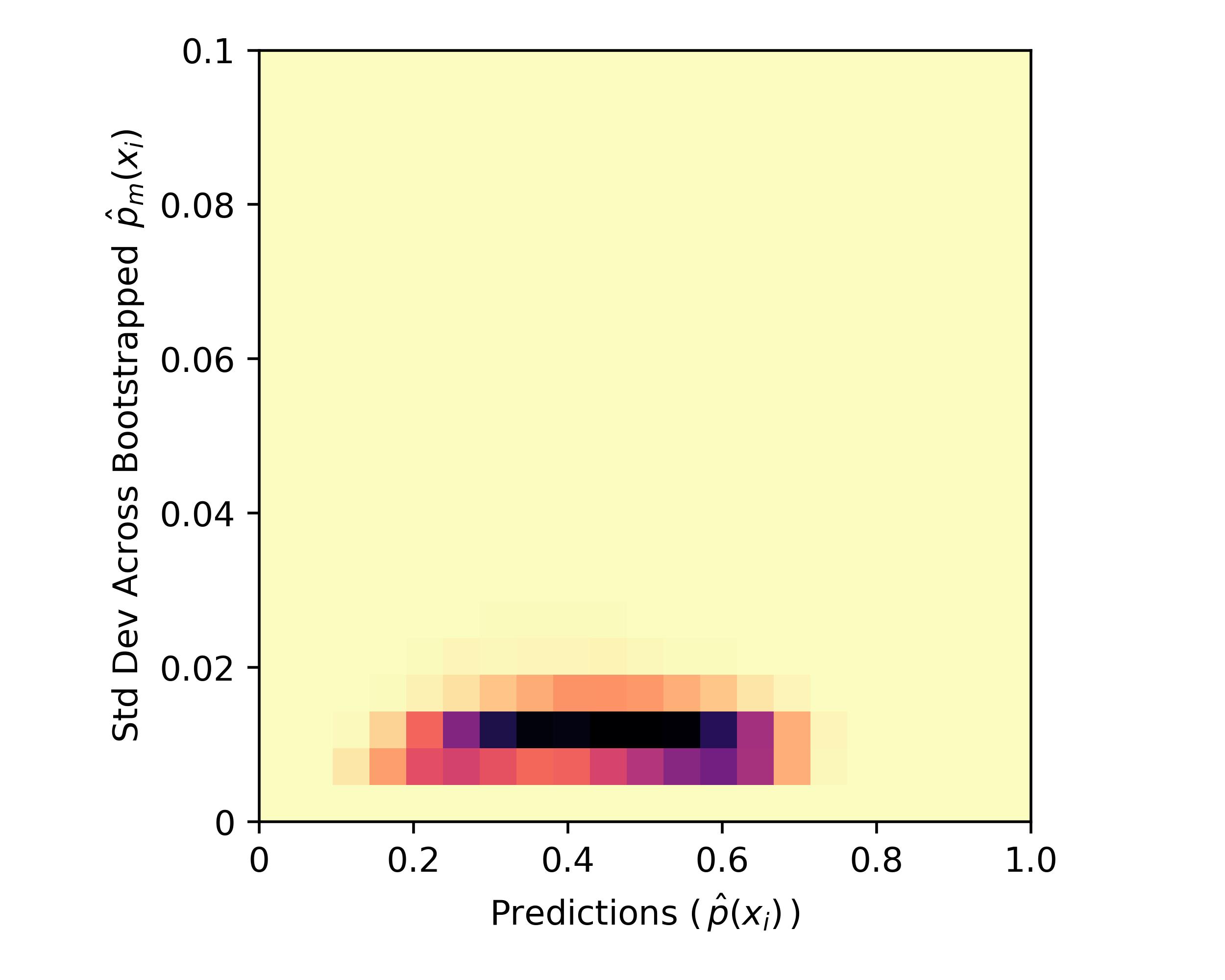}
} 
\subfloat[\centering Random Forest]{
\includegraphics[width=0.30\columnwidth]{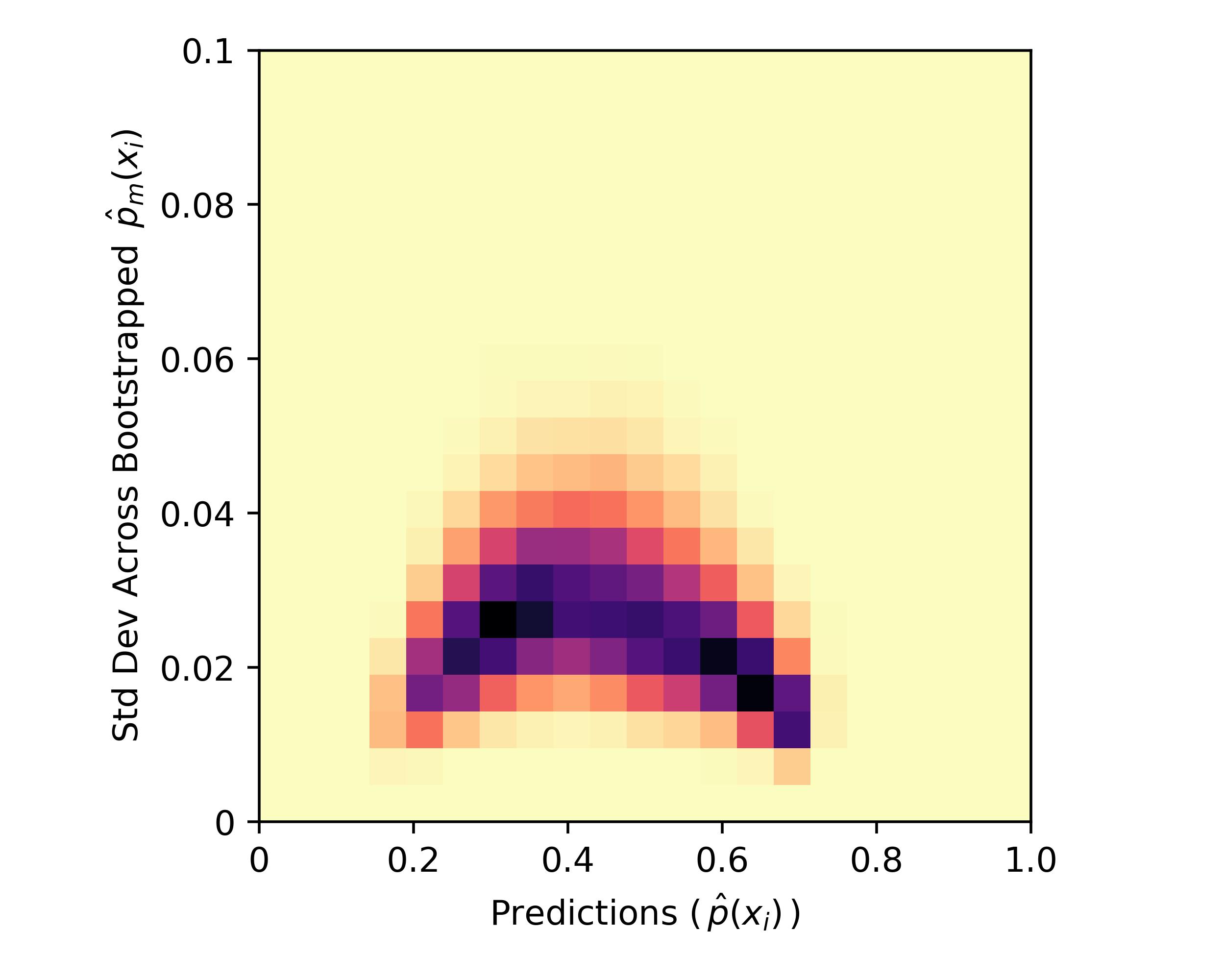}
}
\subfloat[\centering Decision Tree]{
\includegraphics[width=0.30\columnwidth]{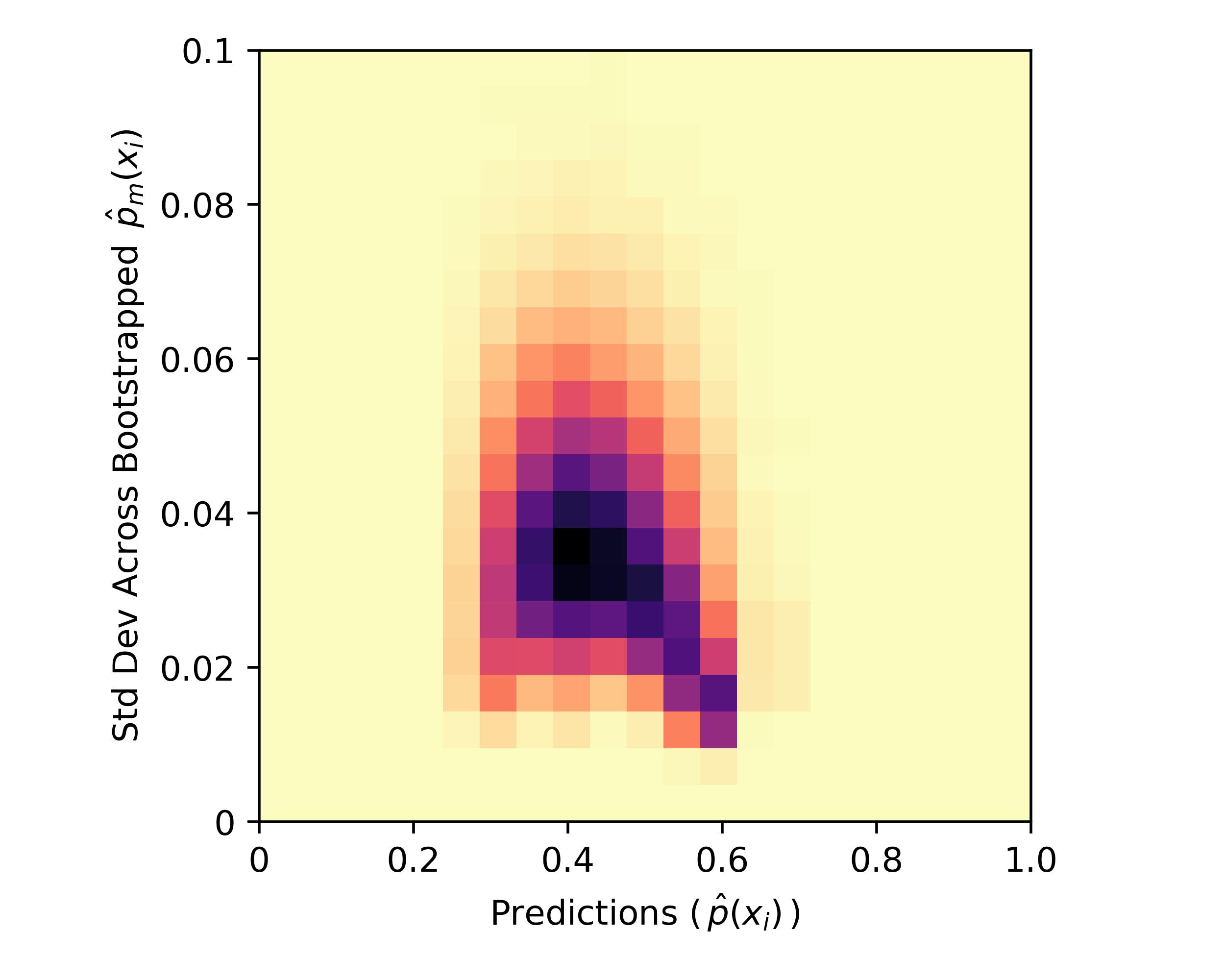}
}
\\
\text{(d) - (f) Swiss Unemployment Data -- Predictions x Conformal P-Values for Outlier Detection} \\[2mm]
\subfloat[\centering Logistic Regression]{
\includegraphics[width=0.30\columnwidth]{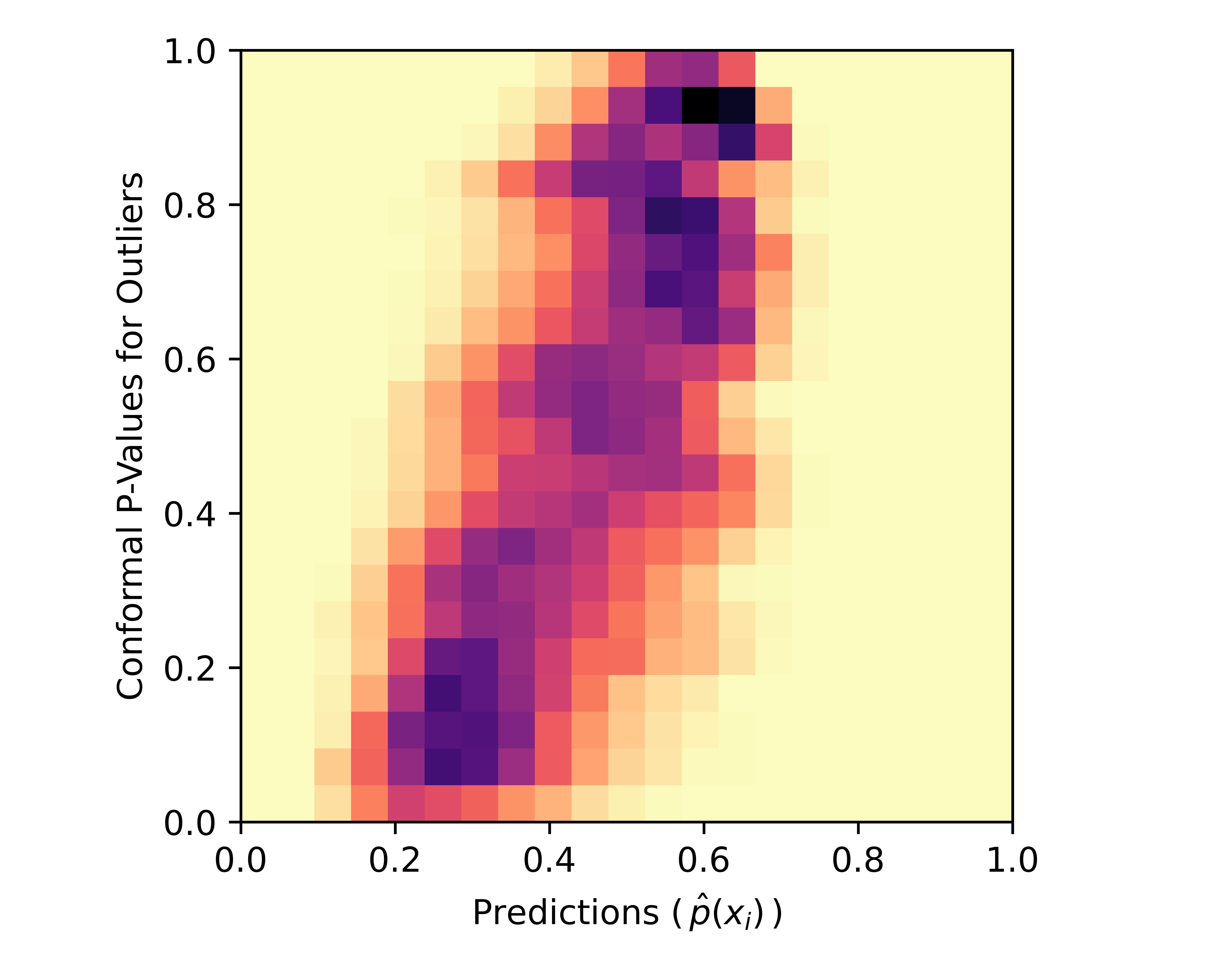}
} 
\subfloat[\centering Random Forest]{
\includegraphics[width=0.30\columnwidth]{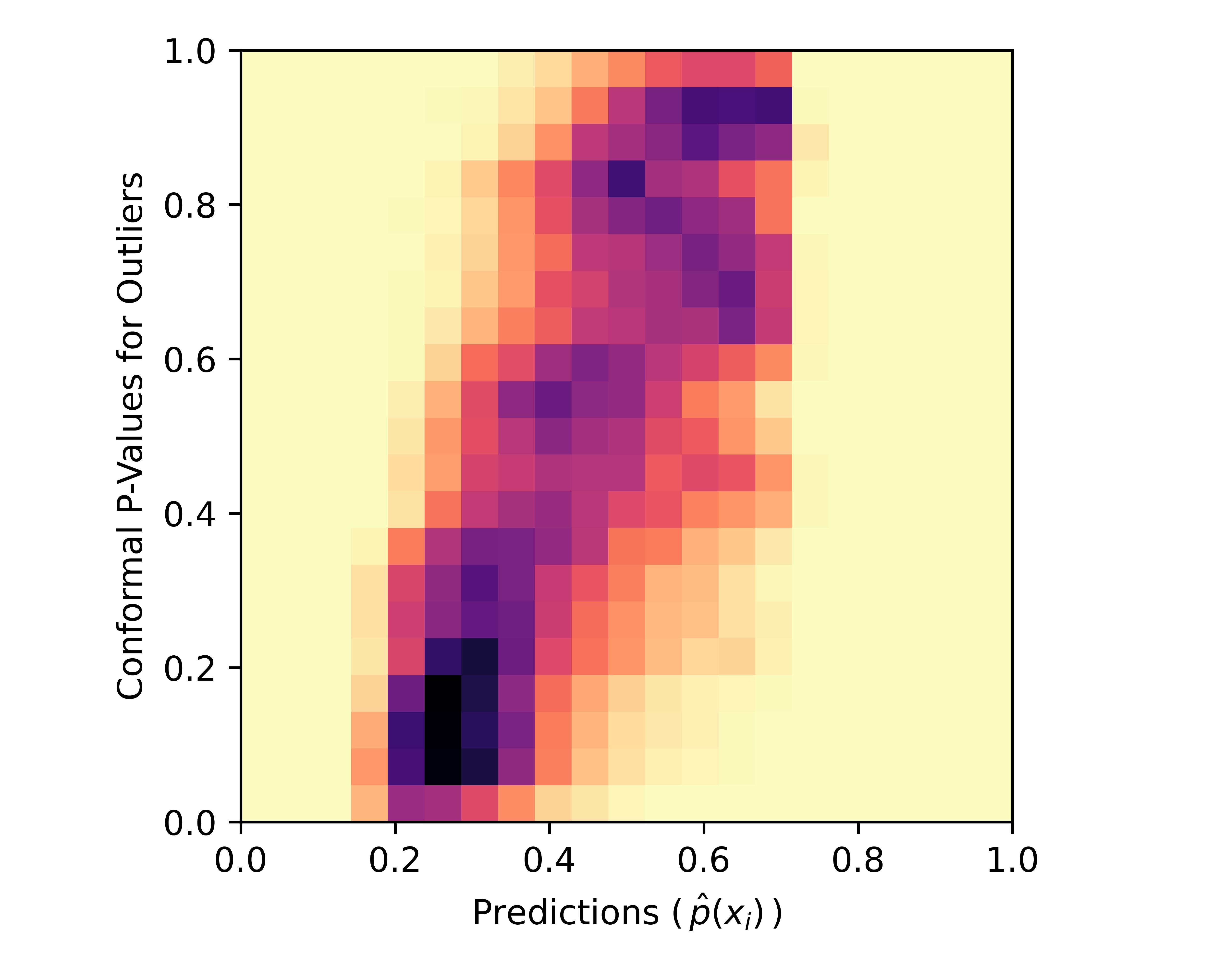}
}
\subfloat[\centering Decision Tree]{
\includegraphics[width=0.30\columnwidth]{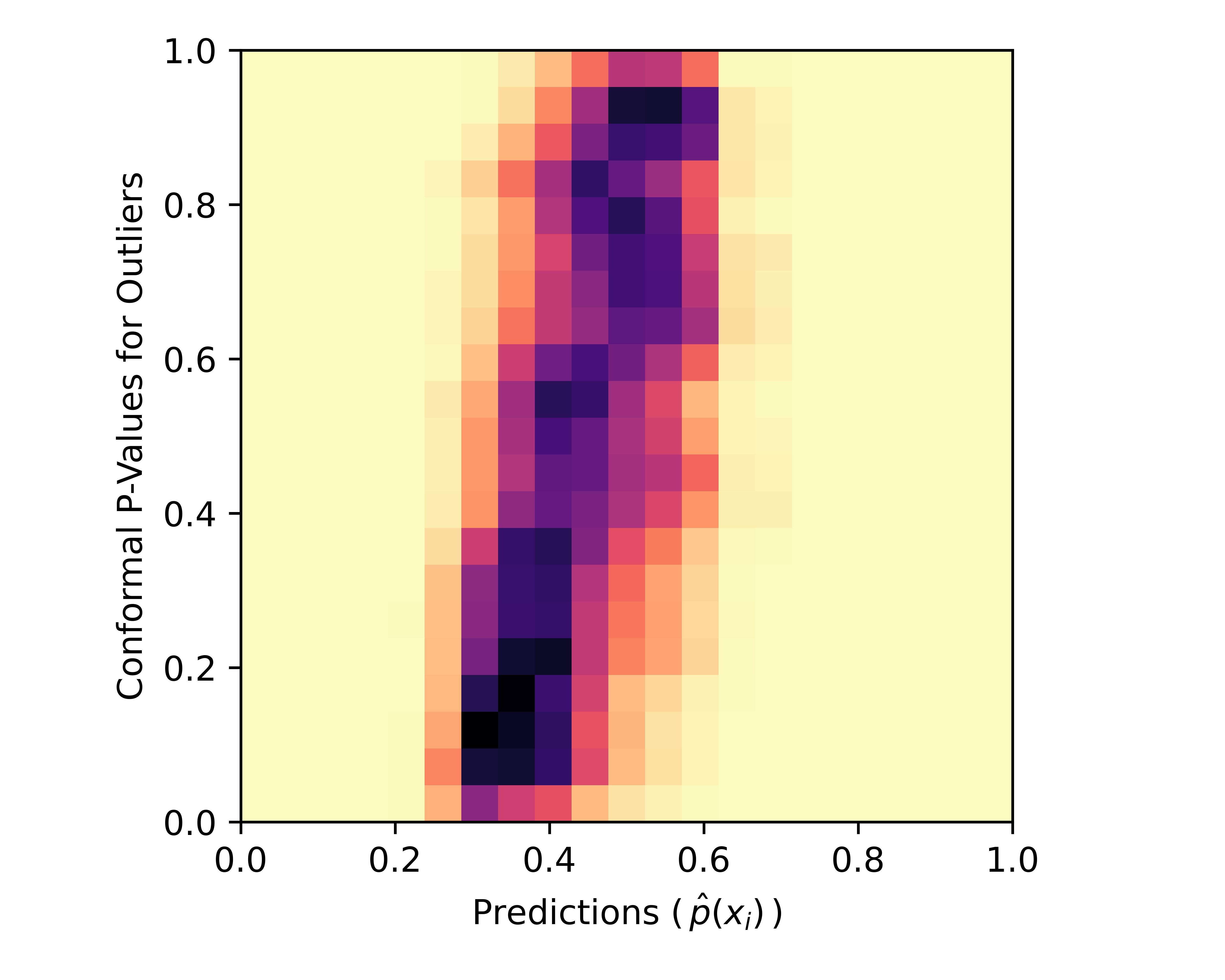}
}
\\
\text{(g) - (i) Census Income Data -- Predictions x Std Dev Across Bootstrapped Predictions} \\[2mm]
\subfloat[\centering Logistic Regression]{
\includegraphics[width=0.30\columnwidth]{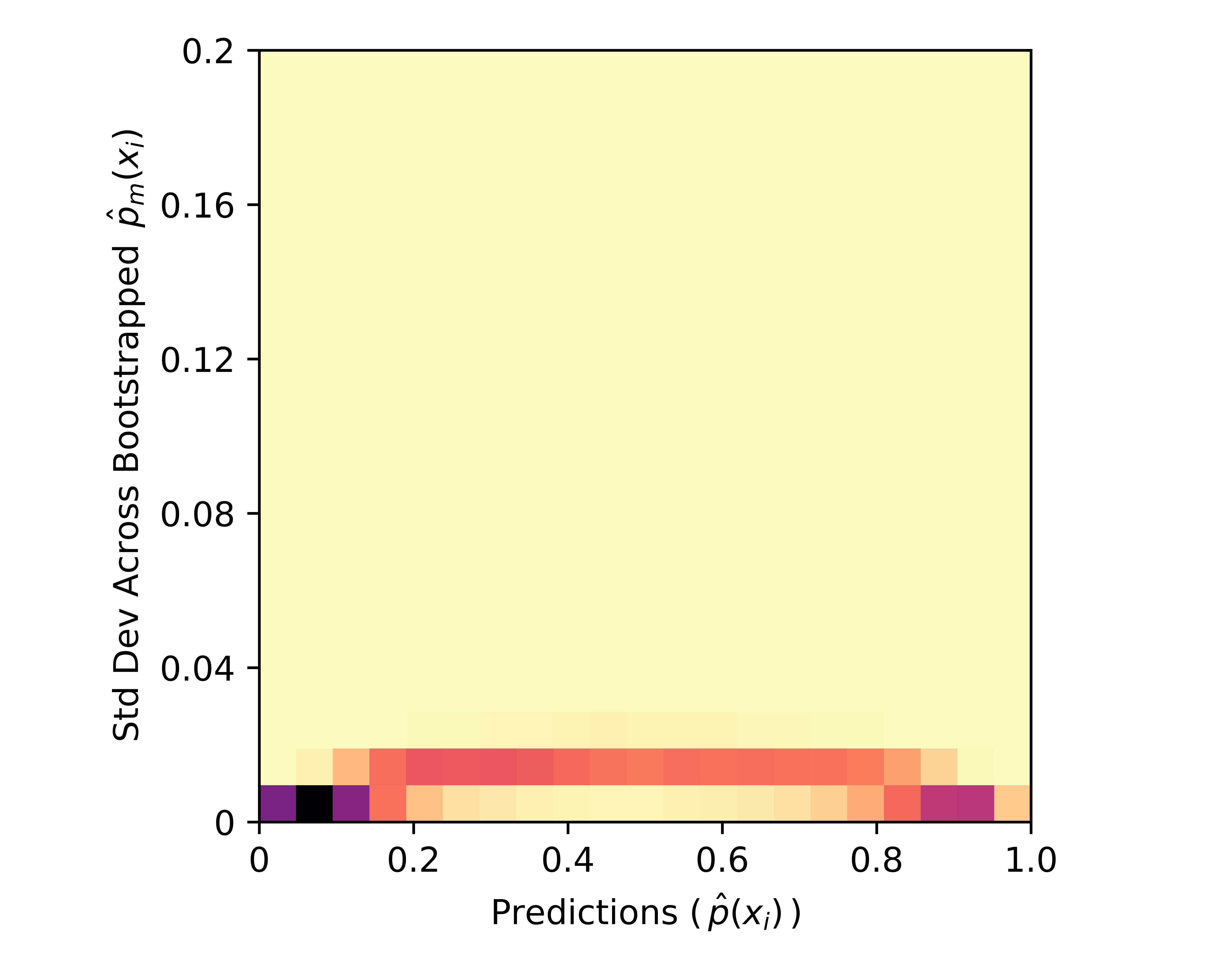}
} 
\subfloat[\centering Random Forest]{
\includegraphics[width=0.30\columnwidth]{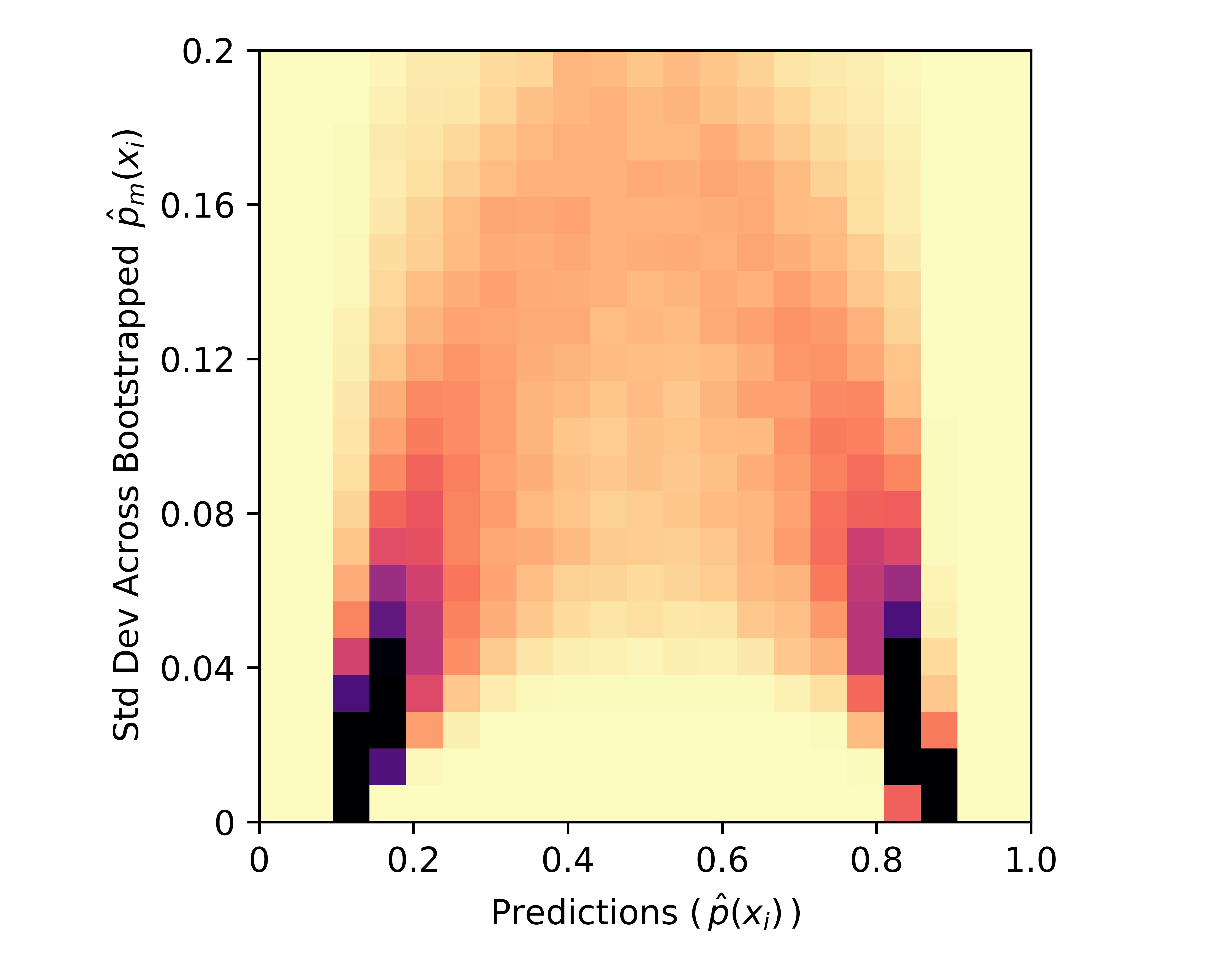}
}
\subfloat[\centering Decision Tree]{
\includegraphics[width=0.30\columnwidth]{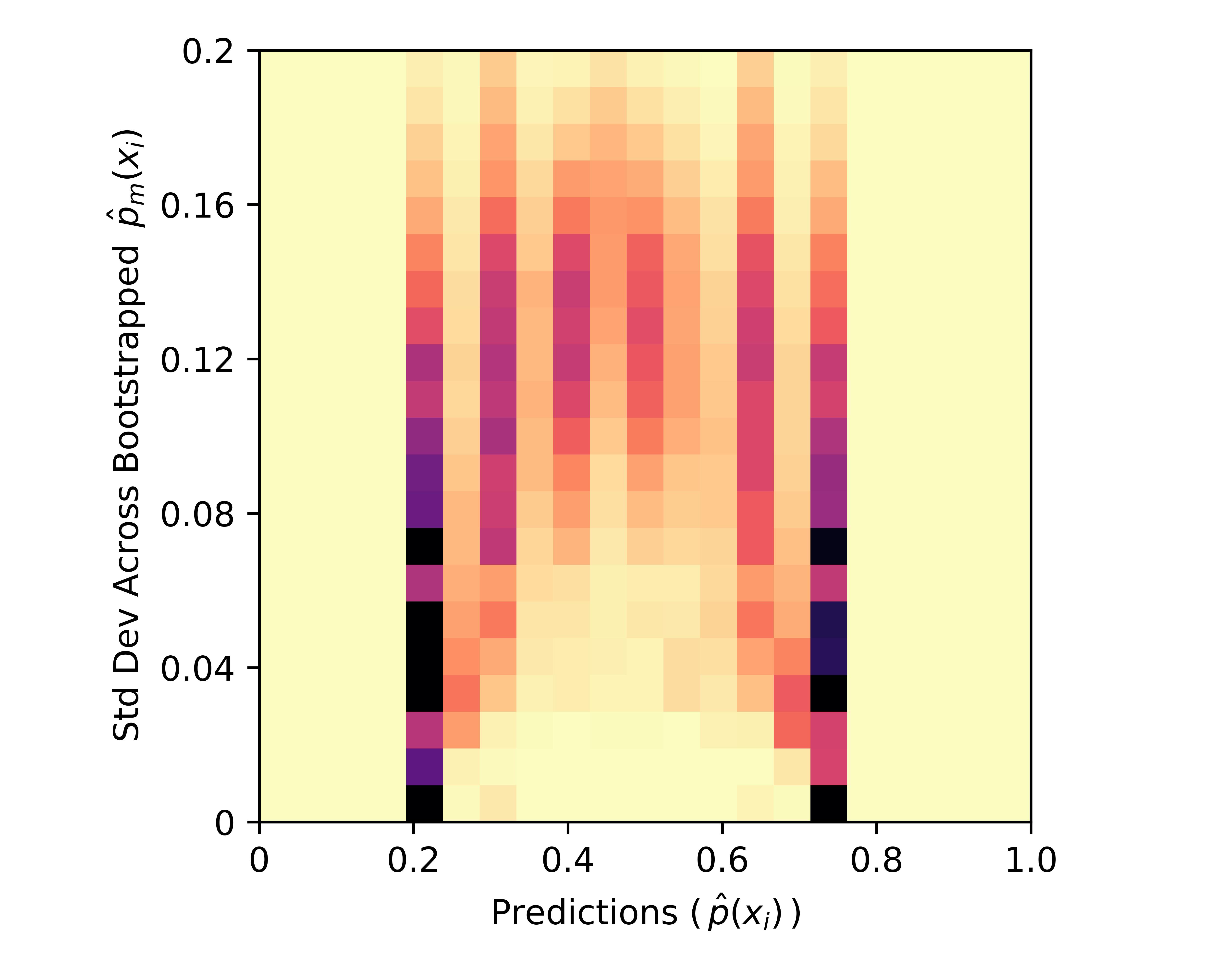}
}
\\
\text{(j) - (l) Census Income Data -- Predictions x Conformal P-Values for Outlier Detection} \\[2mm]
\subfloat[\centering Logistic Regression]{
\includegraphics[width=0.30\columnwidth]{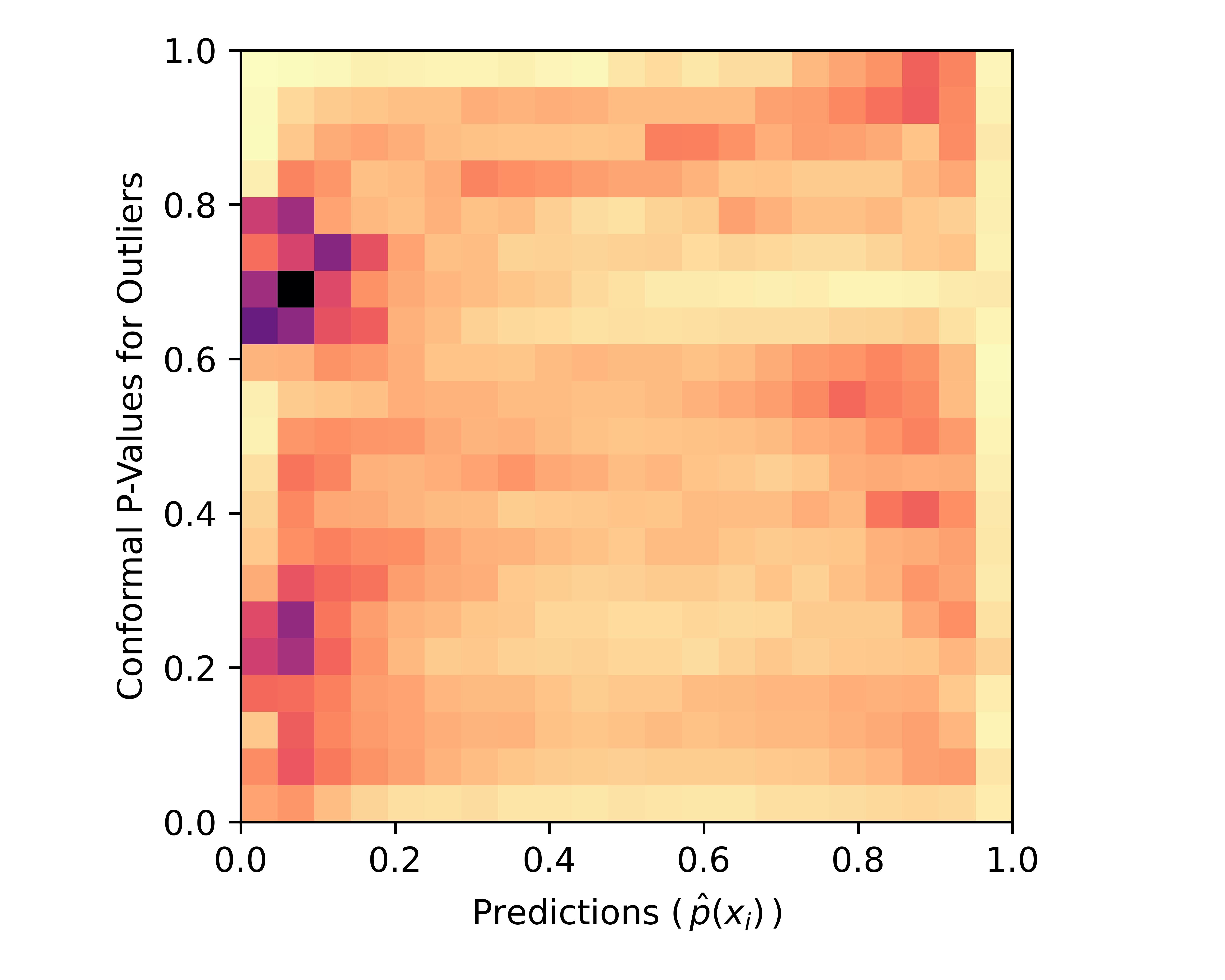}
} 
\subfloat[\centering Random Forest]{
\includegraphics[width=0.30\columnwidth]{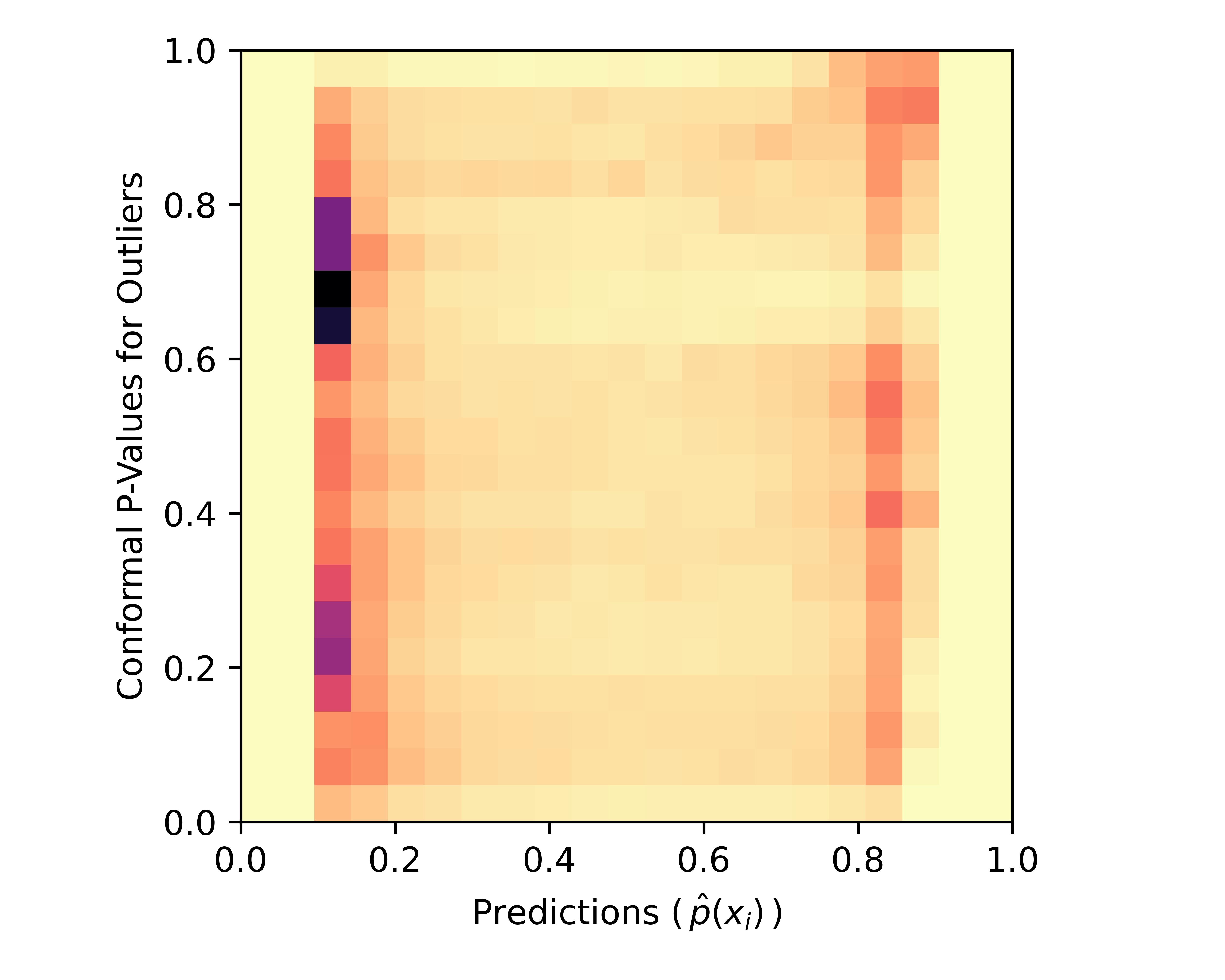}
}
\subfloat[\centering Decision Tree]{
\includegraphics[width=0.30\columnwidth]{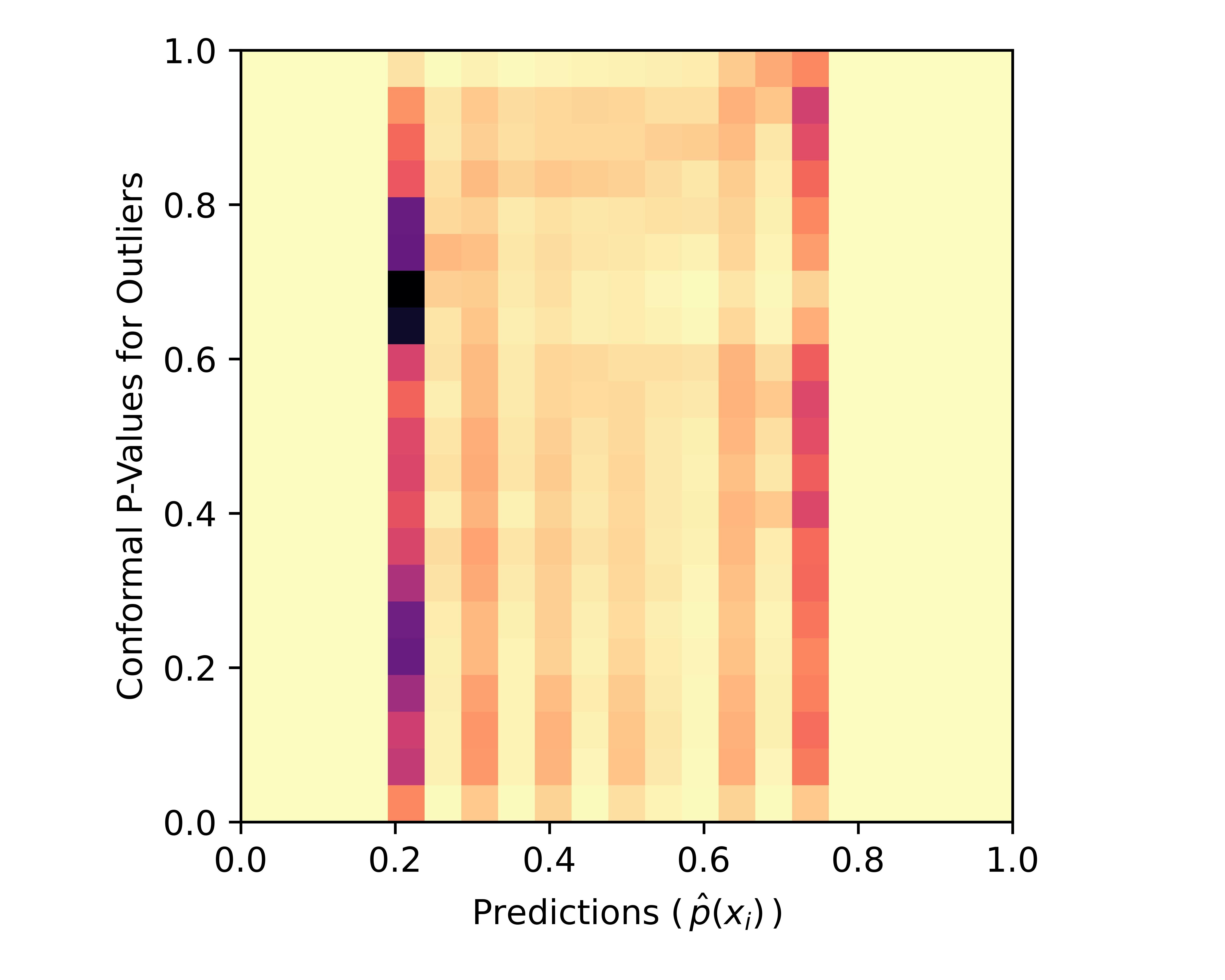}
}
\\
\label{fig:risk_scores_vs_uncertainty}
\end{figure*}

We test our randomization proposals on 2 datasets: (1) Swiss Unemployment Data~\citep{swissData}, and (2) Census Income Data~\citep{ding2021retiring}. For each dataset, we test our 3 randomization methods: randomizing near the decision-boundary, randomizing using variance, and randomizing outliers. We report results for 3 different model classes (logistic regression, random forests, and decision trees) and 3 different selection rates (0.1, 0.25, and 0.5). Specifically, we analyze how each randomization method changes how many resources ($k'$) and what kinds of people ($n'$) are randomized, for some (often minimal) loss in utility. We also compare how much each method reduces the systemic exclusion rate. All our experiments involve an 80-20 train-test split (with 5 repetitions), and we average randomization results over 100 iterations. For each dataset, we provide the following:
\begin{itemize}
    \item Visualization of the distribution of predictions: Figure~\ref{fig:swiss}(a) for Swiss data and Figure~\ref{fig:census}(a) for Census data
    \item Utility and randomization rates for randomizing near the decision boundary: Figure~\ref{fig:swiss}(b)-(d) for Swiss data and Figure~\ref{fig:census}(b)-(d) for Census data
    \item Utility and randomization rates for randomizing using variance: Table~\ref{tab:swiss_variance} for Swiss data and Table~\ref{tab:census_variance} for Census data
    \item Utility and randomization rates for randomizing outliers: Table~\ref{tab:swiss_outliers} for Swiss data and Table~\ref{tab:census_outliers} for Census data
    \item Visualization of the tradeoff between utility and systemic exclusion rate for all randomization methods: Figure~\ref{fig:swiss}(e)-(g) for Swiss data and Figure~\ref{fig:census}(e)-(g) for Census data
    \item Visualization of the density of predictions by point estimate and uncertainty metric: Figure~\ref{fig:risk_scores_vs_uncertainty}
\end{itemize}

\clearpage

\subsection{Pseudocode for Randomization Proposals}

\begin{algorithm}[b!]
  \caption{Partial BF Lottery}
  \label{alg:partial_bf}
  \begin{algorithmic}[1]
    \OUTPUT selected people
    \INPUT people, claims, $k$, $n$, $k'$, $n'$
    \REQUIRE $k' \in (0, k]$, $n' \in (k', n-k+k']$
    \STATE people, claims $\gets$ \textsc{Sort}(people, claims)
    \STATE deterministic selections $\gets$ people[ : $k-k'$]
    \STATE random selections $\gets$ \textsc{Iterative Weighted Selection}($k'$, people[$k-k'$ : $k-k'+n'$], claims[$k-k'$ : $k-k'+n'$])
    \STATE \textbf{return} deterministic selections + random selections
    \end{algorithmic}
\end{algorithm}

\begin{algorithm}[b!]
\caption{Randomization Using Variance}
\label{alg:variance}
  \begin{algorithmic}[1]
    \OUTPUT selected people
    \INPUT people, claims, $k$, $n$, $B$
    \STATE people, claims $\gets$ \textsc{Sort}(people, claims)
    \STATE $\hat{y}_B$ $:=$ list()
    \FOR{i $\in$ 1...$n$}
        \STATE vote $:=$ 0
        \FOR{1...$B$}
            \IF{\textsc{Bootstrapped Claim}(people[i]) $>$ claims[$k$]}
                \STATE vote $\gets$ vote + 1
            \ENDIF
        \ENDFOR
        \STATE $\hat{y}_B$.append(vote / $B$)
    \ENDFOR
    \STATE 
    \STATE deterministic selections $\gets$ people[i] \textbf{for} i $\in$ 1...$n$ \textbf{if} $\hat{y}_B$[i] is 1 
    \STATE uncertain people $\gets$  people[i] \textbf{for} i $\in$ 1...$n$ \textbf{if} $\hat{y}_B$[i] $\in$ (0,1)
    \STATE uncertain claims $\gets$  claims[i] \textbf{for} i $\in$ 1...$n$ \textbf{if} $\hat{y}_B$[i] $\in$ (0,1)
    \STATE $k' \gets k-$ \textsc{Length}(deterministic selections)
    \STATE random selections $\gets$ \textsc{Iterative Weighted Selection}($k'$, uncertain people, uncertain claims)
    \STATE \textbf{return} deterministic selections + random selections
    \end{algorithmic}
\end{algorithm}

\begin{algorithm}[b!]
\caption{Randomization Using Outliers}
\label{alg:variance}
  \begin{algorithmic}[1]
    \OUTPUT selected people
    \INPUT people, claims, $k$, $n$, $\alpha$
    \STATE people, claims $\gets$ \textsc{Sort}(people, claims)
    \STATE deterministic selections $\gets$ people[i] \textbf{for} i $\in$ 1...$n$ \textbf{if} \textsc{Outlier P-Value}(people[i]) $>\alpha$ 
    \STATE uncertain people $\gets$  people[i] \textbf{for} i $\in$ 1...$n$ \textbf{if} \textsc{Outlier P-Value}(people[i]) $\leq\alpha$ 
    \STATE uncertain claims $\gets$  claims[i] \textbf{for} i $\in$ 1...$n$ \textbf{if} \textsc{Outlier P-Value}(people[i]) $\leq\alpha$ 
    \STATE $k' \gets k-$ \textsc{Length}(deterministic selections)
    \STATE random selections $\gets$ \textsc{Iterative Weighted Selection}($k'$, uncertain people, uncertain claims)
    \STATE \textbf{return} deterministic selections + random selections
    \end{algorithmic}
\end{algorithm}

Notes:
\begin{itemize}[topsep=0ex,itemsep=-0.5ex]
    \item The \textsc{Iterative Weighted Selection} refers to Ex~\ref{ex:bf_lottery} and can be performed using numpy's \textit{random.choice} method.
    \item $k'$ denotes how many resources (out of $k$) are randomized, where the randomization occurs over $n'$ people (out of $n$).
    \item Array indexing is zero-based. \textsc{Sort} orders the arrays in descending order from the strongest to weakest claim.
    \item We use $B=11$ for randomization using variance, and train bootstrapped models on a 50\% subset of the training data. 
    \item We provide additional details on how we compute the \textsc{Outlier P-Value} in the next section.
\end{itemize}

\subsection{Conformal Prediction Methodology}\label{app:conformal_prediction}

We use conformal prediction to assign a confidence measure to outlier detection methods~\citep{angelopoulos2021conformal}. Specifically, we use the following procedure:
\begin{enumerate}[topsep=0ex,itemsep=0ex]
    \item Suppose we have a novelty score $s: \X \rightarrow \mathbb{R}$, where larger values indicate more abnormality from the training data. For example, we use average Euclidean distance to the training data. 
    \item We want to find $q: \mathbb{P}(s(x) > q) \leq \alpha$ if $x \sim \X_{\text{train}}$, where $\alpha$ represents the bound on the false positive rate
    \item Reserve a calibration dataset $\X_{\text{cal}}$. For each $x_{\text{cal}}^{j} \in \X_{\text{cal}}$, compute the novelty score $s(x_{\text{cal}}^{j})$ with respect to $\X_{\text{train}}$.
    \item Compute $\hat{q}=\text{quantile}\left(s_{\text{cal}}^{1}\ldots s_{\text{cal}}^{n};\frac{\lceil (n_{\text{cal}}+1)(1-\alpha) \rceil}{n_{\text{cal}}}\right)$
    \item If $s(x_i) > \hat{q}$, then we consider individual $i$ to be an outlier.
    \item Specifically, we take the p-value associated with outlier detection to be: $\frac{1}{n_{\text{cal}+1}}\cdot (1 + \sum_{i=1}^{n_{\text{cal}}} \mathds{1}\{s(x_i) \leq s_{\text{cal}}^{i} \})$
\end{enumerate}


\clearpage

\clearpage

\begin{table*}[t!]
\small
\centering
\caption{Swiss Unemployment Data -- Randomizing Using Variance}
\begin{tabular}{cccccccc} 
\toprule
\multirow{3}{*}{$k/n$} & \multirow{3}{*}{Model} & \multicolumn{2}{c}{Random Rate} & \multicolumn{3}{c}{Utility} \\
\cmidrule(lr){3-4} \cmidrule(lr){5-7}
& & \multirow{2}{*}{$k'/k$} & \multirow{2}{*}{$n'/n$} & \multirow{2}{*}{Variance} & Decision- & \multirow{2}{*}{Top $k$}\\
& & & & & Boundary & \\
\toprule
\multirow{3}{0.75cm}{0.10} & Log. Regression  & 25.7\% & 5.0\% & 69.2\% & 69.2\% & 69.3\% \\
& Random Forest  & 57.1\% & 9.2\% & 70.8\% & 69.9\% & 71.3\% \\
& Decision Tree & 97.4\% & 12.3\% & 64.6\% & 68.3\% & 69.5\% \\
\midrule
\multirow{3}{0.75cm}{0.25} & Log. Regression &  14.0\% & 6.8\% & 62.9\% & 62.8\% & 63.1\% \\
& Random Forest &  32.2\% & 15.0\% & 64.1\% & 63.7\% & 64.3\% \\
& Decision Tree & 73.7\% & 39.0\% & 61.5\% & 58.9\% & 62.9\% \\
\midrule
\multirow{3}{0.75cm}{0.50} & Log. Regression & 7.2\% & 7.0\% & 55.7\% & 55.7\% & 55.8\% \\
& Random Forest  & 19.7\% & 20.1\% & 56.3\% & 56.0\% & 56.5\% \\
& Decision Tree & 50.3\% & 57.6\% & 54.2\% & 52.5\% & 55.5\% \\
\bottomrule
\end{tabular} 
\label{tab:swiss_variance}
\end{table*}

\begin{table*}[t!]
\small
\centering
\caption{Swiss Unemployment Data -- Randomizing Outliers}
\begin{tabular}{ccccccccc} 
\toprule
\multirow{3}{*}{$\alpha$} & \multirow{3}{*}{$k/n$} & \multirow{3}{*}{Model} & \multicolumn{2}{c}{Random Rate} & \multicolumn{3}{c}{Utility} \\
\cmidrule(lr){4-5} \cmidrule(lr){6-8}
& & & \multirow{2}{*}{$k'/k$} & \multirow{2}{*}{$n'/n$} & \multirow{2}{*}{Outliers} & Decision- & \multirow{2}{*}{Top $k$}\\
& & & & & & Boundary & \\
\toprule
\multirow{3}{0.75cm}{0.20} & \multirow{3}{0.75cm}{0.10} & Log. Regression  & 0.5\% & 20.1\% & 69.1\% & 69.3\% & 69.3\% \\
& & Random Forest & 0.4\% & 20.1\% & 71.1\% & 71.2\% & 71.3\% \\
& & Decision Tree & 1.6\% & 20.1\% & 69.1\% & 69.4\% & 69.5\% \\
\midrule
\multirow{3}{0.75cm}{0.20} & \multirow{3}{0.75cm}{0.25} & Log. Regression & 1.2\% & 20.1\% & 62.7\% & 63.0\% & 63.1\% \\ 
& & Random Forest & 1.0\% & 20.1\% & 64.0\% & 64.3\% & 64.4\% \\
& & Decision Tree & 3.0\% & 20.1\% & 62.2\% & 62.8\% & 62.9\% \\
\midrule
\multirow{3}{0.75cm}{0.20} & \multirow{3}{0.75cm}{0.50} & Log. Regression & 3.3\% & 20.1\% & 55.2\% & 55.7\% & 55.8\%\\
& & Random Forest  & 2.8\% & 20.1\% & 55.9\% & 56.4\% & 56.5\%\\
& & Decision Tree & 5.9\% & 20.1\% & 54.8\% & 55.3\% & 55.5\%\\
\midrule
\midrule
\multirow{3}{0.75cm}{0.10} & \multirow{3}{0.75cm}{0.25} & Log. Regression  & 0.3\% & 10.1\% & 62.9\% & 63.1\% & 63.1\%\\
& & Random Forest & 0.3\% & 10.1\% & 64.3\% & 64.4\% & 64.4\%  \\
& & Decision Tree  & 1.2\% & 10.1\% & 62.6\% & 62.9\% & 62.9\% \\
\midrule
\multirow{3}{0.75cm}{0.30} & \multirow{3}{0.75cm}{0.25} & Log. Regression & 4.1\% & 30.0\% & 61.8\% & 62.8\% & 63.1\% \\
& & Random Forest & 3.7\% & 30.0\% & 63.1\% & 64.1\% & 64.4\% \\
& & Decision Tree & 7.2\% & 30.0\% & 61.1\% & 62.5\% & 62.9\% \\
\bottomrule
\end{tabular} 
\label{tab:swiss_outliers}
\end{table*}

\clearpage

\clearpage

\begin{table*}[t!]
\small
\centering
\caption{Census Income Data -- Randomizing Using Variance}
\begin{tabular}{cccccccc} 
\toprule
\multirow{3}{*}{$k/n$} & \multirow{3}{*}{Model} & \multicolumn{2}{c}{Random Rate} & \multicolumn{3}{c}{Utility} \\
\cmidrule(lr){3-4} \cmidrule(lr){5-7}
& & \multirow{2}{*}{$k'/k$} & \multirow{2}{*}{$n'/n$} & \multirow{2}{*}{Variance} & Decision- & \multirow{2}{*}{Top $k$}\\
& & & & & Boundary & \\
\toprule
\multirow{3}{0.75cm}{0.10} & Log. Regression  & 7.2\% & 1.5\% & 91.5\% & 91.5\% & 91.5\% \\
& Random Forest  &  66.9\% & 16.6\% & 90.7\% & 88.6\% & 90.9\%\\
& Decision Tree & 0.0\% & 0.0\% & - & - & 83.3\% \\
\midrule
\multirow{3}{0.75cm}{0.25} & Log. Regression & 3.9\% & 1.9\% & 86.1\% & 86.1\% & 86.1\%  \\
& Random Forest & 48.7\% & 26.4\% & 84.5\% & 81.6\% & 85.2\%  \\
& Decision Tree & 70.6\% & 39.9\% & 81.0\% & 74.2\% & 82.8\% \\
\midrule
\multirow{3}{0.75cm}{0.50} & Log. Regression &  2.3\% & 2.2\% & 72.2\% & 72.1\% & 72.2\% \\
& Random Forest  & 30.0\% & 29.2\% & 70.8\% & 69.4\% & 71.6\% \\
& Decision Tree & 45.9\% & 46.0\% & 67.7\% & 64.5\% & 69.4\% \\
\bottomrule
\end{tabular} 
\label{tab:census_variance}
\end{table*}

\begin{table*}[t!]
\small
\centering
\caption{Census Income Data -- Randomizing Outliers}
\begin{tabular}{ccccccccc} 
\toprule
\multirow{3}{*}{$\alpha$} & \multirow{3}{*}{$k/n$} & \multirow{3}{*}{Model} & \multicolumn{2}{c}{Random Rate} & \multicolumn{3}{c}{Utility} \\
\cmidrule(lr){4-5} \cmidrule(lr){6-8}
& & & \multirow{2}{*}{$k'/k$} & \multirow{2}{*}{$n'/n$} & \multirow{2}{*}{Outliers} & Decision- & \multirow{2}{*}{Top $k$}\\
& & & & & & Boundary & \\
\toprule
\multirow{3}{0.75cm}{0.10} & \multirow{3}{0.75cm}{0.10} & Log. Regression & 10.7\% & 9.9\% & 86.5\% & 91.1\% & 91.5\% \\
& & Random Forest & 3.9\% & 9.9\% & 88.9\% & 90.7\% & 90.9\% \\
& & Decision Tree & 8.5\% & 9.9\% & 79.8\% & 83.2\% & 83.3\% \\
\midrule
\multirow{3}{0.75cm}{0.10} & \multirow{3}{0.75cm}{0.25} & Log. Regression & 9.6\% & 9.9\% & 82.3\% & 85.7\% & 86.1\% \\
& & Random Forest & 7.7\% & 9.9\% & 81.8\% & 84.9\% & 85.2\% \\
& & Decision Tree & 7.5\% & 9.9\% & 79.8\% & 82.0\% & 82.8\% \\

\midrule
\multirow{3}{0.75cm}{0.10} & \multirow{3}{0.75cm}{0.50} & Log. Regression &  9.4\% & 9.9\% & 69.6\% & 71.8\% & 72.2\% \\
& & Random Forest &  9.7\% & 9.9\% & 68.9\% & 71.2\% & 71.6\% \\
& & Decision Tree & 10.3\% & 9.9\% & 67.2\% & 69.1\% & 69.4\% \\
\midrule
\midrule
\multirow{3}{0.75cm}{0.05} & \multirow{3}{0.75cm}{0.25} & Log. Regression  & 5.1\% & 5.0\% & 84.1\% & 86.0\% & 86.1\% \\
& & Random Forest & 3.7\% & 5.0\% & 83.5\% & 85.2\% & 85.2\% \\
& & Decision Tree  & 3.6\% & 5.0\% & 81.3\% & 82.4\% & 82.8\% \\
\midrule
\multirow{3}{0.75cm}{0.20} & \multirow{3}{0.75cm}{0.25} & Log. Regression &  18.6\% & 20.0\% & 78.5\% & 84.4\% & 86.1\% \\
& & Random Forest & 15.6\% & 20.0\% & 78.3\% & 83.8\% & 85.2\%
 \\
& & Decision Tree & 15.1\% & 20.0\% & 76.7\% & 80.8\% & 82.8\% \\
\bottomrule
\end{tabular} 
\label{tab:census_outliers}
\end{table*}

\clearpage

\end{document}